%% file: main.tex
\documentclass[%
 reprint,
superscriptaddress,
 amsmath,amssymb,
 aps,
floatfix,
]{revtex4-2}

\usepackage{graphicx}
\graphicspath{{}}
\usepackage{dcolumn}
\usepackage{bm}
\usepackage{hyperref}
\usepackage{xcolor}

\renewcommand{\figureautorefname}{Fig.}
\renewcommand{\equationautorefname}{Eq.}
\renewcommand{\sectionautorefname}{Sec.}
\renewcommand{\subsectionautorefname}{Sec.}

\newcommand{\Autoref}[1]{\begingroup
  \def\figureautorefname{Figure}%
  \def\sectionautorefname{Section}%
  \def\subsectionautorefname{Section}%
  \def\equationautorefname{Equation}%
  \autoref{#1}%
\endgroup}

\newcommand{\appref}[1]{App.~\ref{#1}}
\usepackage{braket}

\begin{document}

\preprint{APS/123-QED}

\title{Utility-scale quantum experiments using dynamic circuits to address collective dissipation in interacting qubits}
\author{Benjamin Tirado}
\affiliation{Centro de Física de Materiales (CFM-MPC), CSIC-EHU, Donostia-San Sebastián, Spain}
\affiliation{Department of Electricity and Electronics, University of the Basque Country, Leioa, Spain}

\author{Joana Fraxanet}
\affiliation{IBM Quantum, IBM Thomas J. Watson Research Center, Yorktown Heights, NY 10598, USA}

\author{Adrián Juan-Delgado}
\affiliation{Centro de Física de Materiales (CFM-MPC), CSIC-EHU, Donostia-San Sebastián, Spain}
\affiliation{Department of Electricity and Electronics, University of the Basque Country, Leioa, Spain}
\affiliation{Departamento de Física Teórica de la Materia Condensada, Universidad Autónoma de Madrid, Spain.}

\author{Javier Aizpurua}
\affiliation{Department of Electricity and Electronics, University of the Basque Country, Leioa, Spain}
\affiliation{Donostia International Physics Center DIPC, Donostia-San Sebastián, Spain}
\affiliation{IKERBASQUE, Basque Foundation for Science, Bilbao, Spain}

\author{Ruben Esteban}
\affiliation{Centro de Física de Materiales (CFM-MPC), CSIC-EHU, Donostia-San Sebastián, Spain}
\affiliation{Donostia International Physics Center DIPC, Donostia-San Sebastián, Spain}

\begin{abstract}
Open quantum systems are central to quantum optics, condensed matter, and chemistry, yet their simulation remains challenging for both classical and near-term quantum hardware. In this work we implement and execute utility-scale quantum circuits that accurately reproduce the dissipative dynamics of interacting qubits. We consider a one-dimensional chain of many qubits weakly coupled to a common Markovian bath. The Markovian time evolution of the system is implemented through Trotterized evolution with the introduction of ancilla-assisted dissipative channels, including single-qubit and two-qubit dissipators to capture collective decay. Mid-circuit measurements, conditional gates, and hardware-aware transpilation significantly reduce circuit depth. We further implement a biased Clifford data regression (biased CDR), an error mitigation strategy that outperforms the uniform Cliffordization baseline and a variety of zero-noise extrapolation protocols. We execute large-scale quantum experiments of the dynamics of chains comprising up to 86 emitters on the IBM System Two  \texttt{ibm\_basquecountry}. In order to do so, we use 129 total qubits (including ancillas), with the largest circuits contain about 8000 two-qubit gates. To validate these experiments we develop a classical Monte Carlo–Time-Evolving Block-Decimation (MC-TEBD) tensor-network method that incorporates reset operations through stochastic pure-state trajectories, obtaining very good agreement. The approach presented here opens a practical route for utility-scale quantum simulation of dissipative dynamics, enabled by dynamic circuits, targeted error mitigation, and tensor-network validation, and enables to tackle complex dynamics of systems such as quantum emitters in dissipative optical cavities.
\end{abstract}

\maketitle


\section{\label{sec:level1} INTRODUCTION}
The simulation of physical systems is widely regarded as a very promising near-term application of quantum computers \cite{Feynman1982, Lloyd1996}. For example, significant progress has been made during the past years in the simulation of unitary many-body dynamics, with algorithms relying on the Trotter-Suzuki decomposition \cite{trotter1959, suzuki1976} or other hardware-adapted constructions tailored to specific platforms \cite{kandala2017, barends2015, martinez2016}. These approaches have enabled the simulation of closed-system dynamics in spin models \cite{Switzer2026, kim2023, Shtanko2025, Fischer2026}, lattice gauge theories \cite{Cobos2025, Farrell2024, Xu2026, Gonzalez2025, Cochran2025}, and fermionic systems \cite{arute2020, Hartnett2026, Alam2025, Alam2025_2, Shtanko2025}, on near-term devices.

In contrast, the simulation of open quantum systems \cite{breuer2002} remains challenging on near-term quantum devices, where environmental coupling must be introduced as dissipative processes that coexist with coherent evolution. Environment-induced dissipative processes are unavoidable in physical systems and can play a key role in many areas such as quantum optics \cite{scully1997quantum}, condensed matter \cite{weiss2012quantum} and quantum chemistry \cite{Akihito2009}, leading to important phenomena such as dephasing, relaxation or collective emission \cite{Zurek2001, Dicke1954}. From a theoretical perspective, many physically relevant open-system dynamics are well approximated by Markovian processes, i.e., dynamics in which the environment acts as a memory-less or Markovian bath \cite{Rivas2014}. These dynamics are usually described by Markovian master equations \cite{lindblad1976, gorini1976}. However, solving these equations exactly using classical brute-force methods requires storing and evolving the full density matrix, whose number of entries scales exponentially with the number of subsystems \cite{Nielsen_Chuang_2010,breuer2002}. This exponential growth in computational resources highlights the potential advantage of quantum computation.

Several quantum algorithms have been proposed to simulate dissipative systems on quantum computers, including stochastic unravelings \cite{daley2014}, variational methods \cite{yuan2019} and algorithms based on linear combinations of quantum channels \cite{childs2019}. However, these methods often rely on deep circuits and large ancilla registers that place them beyond the reach of current hardware. This limitation has sparked interest in approaches tailored to near-term quantum devices, where circuit depth must be minimized, which requires hardware-aware circuit design, as well as the choice of a mapping strategy.

\begin{figure*}[!htbp]
    \def\svgwidth{\textwidth}
    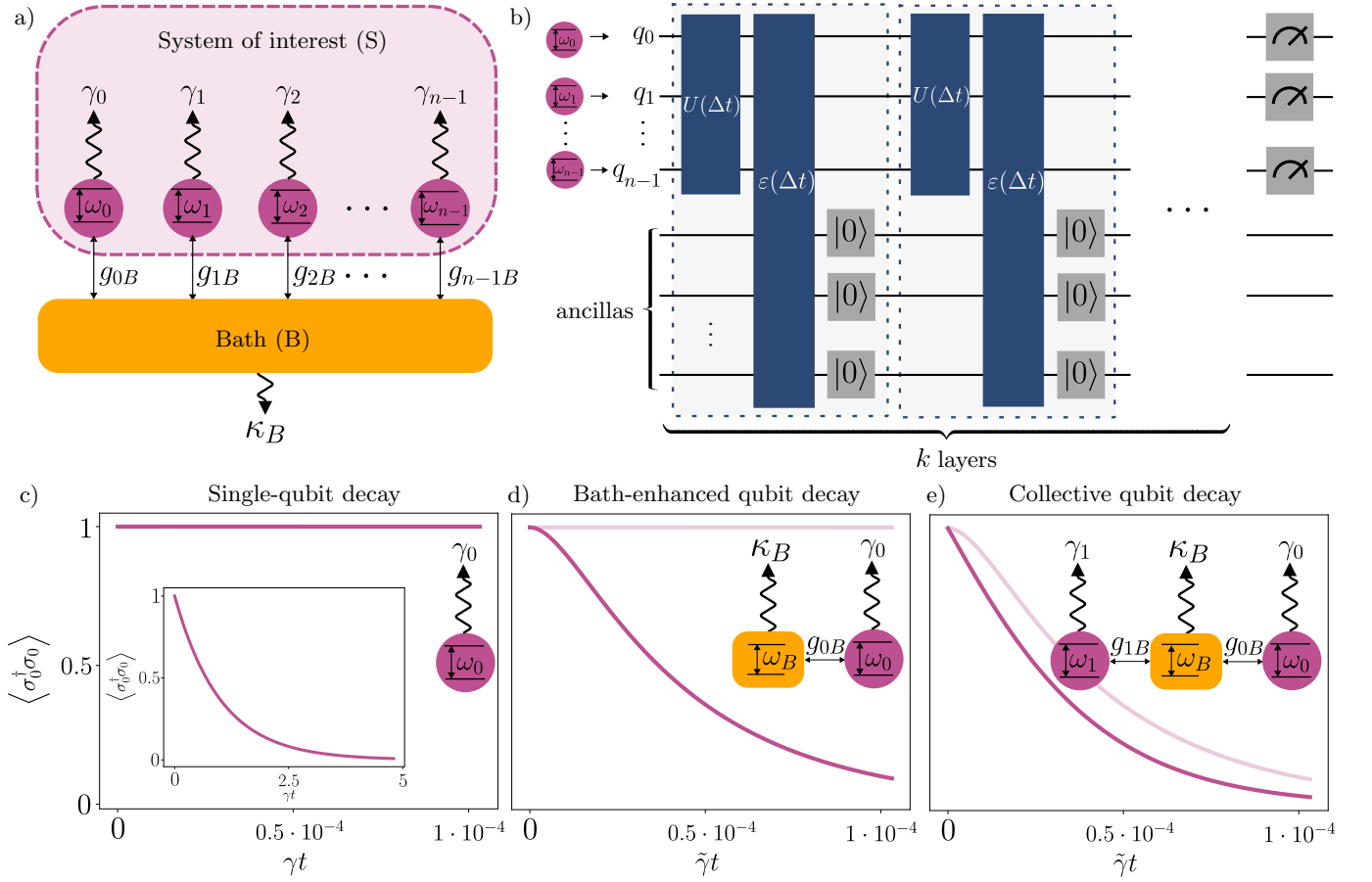
    \caption{Dissipative phenomena in systems of qubits coupled to a common external bath and modeling of the dynamics using quantum circuits. (a) Schematic of the coupling between the system of interest (S) and the external bath (B). The system of interest consists of $n$ qubits, depicted as purple circles and modeled as two-level systems with transition frequencies $\omega_i$ and spontaneous decay rates $\gamma_i$ ($i = 0, 1, ..., n-1$). Each of the qubits is coupled with strength $g_{iB}$ ($i = 0, 1, ..., n-1$) to a common external bath, depicted in yellow, which has a single mode of frequency $\omega_B$ and a dissipation rate $\kappa_B$. (b) Schematic of the Trotterized quantum circuit modelling Markovian time evolution under dissipation. The state of one emitter is encoded using one system qubit under the direct mapping: emitter $0 \to q_0$, emitter $1 \to q_1$, ..., emitter $n-1 \to q_{n-1}$. Additional ancilla qubits are used to model the dissipative part of the evolution. On each of the $k$ steps of the simulation, the system qubits evolve under $U(\Delta t) \approx e^{-i H \Delta t}$ (small blue box) for a short time step $\Delta t = t/k$, with $H$ the system Hamiltonian representing the coherent part of the evolution. Next, we apply a block of operations $\varepsilon (\Delta t)$ (tall blue box) that represents the dissipative decay process of the system qubits, and that is implemented with the ancilla qubits. The ancillas are then reset before starting the next step. (c,d) Time evolution of the population of the excited state of one qubit (c), $\sigma^{\dagger}_0 \sigma_0$, that only decays through spontaneous emission and (d) coupled to an external bath that increases the decay rate. (e) Time evolution of the population of the excited state of one qubit that is coupled through a shared bath to another qubit. The initial states considered are $\ket{\psi (0)} = \ket{1}_S \otimes \ket{0}_B$ for panels c and d and  $\ket{\psi (0)} = \ket{11}_S \otimes \ket{0}_B$ (corresponding to a superradiant state) for panel e. The values of the parameters used are $\omega_0=\omega_1 = 1.2$eV, $g_{0B} = g_{1B} = 0.03$eV, $\gamma_0 = \gamma_1 = 0.8\mu$eV, $\omega_B = 1.1$eV, $\kappa_B = 0.2$eV (from these values, one can obtain the corresponding effective variables $\tilde{\omega}_i$, $\tilde{g}_{i, i+1}$, $\tilde{\gamma}_i$, $\tilde{\gamma}_{i, i+1}$ using the Markovian approximation).}
    \label{fig:presentation_fig}
\end{figure*}

In this work, we investigate the use of quantum circuits to address the collective dissipative dynamics of chains of identical qubits coupled to Markovian baths. We design efficient, low-depth dynamic quantum circuits compatible with near-term hardware, and demonstrate the viability of these circuits in quantum experiments. For concreteness we consider the dynamics of two-level quantum emitters coupled to optical cavities that act as a bath (\autoref{fig:presentation_fig}a), as a representative example of the time evolution of open quantum systems. We emphasize that our study includes complex collective dissipative effects responsible for canonical physical phenomena such as superradiance and subradiance. Therefore, the simulation of such systems requires modeling correlated dissipative channels acting across more than one qubit. In particular, we present circuit constructions that implement both single-qubit dissipators and, more importantly, cross-qubit nearest-neighbor dissipators responsible for collective dissipation.

A key conceptual aspect of our approach is to map local dissipative processes onto quantum channels implemented using small ancilla registers and Trotter-inspired evolution (\autoref{fig:presentation_fig}b) \cite{Nielsen_Chuang_2010, DelRe2020, Cattaneo2023}. By working in an appropriate basis, the collective dissipation can be decomposed into a number of simpler decay channels that admit relatively compact circuit representations. To optimize the circuits for near-term devices, we use dynamic-circuit techniques and tailored transpilation settings that ensure constant two-qubit depth scaling with the system size, even after transpilation to a particular device. This approach relies on recent advances in quantum hardware —such as mid-circuit measurements and conditional gate execution— which enable a reliable implementation of dynamic circuits at utility scale \cite{Buhrman_2024, PRXQuantum.5.030339, Cao2025}. Further, we combine error suppression and error mitigation techniques to improve the accuracy of the results obtained from the execution of quantum experiments in quantum devices. Concretely, we consider dynamical decoupling \cite{Viola_1998}, zero-noise extrapolation \cite{Temme_2017, Giurgica2020} and Clifford Data regression \cite{Czarnik2021}. We demonstrate dissipative dynamics in quantum experiments on IBM quantum computers comprising up to 86 emitters and 129 total qubits, corresponding to system sizes well beyond classical exact brute-force capabilities. Last, to validate the quantum experiments, we develop a hybrid procedure that combines tensor networks and Monte Carlo techniques \cite{Sander2025}. This approach allows us to model the losses of the system, which is challenging using the standard time-evolving block decimation algorithm based on tensor networks \cite{Vidal2004, Schollwoeck2011}.

The structure of the paper is as follows. \Autoref{sec:level2} introduces the dissipative model and the circuit construction used to implement the Markovian time evolution of the system, with emphasis on the modeling of collective nearest-neighbor decay. \Autoref{sec:level3} describes the transformation of these circuits into hardware-efficient dynamic circuits and the transpilation strategy used to preserve a constant two-qubit depth scaling after transpilation. \Autoref{sec:level4} presents a variety of experimental executions on IBM quantum processors, the comparison of error-suppression and error-mitigation methods, and the validation of large-scale experiments using the Monte Carlo tensor-network approach. \Autoref{sec:level5} concludes with a summary of the main contributions.

\section{\label{sec:level2} QUANTUM CIRCUITS IMPLEMENTATION}

We consider the system shown in \autoref{fig:presentation_fig}a. The total system consists of a quantum system, labeled $S$, composed of many qubits, each of which behaves as a two-level system, coupled to an external Markovian bath (initially in the ground state), labeled $B$. In the weak coupling limit, the system-bath interaction is small compared to the intrinsic losses of the total system and the bath can be effectively removed by means of the Markovian approximation. Neglecting long-range interaction terms beyond nearest-neighbor coupling leads to the following effective Markovian master equation for the reduced density operator of the system $\rho$, after tracing out the degrees of freedom of the bath \cite{breuer2002},
\begin{align}
\frac{d\rho}{dt} = -i [\tilde{H}, \rho] +\sum_{i=0}^{n-1} \tilde{\gamma}_i\mathcal{D} (\sigma_i) [\rho] + \nonumber \\
+ \sum_{i=0}^{n-2} \tilde{\gamma}_{i,i+1}\mathcal{D} (\sigma_i, \sigma_{i+1}) [\rho] + \sum_{i=0}^{n-2} \tilde{\gamma}_{i+1,i}\mathcal{D} (\sigma_{i+1}, \sigma_{i}) [\rho].
\label{eqn:gksl_equation}
\end{align}

Throughout this work, we set $\hbar = 1$. $\tilde{H}$ is the (effective) system Hamiltonian, 

\begin{equation}
\tilde{H} =  \sum_{i=0}^{n-1} \tilde{\omega}_i \sigma^{\dagger}_i \sigma_i + \sum_{i=0}^{n-2}  \tilde{g}_{i,i+1} \left(\sigma^{\dagger}_i\sigma_{i+1} + \sigma_i\sigma_{i+1}^{\dagger} \right),
\label{eqn:n_emitters_hamiltonian}
\end{equation}
where $\tilde{\omega}_i$ is the effective frequency of the qubit $i$, shifted under the presence of the bath, $\tilde{g}_{i,i+1}$ is the effective coupling between qubits $i$ and $i+1$, arising from their coupling to the same bath, and  $\sigma_i = \ket{0}_i \bra{1}_i$ is the Pauli lowering operator acting on qubit $i$.  The dissipators $\mathcal{D}$ are defined as 

\begin{equation}
 \tilde{\gamma}_{i,j} \mathcal{D} (\sigma_i, \sigma_j) [\rho]  = \tilde{\gamma}_{i,j} \left( \sigma_i \rho \sigma^{\dagger}_j - \frac{1}{2} \left\lbrace \sigma^{\dagger}_j \sigma_i, \rho \right\rbrace\right),
\label{eqn:n_emitters_single_body_dissipators}
\end{equation}
where $\tilde{\gamma}_{i,i} \equiv \tilde{\gamma}_i$ is the effective decay rate of the qubit $i$ and $\tilde{\gamma}_{i,i+1} = \tilde{\gamma}_{i+1,i}$ are the cross-decay rates associated with collective decay processes between adjacent qubits. Throughout this work, we use $\mathcal{D} (\sigma_i) \equiv \mathcal{D} (\sigma_i, \sigma_i)$.

The evolution of the density operator $\rho$ can be obtained using quantum circuits by splitting the coherent evolution, represented by $U(t) \approx e^{-i \tilde{H} t}$, and the dissipative evolution, represented by $\varepsilon (t) \approx e^{\mathcal{D} t}$ using ancilla qubits. This structure is repeated $k$ times over small time slices $t/k$, with $k$ the number of Trotter steps, to approximately reproduce the time dynamics. This technique is depicted schematically in \autoref{fig:presentation_fig}b and discussed in more detail over the next sections.

We sketch in panels c, d and e of \autoref{fig:presentation_fig} the different dissipative phenomena that can occur in the systems described by \autoref{eqn:gksl_equation}. Concretely, the decay of a single emitter induced by dissipative processes (\autoref{fig:presentation_fig}c and inset) can be enhanced under the presence of the electromagnetic environment (\autoref{fig:presentation_fig}d). When more than one emitter couple to the same electromagnetic environment (\autoref{fig:presentation_fig}e), more complex dissipative effects like superradiance and subradiance —collective phenomena where spontaneous decay is enhanced or suppressed, respectively \cite{Lehmberg1970, Gross1982}— arise. 

This model can be applied to describe a variety of physically-relevant open quantum system dynamics. The qubits can represent, for example, quantum emitters such as molecules or quantum dots. For specificity, we center the discussion on a one-dimensional chain of quantum emitters weakly coupled through the same electromagnetic environment, constituted by an array of optical cavities (or a photonic waveguide) that mediate photon exchange between neighboring emitters, while ignoring long-range interactions for simplicity.  


\subsection{\label{sec:level2a} Single-qubit dissipators}
\begin{figure}[!htbp]
    \centering
     \def\svgwidth{\columnwidth}
    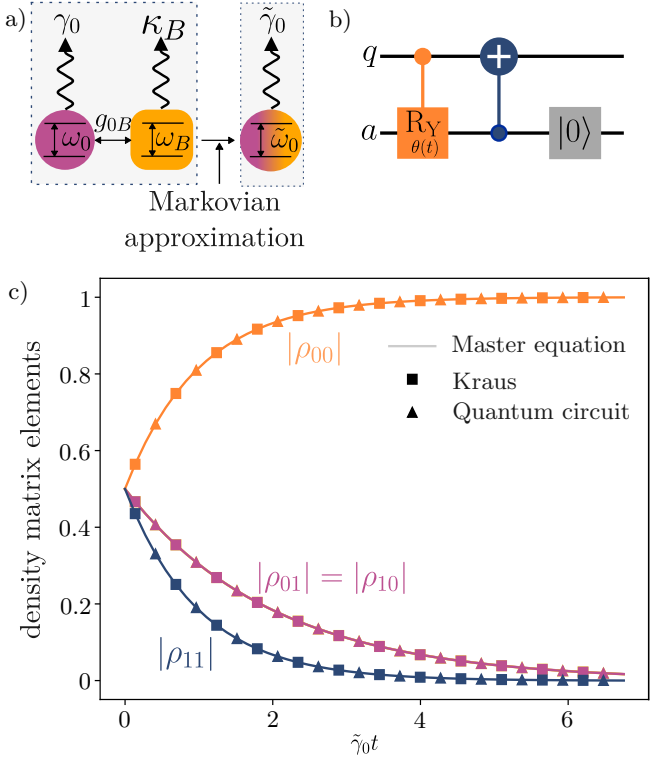
    \caption{Modeling of the dissipative dynamics of a qubit. (a) Schematic of the effective removal of the cavity through the Markovian approximation. The quantum emitter (purple sphere) is considered to be a two-level system with transition frequency $\omega_0$ and spontaneous decay rate $\gamma_0$ and the cavity (blue sphere) has a single mode of frequency $\omega_A$ and a dissipation rate of $\kappa_A$. Both the quantum emitter and the cavity are originally coupled with coupling strength $g_{0B}$. After removing the cavity, the properties of the emitter (now pictured with a color gradient) are modified, resulting in a new effective frequency $\tilde{\omega}_0$ and effective decay rate $\tilde{\gamma}_0$. (b) Time-parametric quantum circuit representing one Trotter step of the exclusively-dissipative dynamics of the single quantum emitter after effectively removing the cavity and moving to the rotating frame. The system qubit is labeled $q$ and the ancilla qubit is labeled $a$. The rotation angle $\theta(t)$ in the $CRy$ gate is connected to the emitter effective decay rate as $\theta(t) = 2 \arcsin \left( \sqrt{1 - e^{-\tilde{\gamma}_0 t}} \right)$. The ancilla is reset before starting another Trotter step. (c) Comparison of the norm of each of the density matrix elements —$|\rho_{00}|$ (red), $|\rho_{11}|$ (purple) and $|\rho_{01}| = |\rho_{10}^{*}|$ (blue)— under the dissipative master equation solved in \autoref{eqn:single_qubit_ME_evo} (solid lines), Kraus operators representing the decay channel in \autoref{eqn:kraus_decomposition} (squares) and a (classical) noiseless simulation of one Trotter step of the quantum circuit in \autoref{fig:one_emitter_model}b (triangles). The initial state is $\ket{\psi (0)}_S = \frac{1}{\sqrt{2}} \left( \ket{0}_S + \ket{1}_S \right)$ and the values of the effective parameters used are $\tilde{\omega}_0 = 1.2045$eV, $\tilde{\gamma}_0 = 9$meV.}
    \label{fig:one_emitter_model}
\end{figure}

We begin by analyzing a single two-level quantum emitter (qubit) coupled to a lossy single-mode optical cavity (external bath) (\autoref{fig:one_emitter_model}a) \cite{Shore1993}. After performing the Markovian approximation and effectively removing the cavity, the system Hamiltonian is simply $\tilde{H} = \tilde{\omega}_0 \sigma^{\dagger} \sigma$, with $\tilde{\omega}_0$ the effective frequency of the quantum emitter. The only dissipator appearing in \autoref{eqn:gksl_equation} is $\tilde{\gamma}_0 \mathcal{D} (\sigma_0)$, with $\tilde{\gamma}_0$ the effective decay rate of the quantum emitter, boosted from its original decay rate due to the presence of the cavity \cite{Purcell1946}. 

In the rotating frame at frequency $\tilde{\omega}_0$, the single-qubit Hamiltonian $\tilde{H}$ can be eliminated, yielding purely-dissipative dynamics. The evolution of the  reduced density operator can then be expressed in matrix form in the computational basis $\left\lbrace \ket{0}, \ket{1} \right\rbrace$, which represents the ground and excited states $\left\lbrace \ket{g}, \ket{e} \right\rbrace$ of the quantum emitter,
\begin{align}
\rho(t)  = \begin{pmatrix}
\rho_{00}(0) + (1 - e^{-\tilde{\gamma}_0 t} )\rho_{11}(0) & e^{-\frac{\tilde{\gamma}_0}{2} t}  \rho_{01}(0) \\
e^{-\frac{\tilde{\gamma}_0}{2} t}  \rho_{10}(0)  & e^{-\tilde{\gamma}_0 t} \rho_{11}(0) 
\end{pmatrix},
\label{eqn:single_qubit_ME_evo}
\end{align} 
where $\rho_{ij}$ ($i,j = 0,1$) are the matrix elements of the reduced density operator in the computational basis. This evolution can be achieved with the circuit in \autoref{fig:one_emitter_model}b, as we show next. First, we represent the circuit in \autoref{fig:one_emitter_model}b through an amplitude damping channel (ADC) \cite{Nielsen_Chuang_2010} $\varepsilon: \mathcal{B}(\mathcal{H_S}) \to \mathcal{B}(\mathcal{H_S})$, where $\mathcal{H}_S$ denotes the Hilbert space of the system of interest $S$, i.e., the quantum emitter, and $\mathcal{B}(\mathcal{H}_S)$ denotes the space of bounded linear operators on $\mathcal{H}_S$. The ADC is a completely-positive trace-preserving (CPTP) map \cite{Choi1975} that can be defined in terms of the transformation equations that it induces on the computational basis states of some dilated Hilbert space $\mathcal{H}_S \otimes \mathcal{H}_{A}$  \cite{Stinespring1955}
\begin{align}
\left| 0 \right>_S \otimes \left|0 \right>_{A} &\to \left| 0 \right>_S \otimes \left|0 \right>_{A}, \nonumber \\
\left| 1 \right>_S \otimes \left|0 \right>_{A} &\to \sqrt{1- p} \left| 1 \right>_S \otimes \left|0 \right>_{A} +  \sqrt{p} \left| 0 \right>_S \otimes \left|1 \right>_{A},
\label{eqn:adc_transformation_equations}
\end{align}
where $\mathcal{H}_{A}$ is an auxiliary subsystem and $p$ is the decay probability. In the circuit in \autoref{fig:one_emitter_model}b, $S$ is represented by a system qubit labeled $q$, and $A$ by an ancilla qubit, labeled $a$. The equivalence between \autoref{eqn:adc_transformation_equations} and the circuit in \autoref{fig:one_emitter_model}b can be verified by inserting the different input states into the latter and identifying $p = \sin^2 \theta$. 

\par Next, we connect the ADC in \autoref{eqn:adc_transformation_equations} with the evolution of the reduced density operator in \autoref{eqn:single_qubit_ME_evo}. To this purpose, we define the isometry $V:\mathcal{H}_S \to \mathcal{H}_S \otimes \mathcal{H}_{A}$, based on the set of transformation equations in \autoref{eqn:adc_transformation_equations}, as \cite{Wilde_2017},
\begin{align}
V \left| 0 \right>_S &= \left|0 \right>_S \otimes \left|0 \right>_{A}, \nonumber \\
V \left| 1 \right>_S &= \sqrt{1- p} \left| 1 \right>_S \otimes \left|0 \right>_{A} +  \sqrt{p} \left| 0 \right>_S \otimes \left|1 \right>_{A}.
\end{align}
$V$ admits a matrix representation in the computational basis $\left\lbrace \ket{0}, \ket{1} \right\rbrace \in \mathcal{H}_S$ and $\left\lbrace \ket{00}, \ket{01}, \ket{10}, \ket{11}\right\rbrace \in \mathcal{H}_S \otimes \mathcal{H}_{A}$ of the input and output space, respectively,
\begin{equation}
V =
\begin{pmatrix} 
1 & 0 \\
0 & \sqrt{p} \\
0 & \sqrt{1-p} \\
0 & 0 
\end{pmatrix}.
\end{equation}

This isometry $V$ can be written as $V = \sum_{j=0}^{1} K_j \otimes \ket{e_j}_{A}$, where  $\left\lbrace \ket{e_j} \right\rbrace_{j=0}^{1} = \left\lbrace \ket{0}, \ket{1} \right\rbrace$ is the computational basis in $\mathcal{H}_{A}$ and $\left\lbrace K_i \right\rbrace$ are the Kraus operators representing the channel $\varepsilon$ \cite{Kraus1983}. From the definition of the isometry it follows that $K_j = (\mathbf{1}_S \otimes \bra{e_j}_A)V$ ($j=0,1$), with $\mathbf{1}_S$ the identity operator in $\mathcal{H}_S$, which gives 

\begin{equation}
K_0 = \begin{pmatrix}
1 & 0 \\
0 & \sqrt{1-p}
\end{pmatrix}, \hspace{20mm}
K_1 = \begin{pmatrix}
0 & \sqrt{p} \\
0 & 0
\end{pmatrix}.
\label{eqn:kraus_operators_adc}
\end{equation}
Since $V$ is an isometry, $V^{\dagger}V = \mathbf{1}_S$, and the Kraus operators verify the completeness relation $\sum_j K_j^{\dagger}K_j = \mathbf{1}_S$. \par According to Ref. \cite{Nielsen_Chuang_2010}, we can use the Kraus decomposition of the channel $\varepsilon$ to obtain the evolution of the reduced density operator  
\begin{align}
\rho(t) = \varepsilon[\rho (0)] = \sum_{i=0}^{1}  K_i \rho(0) K_i^{\dagger} = \nonumber \\
= \begin{pmatrix}
\rho_{00} (0) + p \rho_{11}(0) & \sqrt{1-p} \rho_{01}(0) \\
\sqrt{1-p} \rho_{10} (0) & (1-p) \rho_{11} (0)
\end{pmatrix}.
\label{eqn:kraus_decomposition}
\end{align}

This evolution coincides with the solution of the master equations (\autoref{eqn:gksl_equation} and \autoref{eqn:single_qubit_ME_evo}) under the single-qubit dissipator with $p = 1 - e^{- \tilde{\gamma}_0  t}$, which confirms the equivalence of these approaches. To further illustrate this equivalence, we show in \autoref{fig:one_emitter_model}c the evolution of the absolute value of the density matrix elements obtained from the direct solution of the  dissipative master equation in \autoref{eqn:gksl_equation} (solid lines), from the Kraus representation of the channel $\varepsilon$ in \autoref{eqn:kraus_decomposition} (squares) and from a (classical) noiseless simulation of the quantum circuit in \autoref{fig:one_emitter_model}b (triangles) for the initial state $\ket{\psi (0)}_S = \frac{1}{\sqrt{2}} (\ket{0}_S + \ket{1}_S)$. The noiseless simulation of the quantum circuit involves $k  =1$ Trotter step (for a single emitter there is no Trotter error) and $N_{shots} = 2 \cdot 10^{4}$ shots. All results are identical (up to shot noise introduced by reset operations), as expected, and show a clear decay of the excited-state population (term $|\rho_{11}|$, blue) to the ground state (term $|\rho_{00}|$, yellow), as well as the decay of the coherences (term $|\rho_{01}| = |\rho_{10}|$, purple). 


\subsection{\label{sec:level2b} Two-qubit collective dissipators}
\label{two_qubit_disspators}

\begin{figure}[!htbp]
    \centering
     \def\svgwidth{\columnwidth}
    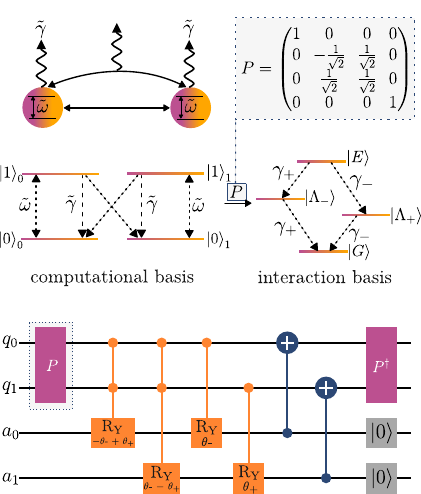
    \caption{Modeling of the collective dissipative dynamics of two qubits (representing two quantum emitters) coupled through a common environment (an optical cavity). (a) Schematic of the system after performing the effective removal of the cavity through the Markovian approximation, resulting in two interacting  two-level systems labeled $q_0$ and $q_1$ and pictured as color-graded spheres. The Markovian approximation results in the modification of the properties of the quantum emitters (with new transition frequency $\tilde{\omega}$ and decay rate $\tilde{\gamma}$), and also in the emergence of a new effective coupling $\tilde{g}_{0,1}$ and a cross-dissipation term $\tilde{\gamma}_{0,1}$. (b) Energy level schemes of the computational (left) and interaction (right) basis. The latter has been obtained after diagonalizing the two-qubit Hamiltonian $\tilde{H} = \tilde{\omega} \sum_{i=0}^1 \sigma_i^{\dagger} \sigma_i + \tilde{g}_{0,1} \left(\sigma_0^{\dagger} \sigma_1 + \sigma_0 \sigma_1^{\dagger} \right)$. The decay rates on the interaction basis $\gamma_{\pm}$ obey the relationship $\gamma_{\pm} \equiv \tilde{\gamma} \pm \tilde{\gamma}_{0,1}$. The basis change matrix $P$ that accomplishes the transformation from the computational to the interaction basis is shown in the top-right inset. (c) Quantum circuit implementing one Trotter step of the dissipative part of the dynamics of the system in (a). The two two-level systems that represent the quantum emitters correspond to qubits $q_0$ and $q_1$, while the ancilla register is represented by qubits $a_0$ and $a_1$. The blocks labeled by $P$ and $P^{\dagger}$ represent the unitary operation necessary to perform the basis change in \autoref{eqn:basis_transformation} (shown in matrix form in the top-right inset in \autoref{fig:two_emitters_model}b) and its inverse, respectively. The dissipation is implemented through the ancilla qubits, the two double-controlled $R_y$ gates ($CCR_y$), the two $CR_y$ gates and the two $CX$ gates, as discussed in the main text.  After the dissipation channel, the basis change is undone through the $P^{\dagger}$ gate. For more than one Trotter step, both ancillas need to be reset before starting another step.}
    \label{fig:two_emitters_model}
\end{figure}

We now extend our previous example to two quantum emitters, which are assumed to be identical for simplicity, coupled to a common cavity mode as the dissipative bath (see \autoref{fig:two_emitters_model}a). Assuming again weak coupling between the emitters and the cavity, we perform the Markovian approximation to effectively remove the cavity. This approximation results in an effective master equation for the reduced density matrix of the two quantum emitters $\rho$ of similar form as in \autoref{eqn:gksl_equation}, with the effective Hamiltonian $\tilde{H} = \tilde{\omega} \sum_{i=0}^1 \sigma_i^{\dagger} \sigma_i + \tilde{g}_{0,1} \left(\sigma_0^{\dagger} \sigma_1 + \sigma_0 \sigma_1^{\dagger} \right)$. Here, $\tilde{\omega}$ is the effective frequency of the quantum emitters and $\tilde{g}_{0,1}$ the effective coupling strength between the quantum emitters, which is real and positive. The implementation of this Hamiltonian as a sequence of quantum gates is standard and is discussed for the general case of an $n$-qubit chain in \appref{unitary_dynamics_appendix}.

The dissipative part of the dynamics after the effective removal of the cavity consists of two single-qubit dissipators $\mathcal{D}(\sigma_i)$ ($i=0,1$) as well as two effective cross-dissipation terms, $\mathcal{D}(\sigma_0, \sigma_1)$ and $\mathcal{D}(\sigma_1, \sigma_0)$, of the form introduced in \autoref{eqn:n_emitters_single_body_dissipators}, with cross-dissipation rates $\tilde{\gamma}_{0,1} = \tilde{\gamma}_{1,0}$. The dissipators $\mathcal{D}(\sigma_i)$ can be implemented as two independent amplitude damping channels, as discussed in \autoref{sec:level2a}. The cross dissipators introduce collective decay channels that require a more complex implementation. To address this challenge, we use a representation in which the cross-dissipative terms decouple into simpler decay channels. Specifically, we use the basis $\left\lbrace \ket{G}, \ket{\Lambda_-}, \ket{\Lambda_+}, \ket{E} \right\rbrace$ (see \autoref{fig:two_emitters_model}b), which diagonalizes the effective Hamiltonian (including the emitter-emitter interaction) and which can be expressed in terms of the two-qubit computational basis states $\left\lbrace \ket{00}, \ket{01}, \ket{10}, \ket{11} \right\rbrace$ as follows \cite{ElGordo2024, ElGordo2025},

\begin{align}
&\ket{G} = \ket{00}, \nonumber \\
&\ket{\Lambda_-} = \frac{1}{\sqrt{2}} \left( \ket{10} - \ket{01} \right), \nonumber \\
&\ket{\Lambda_+} = \frac{1}{\sqrt{2}} \left( \ket{01} + \ket{10} \right),  \nonumber \\
&\ket{E} = \ket{11}.
\label{eqn:basis_transformation}
\end{align}

In the interaction basis $\left\lbrace \ket{G}, \ket{\Lambda_-}, \ket{\Lambda_+}, \ket{E} \right\rbrace$, the dissipators can be rewritten as 
\begin{align}
\tilde{\gamma}_0 \mathcal{D} (\sigma_0)  [\rho] &+ \tilde{\gamma}_1 \mathcal{D} (\sigma_1) [\rho] + \nonumber \\ 
&+ \tilde{\gamma}_{0,1} \left( \mathcal{D} (\sigma_0, \sigma_1) [\rho] + \mathcal{D} (\sigma_1, \sigma_0) [\rho] \right) \equiv \nonumber \\
&\equiv \gamma_+ \left( \mathcal{D} (\sigma_{{G+}}) [\rho] + \mathcal{D} (\sigma_{{+E}}) [\rho] \right) + \nonumber \\ &+ 
\gamma_- \left( \mathcal{D} (\sigma_{{G-}}) [\rho]  + \mathcal{D} (\sigma_{{-E}}) [\rho] \right) + \nonumber \\ &+
\gamma_+ \left( \mathcal{D} (\sigma_{{G+}}, \sigma_{{+E}}) [\rho] + \mathcal{D} (\sigma_{{+E}}, \sigma_{{G+}}) [\rho] \right) - \nonumber \\ &- \gamma_- \left( \mathcal{D} (\sigma_{{G-}}, \sigma_{{-E}}) [\rho] + \mathcal{D} (\sigma_{{-E}}, \sigma_{{G-}}) [\rho]\right),
\label{eqn:dissipators_in_coupled_basis}
\end{align} 
where we have defined the lowering operators in the interaction basis $\sigma_{G+} \equiv \ket{G} \bra{\Lambda_+}$, $\sigma_{G-} \equiv \ket{G} \bra{\Lambda_-}$, $\sigma_{-E} \equiv \ket{\Lambda_-} \bra{E}$, $\sigma_{+E} \equiv \ket{\Lambda_+} \bra{E}$. $\gamma_{\pm} \equiv \tilde{\gamma} \pm \tilde{\gamma}_{0,1}$ represent the decay rates of the transitions ending or starting at $\ket{\Lambda_{\pm}}$.


The four ``diagonal'' dissipators $\mathcal{D} (\sigma_{{G+}})$, $\mathcal{D} (\sigma_{{+E}})$, $\mathcal{D} (\sigma_{{G-}})$, $\mathcal{D} (\sigma_{{-E}})$ in \autoref{eqn:dissipators_in_coupled_basis} represent the four possible decay paths in the interaction basis, namely $\left| \Lambda_+ \right> \to \left| G \right>$, $\left| E \right> \to \left| \Lambda_+ \right>$, $\left| \Lambda_- \right> \to \left| G \right>$, $\left| E \right> \to \left| \Lambda_- \right>$, respectively. As a result, we apply the same procedure as for the case of a single quantum emitter and define four (decay) channels, $\varepsilon_{{G+}}$,  $\varepsilon_{{G-}}$, $\varepsilon_{{+E}}$, $\varepsilon_{{-E}}$, in an enlarged Hilbert space $\mathcal{H}_{S} \otimes \mathcal{H}_{A}$ with transformation equations
\begin{align*}
\varepsilon_{{G}\pm} 
\begin{cases}
\left| G \right>_S \otimes \left|0 \right>_A & \to \left| G \right>_S \otimes \left|0 \right>_A, \\ 
\left| \Lambda_{\pm} \right>_S \otimes \left|0 \right>_A & \to \sqrt{1- p_{\pm}} \left| \Lambda_{\pm} \right>_S \otimes \left|0 \right>_A + \nonumber \\ &+  \sqrt{p_{\pm}} \left| G \right>_S \otimes \left|1 \right>_A,
\end{cases}
\end{align*}
\begin{align*}
\varepsilon_{\pm {E}} 
\begin{cases}
\left| \Lambda_{\pm} \right>_S \otimes \left|0 \right>_A & \to \left| \Lambda_{\pm}  \right>_S \otimes \left|0 \right>_A, \\
\left| E \right>_S \otimes \left|0 \right>_A & \to \sqrt{1- p_{\pm}} \left| E \right>_S \otimes \left|0 \right>_A + \nonumber \\ &+ \sqrt{p_{\pm}} \left| \Lambda_{\pm} \right>_S \otimes \left|1 \right>_A,
\end{cases}
\end{align*}
with $p_{\pm} \approx 1-\exp(-\gamma_{\pm}t)$ the decay probabilities of the transitions ending or starting at $\ket{\Lambda_{\pm}}$.
Each of these sets of transformation equations corresponds to a decay process that, when restricted to the relevant two-level transition subspace, is mathematically equivalent to a single-qubit amplitude damping channel, defined in \autoref{eqn:adc_transformation_equations}, under the identification $\left| G \right> \to \ket{\bar{00}}$, $\left| \Lambda_- \right> \to \ket{\bar{01}}$, $\left| \Lambda_+ \right> \to \ket{\bar{10}}$, $\left| E \right> \to \ket{\bar{11}}$. However, unlike in the single-qubit case, the decay processes act on overlapping subspaces of a larger four-level system and therefore cannot be implemented as totally independent amplitude damping channels. In particular, we can define the conjoined channel $\varphi$ that encodes all four decay paths, using a two-qubit ancilla register, through the set of transformation equations
\begin{align}
\left| G \right>_S \otimes \left|00 \right>_A &\to \left| G \right>_S \otimes \left|00 \right>_A, \nonumber \\
\left| \Lambda_- \right>_S \otimes \left|00 \right>_A & \to \sqrt{1- p_{-}} \left| \Lambda_- \right>_S \otimes \left|00 \right>_A + \nonumber \\ &+  \sqrt{p_{-}} \left| G \right>_S \otimes \left|01 \right>_A, \nonumber \\
\left| \Lambda_+ \right>_S \otimes \left|00 \right>_A &\to \sqrt{1- p_{+}} \left| \Lambda_+ \right>_S \otimes \left|00 \right>_A + \nonumber\\ &+  \sqrt{p_{+}} \left| G \right>_S \otimes \left|10 \right>_A, \nonumber \\
\left| E \right>_S \otimes \left|00 \right>_A &\to \sqrt{1- p_{-}}\sqrt{1- p_{+}} \left| E \right>_S \otimes \left|00 \right>_A + \nonumber \\ &+  \sqrt{p_{+}}\sqrt{1- p_{-}} \left| \Lambda_+ \right>_S \otimes \left|01 \right>_A +  \nonumber \\ &+ \sqrt{p_{-}}\sqrt{1- p_{+}} \left| \Lambda_- \right>_S \otimes \left|10 \right>_A + \nonumber \\ &+  \sqrt{p_{-}}\sqrt{p_{+}} \left| G \right>_S \otimes \left|11 \right>_A.
\label{eqn:conjoined_channel_transformation_equations}
\end{align}
At this stage, two ancilla qubits are used instead of just one to uniquely encode each of the four possible decay paths. This conjoined channel $\varphi$ can be represented by the circuit in \autoref{fig:two_emitters_model}c, where now $\theta_{\pm} = 2\arcsin(\sqrt{p_{\pm}})$ and $P$ is a two-qubit gate that performs the basis change from the computational to the interaction basis, as defined in \autoref{eqn:basis_transformation}. 

The circuit in \autoref{fig:two_emitters_model}c implements the conjoined decay channel described in \autoref{eqn:conjoined_channel_transformation_equations}. Similar to the single-emitter circuit in \autoref{fig:one_emitter_model}b, this circuit uses conditional gates to apply the corresponding dissipative transitions. However, in addition to single-controlled rotations, the circuit includes double-controlled $R_y$ gates ($CCR_y$), i.e., rotations acting on an ancilla qubit that are applied only when both control qubits are in specified states, allowing the implementation of state-dependent decay probabilities. A more rigorous proof of the validity of this circuit using Kraus operators can be found in \appref{app:justification_circuit}.

The circuit works as follows: first, we apply a custom two-qubit gate $P$ that perform a basis change from the two-qubit computational to the interaction basis, following the transformation equations in \autoref{eqn:basis_transformation}. This basis change allows the identification $\left| G \right> \to \ket{\bar{00}}$, $\left| \Lambda_- \right> \to \ket{\bar{01}}$, $\left| \Lambda_+ \right> \to \ket{\bar{10}}$, $\left| E \right> \to \ket{\bar{11}}$. After the basis change, if the system qubits are found to be in $\left| \Lambda_{-} \right>$ or  $\left| \Lambda_{+} \right>$, qubit $q_0$ decays with probability $p_{-}$ and qubit $q_1$, with probability $p_{+}$. In either of these cases, the control conditions of the first two double-controlled $R_y$ gates ($CCR_y$) are not satisfied, and the single-controlled $R_y$ gates together with the $CX$ gates implement the corresponding decay on the system qubits. However, if the system qubits are found in the doubly excited state  $\left| E\right>$, the decay probabilities are reversed, i.e., qubit $q_0$ now decays with probability $p_{+}$ and $q_1$, with probability $p_{-}$. In this case, the control conditions of the $CCR_y$ gates are satisfied and the gates perform a rotation of angle $\theta_+ - \theta_-$ to the ancilla qubit $a_0$ and a rotation of angle $\theta_- - \theta_+$ to $a_1$. Part of these rotations are then undone by the next $CR_y$ gates in a way so that $a_0$ ends up being rotated by an angle $\theta_+$ and $a_1$, by $\theta_-$ (i.e., the opposite values than when the system is not in the $\ket{E}$ state). After applying the $CX$ gates, this results in the qubit $q_0$ decaying with probability $p_{+}$ and $q_1$, with probability $p_{-}$, as desired (see right panel in \autoref{fig:two_emitters_model}b). Last, the system qubits are brought back to the computational basis through the inverse of the basis change gate $P$ and the ancillas are reset before starting the next Trotter step. A formal justification of this circuit and its connection to the master equation in \autoref{eqn:gksl_equation} is shown in \appref{app:justification_circuit}.

We note that the mapping of the other four ``off-diagonal'' dissipators in \autoref{eqn:dissipators_in_coupled_basis}, namely $\mathcal{D} (\sigma_{{G+}}, \sigma_{{+E}})$, $\mathcal{D} (\sigma_{{+E}}, \sigma_{{G+}})$, $\mathcal{D} (\sigma_{{G-}}, \sigma_{{-E}})$,$\mathcal{D} (\sigma_{{-E}}, \sigma_{{G-}})$ is more complex. Nonetheless, we show in \appref{appendix:omission} that these dissipators only affect weakly the population dynamics.

\subsection{\label{sec:level2c} Generalization to $n$-qubit chains}

We use the circuit in \autoref{fig:two_emitters_model}c as the core building block for constructing the circuit that models the population dynamics of a chain composed by arbitrary number $n$ of identical emitters with nearest-neighbor couplings. The dynamics of this chain of identical emitters are described by the master equation with Hamiltonian \autoref{eqn:n_emitters_hamiltonian} and dissipators in \autoref{eqn:n_emitters_single_body_dissipators} by setting equal frequencies and couplings ($\tilde{\omega}_i = \tilde{\omega}_j \equiv \tilde{\omega} = 1.2045$eV, $\tilde{\gamma}_i = \tilde{\gamma}_j \equiv \tilde{\gamma} = 9$meV, $i,j = 0,1,...,n-1$ and $\tilde{g}_{i,i+1} = \tilde{g}_{j,j+1} \equiv \tilde{g} = 4.5$meV, $\tilde{\gamma}_{i,i+1} = \tilde{\gamma}_{j,j+1} \equiv \tilde{\gamma}_{0,1} = 9$meV, $i,j = 0,1,...,n-2$).

To achieve this generalization, we employ a brick-wall construction \cite{Vidal2004}, in which two-emitter dissipative blocks are applied sequentially to neighboring pairs of emitters and ancillas in alternating layers (see \appref{appendix:generalization_to_n_qubits}). In each layer, the blocks act on disjoint qubit pairs and can therefore be executed in parallel, enabling the efficient implementation of nearest-neighbor dissipation across the entire chain. This brick-wall construction results in a constant scaling of the virtual circuit depth with the number of emitters, independent of the system size for $n > 2$ (for a single Trotter step), as shown by the yellow dashed lines in \autoref{fig:dynamic_circuits}a. 

To simulate longer evolution times and reduce Trotterization errors, the circuit can be concatenated $k$ times, with the parameters of the parametric gates rescaled according to $t \rightarrow t/k$ at each step. In this way, the full evolution is approximated as a product of $k$ identical circuit layers, each implementing a short-time evolution. However, while increasing $k$ improves the accuracy of the approximation, it also increases the total circuit depth of the virtual circuit linearly. The Trotter error and the convergence of the results are analyzed numerically in \appref{trotterization_analysis}.

Importantly, ensuring constant scaling for the transpiled circuit, or Instruction Set Architecture (ISA) circuit, using preset transpilation options in \texttt{Qiskit} \cite{Javadi-Abhari2024} is not straightforward, as shown by the yellow solid line in \autoref{fig:dynamic_circuits}a. Moreover, even when the scaling is favorable, the two-qubit-depth overhead associated with the core two-emitter block can be substantial. This overhead is largely due to the two controlled-controlled-$R_y$ ($CCR_y$) gates in the core virtual circuit. Upon transpilation, each $CCR_y$ gate typically decomposes into single $R_y$ rotations and two $CCX$ gates. Each of the latter involves six or more $CX$ gates, resulting in a total of roughly 24 two-qubit gates for the two $CCR_y$ gates— about half of the post-transpilation two-qubit depth for $n=2$.

To overcome these challenges while preserving the constant scaling of the virtual circuit, we introduce in \autoref{sec:level3} circuit adaptations based on dynamic circuits \cite{Riste2013, fossfeig2023} and custom transpilation techniques.



%

\section{\label{sec:level3} CIRCUIT ADAPTATION FOR EFFICIENT HARDWARE EXECUTION}

\begin{figure*}[!htbp]
    \centering
    \def\svgwidth{\textwidth}
    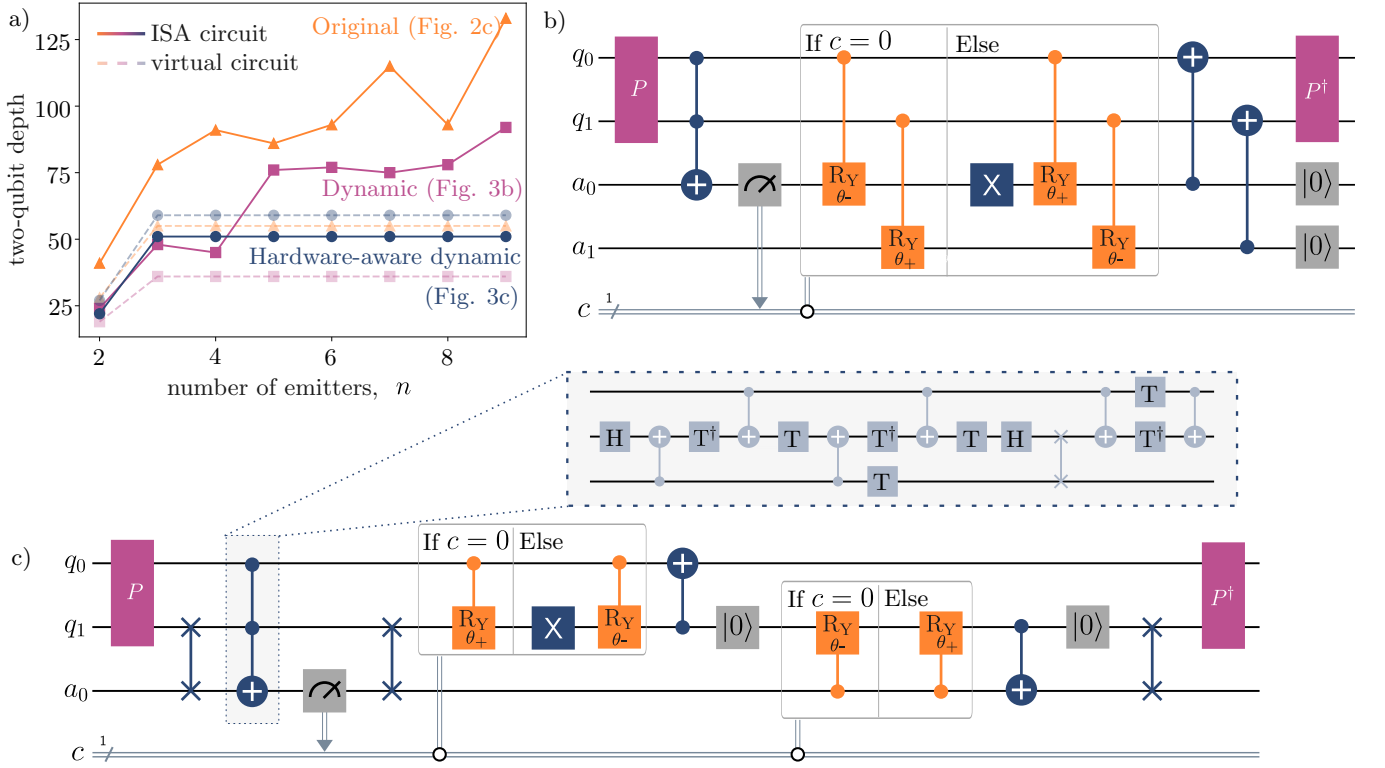
    \caption{Near-term hardware-aware transformation of the circuit in \autoref{fig:two_emitters_model}c. (a) Scaling with the number of emitters $n$ of the two-qubit depth of the virtual (dashed lines) and ISA circuits obtained after transpilation (solid lines) for the original circuit in \autoref{fig:two_emitters_model}c (yellow triangles), the dynamic circuit in \autoref{fig:dynamic_circuits}b (purple squares) and the hardware-aware dynamic circuit in \autoref{fig:dynamic_circuits}c (blue circles). (b) Dynamic circuit that models the collective dissipation of two emitters. This circuit is the first adaptation of the original circuit in  \autoref{fig:two_emitters_model}c into a dynamical circuit, which achieves a significant two-qubit depth reduction. Only two two-level systems representing the quantum emitters are considered for simplicity, corresponding to qubits $q_0$ and $q_1$. The corresponding ancilla register is represented by qubits $a_0$ and $a_1$. The blocks labeled by $P$ and $P^{\dagger}$ represent the unitary operation necessary to perform the basis change in \autoref{eqn:basis_transformation} and its inverse, respectively. After this basis change, a $CCX$ gate is applied to $a_0$ with $q_0$ and $q_1$ acting as control qubits. The circuit then follows with a mid-circuit measurement of the ancilla qubit $a_0$, with the result stored in the classical register $c$, and a conditional block. If the outcome of the measurement is zero ($c=0$), the first block is applied. On the contrary, if $c=1$, the second block is applied. To finish the decay channel, we apply a parallel sequence of $CX$ gates after the conditional blocks and undo the basis change with the $P^{\dagger}$ gate. For more than one Trotter step, the ancilla qubits must be reset. (c) Hardware-aware dynamic circuit that models the collective dissipation of two emitters. This version of the dynamic circuit ensures a constant scaling of the two-qubit depth, even in the ISA circuit, for $n>2$ emitters. First, $SWAP$ gates have been added to the circuit every time that two non-adjacent qubits needed to interact. The addition of $SWAP$s at the virtual-circuit level required decomposing the $CCX$ gate into simpler gates as shown in the light grey box above the main circuit in \autoref{fig:dynamic_circuits}c. Further, the ancilla qubit reduction requires dividing the conditional block in \autoref{fig:dynamic_circuits}b into two, smaller conditional blocks and adjusting the control and targets. Between these two conditional blocks, an extra reset is needed to ensure the ancilla qubit always starts in the state $\ket{0}$. \autoref{fig:dynamic_circuits}b and \autoref{fig:dynamic_circuits}c are displayed for the case of two emitters (and one Trotter step) and they follow the same generalization for $n>2$ emitters as the original circuit, discussed in \autoref{sec:level2c} and shown schematically for $n=4$ in \autoref{fig:generalization_schematic}a (\appref{appendix:generalization_to_n_qubits}).}
    \label{fig:dynamic_circuits}
\end{figure*}

We show next that a significant reduction in two-qubit gate depth can be achieved by using dynamic circuits \cite{Corcoles2021}. Dynamic circuits have been applied to constant-depth long-range entanglement  \cite{PRXQuantum.5.030339} and adaptive feedforward-based primitives such as the semiclassical quantum Fourier transform \cite{Griffiths1996, Baumer2024} but here they serve as a resource-efficient substitute to reduce the circuit depth. In particular, we replace both $CCR_y$ gates with a single Toffoli ($CCX$) gate, followed by a mid-circuit measurement and some conditional blocks, as observed in \autoref{fig:dynamic_circuits}b. The result of the measurement determines which block to execute. Following this approach, it is possible to include only a single $CCX$, which depending on the measurement outcome will implement the decay path $\ket{E} \to \ket{\Lambda_{\pm}}$ (qubit $q_0$ decays with probability $p_+$ and $q_1$, with probability $p_-$) or $\ket{\Lambda_{\pm}} \to \ket{G}$ (qubit $q_0$ decays with probability $p_-$ and $q_1$, with probability $p_+$) corresponding to the transformation in \autoref{eqn:conjoined_channel_transformation_equations}. The scaling of the two-qubit depth of the proposed dynamic circuit with the number of emitters before and after transpilation is shown in purple dashed and solid lines, respectively, in \autoref{fig:dynamic_circuits}a, showing in both cases a clear depth reduction compared to the scaling of the original circuit in \autoref{fig:two_emitters_model}c (yellow lines in \autoref{fig:dynamic_circuits}a).

On the other hand, the two-qubit depth still increases with the number of emitters in the ISA circuit obtained after transpilation (solid purple line in \autoref{fig:dynamic_circuits}a). To ensure constant-depth scaling post transpilation, we apply three techniques, in the following order:
\begin{enumerate}
\item Reduce the number of ancilla qubits from $n$ to $\lfloor n/2 \rfloor$ ($n \geq 2$). Assigning one ancilla qubit per system qubit pair reduces the total number of qubits from $2n$ to $n + \lfloor n/2 \rfloor$ and lowers connectivity constraints. This assignment is beneficial for the brick-wall construction, where each ancilla only needs to mediate the decay of one emitter pair at a time. A drawback is a slight reduction of the parallelizability of some operations, increasing the virtual circuit two-qubit depth. Nonetheless, we consider that the decrease in the number of qubits and the improvement in connectivity compensates for this drawback, specially when considering the follow-up techniques. 
\item Specify a tailored initial layout that places an ancilla after every pair of system qubits. This is a direct application of the previous feature that significantly reduces the number of SWAP operations required during transpilation. While this layout can be non-optimal for small system sizes, it naturally matches the brick-wall circuit pattern and provides favorable scaling for larger systems.
\item Add layers of $SWAP$ gates into the virtual circuit whenever two non-adjacent qubits need to interact. By adding the $SWAP$ operations at the virtual-circuit level, we can ensure that these operations are executed as desired at the hardware level without relying on the routing used by the transpiler. To preserve this routing strategy, automatic routing during transpilation is disabled, preventing the transpiler from introducing additional $SWAP$ gates beyond the ones we have inserted manually.
\end{enumerate}

The inclusion of these features before transpilation results in the hardware-aware virtual circuit in \autoref{fig:dynamic_circuits}c (displayed for the simple case of two qubits and one Trotter step) and the scaling with the number of emitters shown by the blue solid lines in \autoref{fig:dynamic_circuits}a. Remarkably, the post-transpilation two-qubit depth remains independent of the number of emitters (for $n>2$), and thus exhibits the same constant scaling that characterizes the virtual circuits. For very small system sizes, the default transpilation (solid purple line in \autoref{fig:dynamic_circuits}a) can find better layouts with slightly lower two-qubit depth than our tailored approach. However, this advantage does not persist as the number of emitters grows. In contrast, the final approach performs consistently better across all system sizes and achieves lower two-qubit depth for intermediate to larger systems.


 \section{\label{sec:level4} RESULTS}

This section presents a series of experimental results obtained after executing the hardware-aware circuit (\autoref{fig:dynamic_circuits}c), for different system sizes, on IBM quantum computers. In particular, we use the \texttt{ibm\_basquecountry} quantum computer, which is a Heron r2 device with heavy-hex connectivity. The results obtained in the quantum experiments are benchmarked against classical numerical simulations across different system sizes comprising many quantum emitters.

We start by benchmarking different error suppression and error mitigation techniques on a ten-emitter system (15 total qubits, including the ancilla qubits). Afterwards, we discuss the classical validation of the quantum experiment results obtained in a fifty-emitter system (75 total qubits). To validate these experiments we develop a hybrid method that combines Monte Carlo and tensor-networks techniques. Finally, we show the results obtained for the largest chain size considered, comprising of eighty-six emitters (129 total qubits).

\subsection{\label{sec:level4a} Comparing error suppression and mitigation techniques}
We begin by considering the circuit in \autoref{fig:dynamic_circuits}c for a configuration of ten emitters, which involves ten system qubits and five ancillas (15 total qubits). We execute this circuit on the IBM quantum computer \texttt{ibm\_basquecountry} using $N_{shots} = 2 \cdot 10^4$ shots and $k=1$ Trotter step. The Trotter error is analyzed in \appref{trotterization_analysis}. Solving the Markovian master equation for this particular configuration with classical brute-force methods is already computationally demanding due to the exponentially-large size of the Hilbert space of the system, but still feasible.

\begin{figure}[tbp]
    \centering
    \def\svgwidth{\columnwidth}
    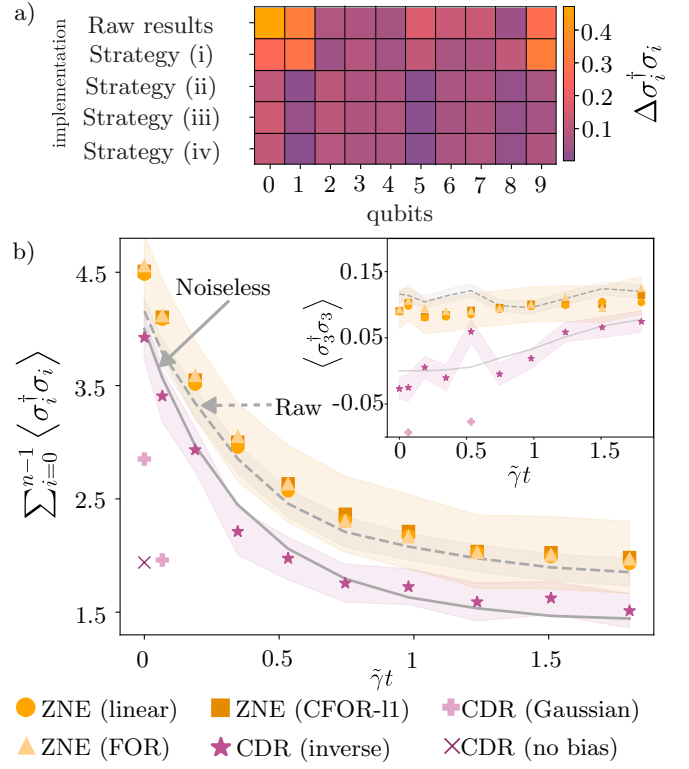
    \caption{Comparison of the performance of error suppression and mitigation techniques in \texttt{ibm\_basquecountry} using the hardware-aware circuit in \autoref{fig:dynamic_circuits}c extended to a chain of ten quantum emitters. (a) Colormap of the absolute error of the excited-state population of each system qubit $q_i$, with $i = 0, ..., 9$ at $\tilde{\gamma}t = 0.45$. The error is obtained without dynamical decoupling (raw results, first row) and for the four different strategies discussed in the main text: (i) XY4 sequence on idle qubits while performing mid-circuit measurements (second row), (ii) XY4 sequence while performing mid-circuit measurements and resets (third row), (iii) XY4 during every operation (fourth row) and (iv) XY8 during every operation (last row). The error shown corresponds to the absolute error between excited-state populations obtained in the quantum experiments, under different dynamical decoupling implementations and the excited-state populations obtained in a noiseless (classical) execution of the quantum circuit. (b) Time-evolution of the total population (main panel) and the excited-state population of the emitter $q_3$ (inset) under different error mitigation techniques: ZNE (linear) with linear extrapolation (circles), ZNE (FOR) with first-order Richardson extrapolation (triangles), ZNE (CFOR-$l1$) with constrained first-order Richardson extrapolation (squares), CDR (inverse) under inverse bias (stars), CDR (Gaussian) under Gaussian bias with mean $\mu = 0$ and standard deviation $\sigma = 1$ (plus signs) and CDR (no bias) without any bias (crosses). The mitigation errors for the best performing ZNE and CDR techniques (linear and inverse, respectively) are shown as shaded regions. The unmitigated raw results correspond to the grey dashed lines and the solid grey lines correspond to the noiseless classical simulation of the circuit. The initial state of the system qubits is the product state $\ket{\psi (0)}_S = \ket{0100010011}_S$. All emitters and couplings are assumed identical and the effective parameters used are $\tilde{\omega} = 1.2045$eV, $\tilde{g}_{i,i+1} = 4.5$meV, $\tilde{\gamma} = 9$meV, $\tilde{\gamma}_{i,i+1} = 9$meV, $\forall i = 0, 1, ..., 8$.}
    \label{fig:error_mitigation_benchmark}
\end{figure}

We analyze in \autoref{fig:error_mitigation_benchmark} the results obtained after using different error mitigation and error suppression techniques. Regarding error suppression, we consider different dynamical decoupling sequences \cite{Viola_1998}, applied on idle qubits during distinct types of operations.  Dynamical decoupling helps to mitigate the effects of decoherence that arise during idle periods within dynamic circuits, i.e., reset operations, mid-circuit measurements and conditional evaluation of control branches. In particular, we follow four strategies: applying an XY4 sequence —$X$-$Y$-$(-X)$-$(-Y)$ gates— to idle qubits during (i) mid-circuit measurements, (ii) mid-circuit measurements and resets, (iii) every operation and (iv) applying an XY8 —two concatenated XY4 sequences—  to idle qubits during every operation.

To assess the effectiveness of these different decoupling strategies, we show in \autoref{fig:error_mitigation_benchmark}a the absolute error in the calculation of the excited-state population between the noiseless results obtained by executing the same circuit on a classical computer and the results retrieved from the quantum processing unit (QPU) for different strategies, for the ten-emitter configuration at $\tilde{\gamma}t = 0.45$. No error mitigation is considered here. For the results shown in \autoref{fig:error_mitigation_benchmark}a, we perform bootstrap resampling \cite{Efron1994} of the output bitstring distribution, i.e., we repeatedly generate resampled datasets from the measured bitstring counts to estimate the statistics of the observable (details in \appref{appendix:bootstrap})

We find that even the simplest strategy —XY4 sequence on idle qubits while performing mid-circuit measurements, Strategy (i) in \autoref{fig:error_mitigation_benchmark}a— yields a significant improvement for most qubits when compared to the unmitigated results (Raw results in \autoref{fig:error_mitigation_benchmark}a). We attribute the convenience of using dynamical decoupling to the use of dynamical circuits, where mid-circuit measurements, conditional operations, and resets introduce relatively long idle periods compared with ordinary gate execution. The second substantial improvement occurs when the same XY4 sequence is applied to both mid-circuit measurements and resets (Strategy (ii)). The remaining strategies —XY4 during every operation (Strategy (iii)) and XY8 during every operation (Strategy (iv))— provide similar results but they standardize the decoupling treatment across the full circuit. Overall, we find that XY4 during every operation provides a good balance between accuracy and implementation overhead for our particular circuits. For this reason, we use this dynamical decoupling strategy in the following (\autoref{fig:error_mitigation_benchmark}b and \autoref{fig:quantum_experiment_results}). Finally, some outliers can be identified in \autoref{fig:error_mitigation_benchmark}a. First, for qubit $q_4$, applying any dynamical decoupling actually yields slightly worse results than not applying any sequence at all. Additionally, the error in the excited-state population of qubit $q_2$ is smallest when using the simplest dynamical decoupling sequence —XY4 sequence only during mid-circuit measurements— instead of the more advanced routines. These outliers likely depend on the specific qubit errors in a particular experiment, so our analysis focuses on the overall behavior rather than these exceptions. 

Dynamical decoupling reduces errors during runtime but further improvement can be expected from using error mitigation techniques during post-processing. We first consider zero-noise extrapolation (ZNE) \cite{Temme_2017}. To retrieve noise-amplified results for ZNE, we use post-transpilation random partial gate folding \cite{Giurgica2020}, exclusively on $CZ$ gates, for low noise-amplification factors $\lambda \in (1, 3)$. Small noise amplification factors are required to ensure that the noise-amplified circuits do not grow excessively. After obtaining the noise-amplified results, we extrapolate the results to $\lambda = 0$ (zero-noise regime) using different extrapolation methods. The techniques that perform best in this analysis are linear extrapolation, first-order Richardson extrapolation \cite{Richardson1911} and constrained Richardson extrapolation with $l1$-norm minimization \cite{Boyd2004}. This last technique is a variation of the first-order Richardson that linearly combines different noise-amplified results, subject to the Richardson constraints and a minimization of the $l1$-norm of the coefficient vector. In this way we obtain a sparse coefficient vector that also indicates which noise-amplified circuits are the most relevant ones for zero-noise result retrieval. A more detailed discussion of this technique can be found in \appref{appendix:constrained_richardson_extrapolation}.

The performance of these extrapolation schemes is shown in \autoref{fig:error_mitigation_benchmark}b in yellow tones. For reference, we also show the raw results obtained in a quantum experiment without mitigation and after bootstrapping (grey dashed lines), the standard deviation of the bootstrap distribution (shaded grey region around raw results), and the noiseless classical simulation of the circuit (grey solid lines). In particular, we show the  time evolution of the total population (\autoref{fig:error_mitigation_benchmark}b) and the excited-state population of the fourth emitter (inset in \autoref{fig:error_mitigation_benchmark}b) when the chain is initialized to the state  $\left| \psi (0) \right>_S = \left| 0100010011 \right>_S $. These results show that in our simulations all three extrapolation methods —linear (circles), first-order Richardson extrapolation (FOR) (triangles) and constrained first-order Richardson extrapolation with $l1$-norm minimization (CFOR-$l1$) (squares)— slightly degrade the total population results when compared to the raw data (grey dashed lines), while only slightly improving the raw data for the single-emitter population. Furthermore, all three extrapolation methods yield numerically similar zero-noise estimates. We conclude that ZNE is unable to improve the obtained dynamics for the longer chain considered here. This failure could occur because the noise-induced error is too large and the noise sources too complex to be fully corrected with ZNE using random partial gate folding.

Motivated by the limitations of ZNE for our circuits, we turn to Clifford data regression (CDR) \cite{Czarnik2021, Lowe2021, Strikis2021, Weaving2025}, which relies on circuit data learning rather than noise scaling. For this error mitigation technique, we start by generating $N = 10^2$ random Cliffordizations of the two-qubit  ISA circuit. After transpiling the hardware-aware circuit (\autoref{fig:dynamic_circuits}c) on IBM Heron devices, the only non-Clifford gates are arbitrary $R_z$ rotations and the ancilla resets. Both are replaced by $R_z (\theta_C)$ gates with a Clifford angle $\theta_C \in \{0,\pi/2,\pi,3\pi/2\}$. As a baseline, we consider an unbiased Cliffordization in which $\theta_C$ is sampled uniformly from $\{0,\pi/2,\pi,3\pi/2\}$. We compare this baseline with biased Cliffordization rules that favor Clifford angles close to the original non-Clifford angle $\theta(t)$ of each parametrized rotation gate $R_z(\theta(t))$. For each simulation instant $t_i$, the replacement angle is sampled from a distribution that decays with the distance $d(\theta (t_i), \theta_C) \equiv |\theta (t_i) - \theta_C |$. We have considered a variety of probability distributions, and we discuss next as representative examples a Gaussian distribution with mean $\mu = 0$ and standard deviation $\sigma$, $\exp(-d(\theta (t_i), \theta_C)^2/\sigma^2)$, the inverse distribution, $1/(d(\theta (t_i), \theta_C) + \epsilon)$ with $\epsilon >0$ a small parameter that avoids divergences at $d(\theta (t_i), \theta_C) = 0$; and the uniform distribution used as the unbiased baseline. A more detailed discussion is included in \appref{appendix:bised_clifford_data_regression}. 

A training dataset of expectation values is obtained from simulations performed on both a classical processing unit (CPU) and a quantum processing unit (QPU) of the Cliffordized circuits $\{(\langle O \rangle^{(i)}_{\mathrm{noiseless}}, \langle O \rangle^{(i)}_{\mathrm{noisy}})\}_{i=0}^{N-1}$ as in usual Clifford data regression. A linear regression model is then trained on this dataset and subsequently applied to correct the noisy expectation value of the observable for the unmodified circuit. For the total population (main panel in \autoref{fig:error_mitigation_benchmark}b) and the population of emitter 3 (inset) of the ten-emitter chain all three types of biases, namely inverse bias (stars), Gaussian bias with $\sigma= 1$ (plus signs) and no bias/uniform bias (crosses), yield very different results. More remarkably, and unlike ZNE, CDR under inverse bias  (stars) closely reproduces the expectation value of the observable obtained in the noiseless simulation of the circuit for almost all time steps, in both \autoref{fig:error_mitigation_benchmark}b and the inset. This improved performance could occur because inverse bias Clifford circuits present minimal distance to the problem circuit, when compared to the other two types of biases presented in this work.


Based on these results, we adopt XY4 dynamical decoupling during all operations combined with inverse-bias CDR as our default error suppression and mitigation strategy for systems constituted of a large number of quantum emitters.

\subsection{\label{sec:level4b} Large-scale dissipative simulations}
After identifying effective error mitigation and suppression strategies using intermediate-size chains, we assess next the scalability of the approach to larger emitter chains. We consider next a fifty-emitter chain, which involves a total of 75 qubits (50 system qubits representing the quantum emitters and 25 ancilla qubits). This system is already beyond the reach of exact brute-force classical simulation and requires using approximate methods such as tensor networks for validation. Standard approaches based on tensor network techniques, such as the time-evolving block decimation algorithm (TEBD) \cite{Vidal2004}, have been developed to treat purely-Hamiltonian dynamics, and thus cannot be directly applied to our circuits.

\begin{figure*}[tbp]
    \centering
    \def\svgwidth{\textwidth}
    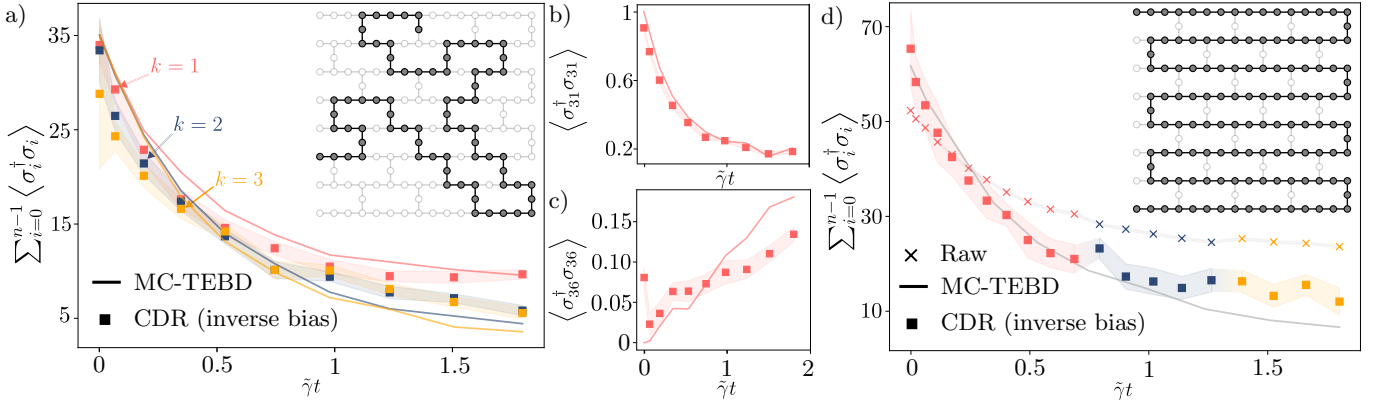
    \caption{Quantum experiments in \texttt{ibm\_basquecountry} of intermediate-size and large chains. (a) Time evolution of the total population for an intermediate chain of $n = 50$ emitters, involving 50 system qubits and 25 ancillas (75 total qubits arranged as shown in the inset) for $k=1$ (red), $k = 2$ (blue) and $k = 3$ (yellow) Trotter steps. The squares represent the post-processed results from the quantum experiment using CDR under inverse bias. This data is compared with the reference data, shown in circles, obtained using the tensor network technique discussed in the main text with a maximum bond dimension $\max \chi = 50$ and $N = 100$ trajectories. The uncertainties on the CDR-corrected data (shaded regions around the squares) are determined by quadratic error propagation of both the regression errors and the bootstrapping error (see \appref{appendix:bised_clifford_data_regression}). The initial state of the system qubits is a product state of thirty five randomly-selected system qubits initialized to the excited state and fifteen to the ground state. (b, c) Time evolution of the excited-state population of two representative quantum emitters for one Trotter step ($k=1$). (d) Time evolution of the total population for a large chain of $n = 86$ emitters, involving 86 system qubits and 43 ancillas (129 total qubits arranged as shown in the inset). The raw data, obtained from the quantum experiments before post-processing, is shown as crosses. This data represents the mean obtained after applying bootstrapping to the output bitstring distribution. The standard deviation of the bootstrapped distribution is represented by shaded regions around the crosses. The initial state of the system qubits is a product state of sixty randomly-selected system qubits initialized to the excited state and twenty-six to the ground state. All emitters and couplings are assumed identical and the effective parameters used are $\tilde{\omega} = 1.2045$eV, $\tilde{g}_{i,i+1} = 4.5$ meV, $\tilde{\gamma} = 9$meV, $\tilde{\gamma}_{i,i+1} = 9$meV, $\forall i = 0, 1, ..., n-2$.}
       \label{fig:quantum_experiment_results}
\end{figure*}

To circumvent this difficulty, we develop an approach that combines matrix product state (MPS) methods \cite{Schollwoeck2011} and Monte Carlo (MC) trajectory sampling and that thus enables the validation of the results obtained from the quantum experiments \cite{Sander2025}. This approach follows closely the circuit implementation of the dissipative dynamics, i.e., these classical calculations also consider separately the unitary evolution of all qubits (system and ancilla registers) and the reset operations on the ancillas, which allows us to perform an efficient stochastic unraveling in terms of pure-state trajectories \cite{Molmer1992}. Concretely, we represent the combined system-ancilla state as a MPS and simulate the circuit layer-by-layer using tensor-network contractions through the time-evolving block decimation algorithm (TEBD). The reset operations are modeled as projective measurements on the ancilla qubits followed by sampling and conditional operations, thereby generating an ensemble of pure-state trajectories \cite{Molmer1993}. In this sense, this Monte Carlo-Time-Evolving Block Decimation (MC-TEBD) is related to recent tensor-network trajectory methods for open-system simulation \cite{Sander2025}, formulated here directly at the circuit level using the ancilla-assisted implementation of the dissipative channels and stochastic sampling of reset operations. Observable expectation values are recovered by averaging over $N_{traj}$ such trajectories. Further implementation details and a detailed description of the algorithm are provided in \appref{app:mcwf-tebd}.

We show in panel a of \autoref{fig:quantum_experiment_results} the time evolution of the total population, as well as the excited-state populations of two representative emitters in panels b, c for the 75-total-qubits circuit in  \autoref{fig:dynamic_circuits}c using $N_{shots} = 2 \cdot 10^4$ and for $k=1$ (red), $k = 2$ (blue) and $k = 3$ (yellow) Trotter steps.  For each observable, we perform a bootstrapping of the raw data—obtained using XY4 dynamical decoupling during every operation— by generating $B = 10^2$ bootstrap samples of size $N_{shots}$ (\appref{appendix:bootstrap}). We show the corrected data (squares) —obtained after applying inverse-bias CDR to the bootstrapped raw data— and the classical MC-TEBD results (circles). The shaded regions around the corrected data represent the correction error, which is estimated using quadratic error propagation on both the regression model errors and the bootstrapped data (see \appref{appendix:bised_clifford_data_regression}). The CDR-corrected results generally track the MC-TEBD simulations closely, often overlapping within regions of error. This agreement supports the validity of both the hardware implementation and the mitigation strategy at a system size already beyond exact classical tractability. Additionally, the comparison between different values of $k$ serves to determine the minimum number of Trotter steps required at each simulation time. We use this criterion in the larger-scale experiments below, selecting at each time instant the minimum $k$ that keeps the Trotter error below a certain threshold. Further details are given in \appref{trotterization_analysis}.

To choose the initial qubit layout (shown in the inset in \autoref{fig:quantum_experiment_results}a), we perform a real-time qubit selection at the time when the quantum experiments are launched. Details about this procedure can be found in \appref{app:device_characterization}.

Finally, we consider in \autoref{fig:quantum_experiment_results}d a one-dimensional chain of 129 physical qubits in the IBM quantum computer \texttt{ibm\_basquecountry} for $N_{shots} = 2 \cdot 10^4$. This chain corresponds to 86 system qubits and 43 ancilla qubits, arranged as shown in the inset in \autoref{fig:quantum_experiment_results}d. As discussed above, we choose the number of Trotter steps at each instant —$k=1$ (red), $k=2$ (blue) and $k=3$ (yellow)— so that the Trotter error always remains under a certain threshold. For the largest Trotter step size, $k = 3$, the corresponding transpiled circuits involve approximately 8000 two-qubit gates and reach a two-qubit depth of 193, placing them in the utility-scale regime. We show in \autoref{fig:quantum_experiment_results}d both the raw total population from the quantum experiment (crosses) and the corresponding post-processed results obtained using inverse-bias CDR (squares), alongside the MC-TEBD reference data (grey circles). Even at this scale, the corrected data captures satisfactorily the general features of the dissipative dynamics, specially for $k=1$ and $k=2$. Most importantly, in almost every instance, the CDR-corrected results substantially improve upon the raw data and closely track the MC-TEBD simulations, overlapping in most cases within error regions. These results indicate the feasibility of large-scale complex open quantum system simulations on current devices. To the best of our knowledge, this is the first large digital quantum computing experiment of Markovian dissipative dynamics with both local and nearest-neighbor collective decay channels.

 \section{\label{sec:level5} CONCLUSIONS}
We present an utility-scale experimental demonstration of collective dissipative dynamics on a 156-qubit quantum processor. We consider a 1-dimensional chain of quantum emitters coupled to a common Markovian bath. The implementation of the quantum circuit requires not only single-qubit amplitude damping, but also cross-qubit dissipators responsible for collective decay. By mapping the dissipative processes to ancilla-assisted circuits, we realize non-unitary evolution on many qubits directly at the circuit level.

Importantly, we construct hardware-aware dynamic circuits that preserve the physics while reducing the resources required for execution on current quantum computers. Mid-circuit measurements, conditional operations, ancilla reuse, tailored qubit layouts, and explicit routing altogether transform the original collective-dissipation circuit into a circuit with favorable two-qubit depth scaling after transpilation. These improvements enable us to execute experiments consisting of 86 emitters and 129 total qubits (including ancillas), reaching circuit sizes beyond exact state-vector or density-matrix simulation.

Furthermore, we adapt Clifford data regression (CDR) to the dissipative dynamic circuits studied here by benchmarking different Cliffordization rules that favor Clifford angles close to the original (non-Clifford) rotations. Within the biased CDR, the inverse-bias variant gives the best performance in our benchmarks, substantially improving the raw hardware results and outperforming the zero-noise extrapolation methods tested in this work. Our results emphasize that physics-informed mitigation strategies in dynamic circuits provide an advantage.

We introduce a classical simulation method, based on Monte Carlo and the Time-Evolving Block Decimation algorithm (MC-TEBD) and designed for circuits containing reset operations capturing dissipation, to validate the results from the large-scale quantum experiments. The MC-TEBD algorithm treats resets as stochastic trajectories, occurring within the standard TEBD simulation, allowing us to obtain reference data for system sizes inaccessible to exact methods. The observed agreement between mitigated quantum experiments and MC-TEBD simulations provides evidence that the dynamics measured in the quantum computer reflect the intended collective dissipative evolution.

Our results extend the scope of near-term quantum simulation from closed-system unitary dynamics to open-system processes involving collective dissipation. This advance is an important step in the use of near-term quantum processors to study realistic physical systems, such as quantum emitters coupled to optical cavities, where coupling to the environment plays a central role in the dynamics. Future work may build on this framework to simulate higher-dimensional emitter arrays, include non-identical emitters or explore more complex observables beyond population dynamics such as two-emitter correlators.

 \section{\label{sec:level6} ACKNOWLEDGEMENTS}
We thank Niall Robertson for helpful discussions about tensor networks. We thank Jessie Yu and Stefan Elrington for their continued assistance with troubleshooting during the dynamic circuits early access. We thank Cristina Sanz and Lena Perennes for their help in coordinating with the computing support teams. We acknowledge financial support through the Grant No. PID2022-139579NB-I00 funded by MICIU/AEI/10.13039/501100011033 and by ERDF/EU, through the Grant No. IT 1526-22 funded by the Department of Science, Universities and Innovation of the Basque Government. Funding for these actions/grants and contracts comes from the European Union's Recovery and Resilience Facility-Next Generation, in the framework of the General Invitation of the Spanish Government’s public business entity Red.es to participate in talent attraction and retention programmes within Investment 4 of Component 19 of the Recovery, Transformation and Resilience Plan. We acknowledge the Road To Utility (R2U) program within the BasQ and IBM Quantum collaboration.

\clearpage
\onecolumngrid
\raggedbottom
\appendix

\section{Analytical study of the core two-qubit circuit}
\label{appendix:analytical_study_of_the_two_qubit_circuit}
Let us consider the exclusively-dissipative master equation (ME) for two identical two-level systems, obtained by setting $\tilde{H}=0$ in \autoref{eqn:gksl_equation} and setting $\tilde{\gamma}_0 = \tilde{\gamma}_1 \equiv \tilde{\gamma}$
\begin{equation}
\frac{d\rho}{dt} = \tilde{\gamma} \sum_{i=0}^{1} \mathcal{D}(\sigma_i) [\rho] + \tilde{\gamma}_{0,1} \sum_{ \substack{i,j=0 \\ i \neq j}}^{1}  \mathcal{D}(\sigma_i, \sigma_j) [\rho],
\label{eqn:ME_two_emitters_app}
\end{equation}
with dissipators defined in \autoref{eqn:n_emitters_single_body_dissipators}
\begin{equation}
 \tilde{\gamma}_{0,1} \mathcal{D} (\sigma_i, \sigma_j) [\rho]  = \tilde{\gamma}_{0,1} \left( \sigma_i \rho \sigma^{\dagger}_j - \frac{1}{2} \left\lbrace \sigma^{\dagger}_j \sigma_i, \rho \right\rbrace\right)
\end{equation}
and $\mathcal{D} (\sigma_i) \equiv \mathcal{D} (\sigma_i, \sigma_i)$. In the interaction basis, $\left\lbrace \ket{G}, \ket{\Lambda_-}, \ket{\Lambda_+}, \ket{E} \right\rbrace$, where
\begin{align}
&\ket{G} = \ket{00}, \nonumber \\
&\ket{\Lambda_-} = \frac{1}{\sqrt{2}} \left( \ket{10} - \ket{01} \right), \nonumber \\
&\ket{\Lambda_+} = \frac{1}{\sqrt{2}} \left( \ket{01} + \ket{10} \right),  \nonumber \\
&\ket{E} = \ket{11},
\label{eqn:app_basis_transformation}
\end{align}
the dissipators can be rewritten as \cite{ElGordo2024}
\begin{align}
&\tilde{\gamma} \left( \mathcal{D}(\sigma_0)[\rho] + \mathcal{D}(\sigma_1)[\rho] \right) + \tilde{\gamma}_{0,1} \left( \mathcal{D}(\sigma_0,\sigma_1)[\rho] + \mathcal{D}(\sigma_1,\sigma_0)[\rho] \right) \equiv
\nonumber \\
&\equiv \gamma_+ \left( \mathcal{D}(\sigma_{{G+}})[\rho] + \mathcal{D}(\sigma_{{+E}})[\rho] \right)
+ \gamma_- \left( \mathcal{D}(\sigma_{{G-}})[\rho] + \mathcal{D}(\sigma_{{-E}})[\rho] \right)
+ \gamma_+ \left( \mathcal{D}(\sigma_{{G+}},\sigma_{{+E}})[\rho]
+ \mathcal{D}(\sigma_{{+E}},\sigma_{{G+}})[\rho] \right) \nonumber \\
&- \gamma_- \left( \mathcal{D}(\sigma_{{G-}},\sigma_{{-E}})[\rho]
+ \mathcal{D}(\sigma_{{-E}},\sigma_{{G-}})[\rho] \right),
\label{eqn:app_dissipators_in_coupled_basis}
\end{align}
where we have defined the lowering operators in the interaction basis $\sigma_{\alpha \beta} \equiv \ket{\alpha} \bra{\beta}$, with $\alpha, \beta = {G, +, -, E}$ and the decay rates of the transitions ending or starting at $\ket{\Lambda_{\pm}}$, $\gamma_{\pm} \equiv \tilde{\gamma} \pm \tilde{\gamma}_{0,1}$. After the basis change, \autoref{eqn:app_dissipators_in_coupled_basis} shows only eight dissipative terms instead of the typical sixteen (e.g., terms such as $\mathcal{D}(\sigma_{{G-}},\sigma_{{+E}})[\rho]$ and $\mathcal{D}(\sigma_{{+E}},\sigma_{{-E}})[\rho]$ do not appear) because we are considering identical emitters.

As mentioned in \autoref{sec:level2b}, under the assumption of identical emitters, the four ``off-diagonal'' dissipators $\mathcal{D} (\sigma_{{G+}}, \sigma_{{+E}})$, $\mathcal{D} (\sigma_{{+E}}, \sigma_{{G+}})$, $\mathcal{D} (\sigma_{{G-}}, \sigma_{{-E}})$,$\mathcal{D} (\sigma_{{-E}}, \sigma_{{G-}})$ only affect weakly population dynamics, i.e., the observables $\sigma_i^{\dagger} \sigma_i$ ($i=0,1$), and are therefore neglected. We justify this approach next. 

\subsection{Omission of off-diagonal dissipators for population dynamics}
\label{appendix:omission}
Let us consider the effects of all eight dissipators on the evolution of the reduced density matrix operator,
\begin{align}
\frac{d \rho}{dt} &= \gamma_{+} \left( \mathcal{D} (\sigma_{{G+}}) + \mathcal{D} (\sigma_{{+E}}) \right)
+ \gamma_{+} \left(  \mathcal{D} (\sigma_{{G+}}, \sigma_{{+E}})  +  \mathcal{D} (\sigma_{{+E}}, \sigma_{{G+}}) \right) \nonumber \\
&+ \gamma_{-} \left( \mathcal{D} (\sigma_{{G-}}) + \mathcal{D} (\sigma_{{-E}}) \right) 
- \gamma_{-} \left( \mathcal{D} (\sigma_{{G-}}, \sigma_{{-E}}) + \mathcal{D} (\sigma_{{-E}}, \sigma_{{G-}}) \right) \left[ \rho \right],
\label{app:dissipative_ME_all_eight_dissipators}
\end{align}
which can be expressed in matrix form in the interaction basis, ordered as $\left\lbrace \ket{G}, \ket{\Lambda_-}, \ket{\Lambda_+}, \ket{E} \right\rbrace$, as 
\begin{align}
\frac{d}{dt} \begin{pmatrix}\rho'_{00} & \rho'_{01} & \rho'_{02} & \rho'_{03}\\\rho'_{10} & \rho'_{11} & \rho'_{12} & \rho'_{13}\\\rho'_{20} & \rho'_{21} & \rho'_{22} & \rho'_{23}\\\rho'_{30} & \rho'_{31} & \rho'_{32} & \rho'_{33}\end{pmatrix} =
\begin{pmatrix}\gamma_{+} \rho'_{22} + \gamma_{-} \rho'_{11} & -\gamma_{-} \left(\frac{\rho'_{01} }{2}+ \rho'_{13}\right) & \gamma_{+} \left(- \frac{\rho'_{02}}{2} + \rho'_{23}\right) & -\frac{ \left(\gamma_{+} + \gamma_{-}\right) }{2} \rho'_{03}\\ -\gamma_{-} \left(\frac{\rho'_{10}}{2} + \rho'_{31}\right) & \gamma_{-} \left(- \rho'_{11} + \rho'_{33}\right) & \frac{-1}{2}\left(\gamma_{+} + \gamma_{-}\right) \rho'_{12} & -\left(\frac{ \gamma_{+} }{2} + \gamma_{-}\right) \rho'_{13}\\\gamma_{+} \left(- \frac{\rho'_{20} }{2}+ \rho'_{32}\right) & \frac{1}{2} \left(- \gamma_{+} - \gamma_{-}\right) \rho'_{21} & \gamma_{+} \left(- \rho'_{22} + \rho'_{33}\right) & -\left(\gamma_{+} + \frac{ \gamma_{-} }{2} \right) \rho'_{23}\\ \frac{-1}{2} \left(\gamma_{+} + \gamma_{-}\right) \rho'_{30} & -\left( \frac{ \gamma_{+} }{2} +  \gamma_{-}\right) \rho'_{31} & -\left( \gamma_{+} + \frac{\gamma_{-} }{2} \right) \rho'_{32} &  - \left(\gamma_{+} + \gamma_{-}\right) \rho'_{33}\end{pmatrix},
\label{app:density_matrix_evo}
\end{align} 
where $\rho' = P^{\dag} \rho P$ is the reduced density operator in the interaction basis and $P$ the operator that changes basis from the computational basis to the interaction basis according to \autoref{eqn:app_basis_transformation}. 

We can compute  the explicit dependence of our observables of interest, namely $\sigma_i^{\dagger} \sigma_i$, $i=0,1$, in terms of the reduced density operator elements,
\begin{align}
\left< \sigma_0^{\dag} \sigma_0 \right> &\equiv \mathrm{Tr} \left(  \sigma_0^{\dag} \sigma_0 \rho \right) = \mathrm{Tr} \left( \left[ \sigma^{\dag}\sigma \otimes \mathbf{1} \right] \rho \right) = 
\mathrm{Tr} \left( \left[ \left| 10 \right> \left< 10 \right| + \left| 11 \right> \left< 11 \right| \right] \rho \right) =\nonumber  \\
&=\sum_{i \in \{ G, \Lambda_-, \Lambda_+, E \}} \left< i \right| \left( \left[ \frac{1}{2} \left( \left| \Lambda_+ \right> \left<\Lambda_+  \right| + \left| \Lambda_+ \right> \left<\Lambda_-  \right| + \left| \Lambda_- \right> \left<\Lambda_+  \right| + \left| \Lambda_- \right> \left<\Lambda_-  \right|\right) + \left| E \right> \left< E \right| \right] \rho \right) \left| i \right> = \nonumber \\
&= \frac{1}{2} \left( \rho'_{22} + \rho'_{12} + \rho'_{21} + \rho'_{11} \right) + \rho'_{33},
\label{eqn:exp_value_sigma0}
\end{align} 
\begin{align}
\left< \sigma_1^{\dag} \sigma_1 \right> &\equiv \mathrm{Tr} \left(  \sigma_1^{\dag} \sigma_1 \rho \right) =  \mathrm{Tr} \left( \left[ \mathbf{1} \otimes \sigma^{\dag}\sigma \right] \rho \right) = 
\mathrm{Tr} \left( \left[ \left| 01 \right> \left< 01 \right| + \left| 11 \right> \left< 11 \right| \right] \rho \right) = \nonumber \\
&=\sum_{i \in \{ G, \Lambda_-, \Lambda_+, E \}} \left< i \right| \left( \left[ \frac{1}{2} \left( \left| \Lambda_+ \right> \left<\Lambda_+  \right| - \left| \Lambda_+ \right> \left<\Lambda_-  \right| - \left| \Lambda_- \right> \left<\Lambda_+  \right| + \left| \Lambda_- \right> \left<\Lambda_-  \right|\right) + \left| E \right> \left< E \right| \right] \rho \right) \left| i \right> = \nonumber \\
&= \frac{1}{2} \left( \rho'_{22} - \rho'_{12} - \rho'_{21} + \rho'_{11} \right) + \rho'_{33}.
\label{eqn:exp_value_sigma1}
\end{align}
Differentiating equations \autoref{eqn:exp_value_sigma0} and \autoref{eqn:exp_value_sigma1} with respect to time allows us to insert the matrix element evolutions in \autoref{app:density_matrix_evo} to obtain
\begin{align}
\frac{d}{dt} \left\langle \sigma_0^{\dag}\sigma_0 \right\rangle
&= \frac{d}{dt} \left[ \frac{1}{2} \left( \rho'_{22} + \rho'_{12} + \rho'_{21} + \rho'_{11} \right) + \rho'_{33} \right] \nonumber\\
&= -\frac{1}{2}\left[
\left(\gamma_{+}+\gamma_{-} \right)\rho'_{33}
+ \gamma_{+}\rho'_{22}
+ \gamma_{-}\rho'_{11}
+ \frac{1}{2}(\gamma_{+}+\gamma_{-})(\rho'_{12}+\rho'_{21})
\right],
\label{eqn:evolution_exp_value_sigma0}
\end{align}
\begin{align}
\frac{d}{dt} \left\langle \sigma_1^{\dag}\sigma_1 \right\rangle
&=  \frac{d}{dt} \left[ \frac{1}{2} \left( \rho'_{22} - \rho'_{12} - \rho'_{21} + \rho'_{11} \right) + \rho'_{33} \right] \nonumber\\
&= -\frac{1}{2}\left[
\left(\gamma_{+} + \gamma_{-}\right)\rho'_{33}
+ \gamma_{+}\rho'_{22}
+ \gamma_{-}\rho'_{11}
- \frac{1}{2}(\gamma_{+} + \gamma_{-})(\rho'_{12} + \rho'_{21})
\right].
\label{eqn:evolution_exp_value_sigma1}
\end{align}
If instead of considering all eight dissipators in \autoref{app:dissipative_ME_all_eight_dissipators} we only consider the four diagonal ones, the matrix elements of the reduced density operator evolve as follows
\begin{align}
\frac{d}{dt} \begin{pmatrix}\rho'_{00} & \rho'_{01} & \rho'_{02} & \rho'_{03}\\\rho'_{10} & \rho'_{11} & \rho'_{12} & \rho'_{13}\\\rho'_{20} & \rho'_{21} & \rho'_{22} & \rho'_{23}\\\rho'_{30} & \rho'_{31} & \rho'_{32} & \rho'_{33}\end{pmatrix} =
\begin{pmatrix}\gamma_{+} \rho'_{22} + \gamma_{-} \rho'_{11} & - \gamma_{-} \frac{\rho'_{01} }{2} & -\gamma_{+} \frac{\rho'_{02}}{2} & \frac{ \left(- \gamma_{+} - \gamma_{-}\right) }{2} \rho'_{03}\\\ -\gamma_{-} \frac{\rho'_{10}}{2} & \gamma_{-} \left(- \rho'_{11} + \rho'_{33}\right) & \frac{1}{2}\left(- \gamma_{+} - \gamma_{-}\right) \rho'_{12} & \left(- \frac{ \gamma_{+} }{2} - \gamma_{-}\right) \rho'_{13}\\ -\gamma_{+}\frac{\rho'_{20} }{2} & \frac{1}{2} \left(- \gamma_{+} - \gamma_{-}\right) \rho'_{21} & \gamma_{+} \left(- \rho'_{22} + \rho'_{33}\right) & \left(-  \gamma_{+} - \frac{ \gamma_{-} }{2} \right) \rho'_{23}\\ \frac{1}{2} \left(- \gamma_{+} - \gamma_{-}\right) \rho'_{30} & \left(- \frac{ \gamma_{+} }{2} -  \gamma_{-}\right) \rho'_{31} & \left(- \gamma_{+} - \frac{\gamma_{-} }{2} \right) \rho'_{32} &  - \left(\gamma_{+} + \gamma_{-}\right) \rho'_{33}\end{pmatrix}.
\label{app:density_matrix_evo_with only diagonals}
\end{align} 
Comparing the total evolution in \autoref{app:density_matrix_evo} with the evolution of the reduced density operator under the exclusive action of the four diagonal dissipators in \autoref{app:density_matrix_evo_with only diagonals}, we find that none of the five density matrix elements on which $\left< \sigma_0^{\dagger} \sigma_0 \right>$ and $\left< \sigma_1^{\dagger} \sigma_1 \right>$ depend -  $\rho'_{11}$,  $\rho'_{12}$, $\rho'_{21}$, $\rho'_{22}$ and  $\rho'_{33}$ - is directly affected by the four off-diagonal dissipators $\mathcal{D} (\sigma_{{G+}}, \sigma_{{+E}})$, $\mathcal{D} (\sigma_{{+E}}, \sigma_{{G+}})$, $\mathcal{D} (\sigma_{{G-}}, \sigma_{{-E}})$,$\mathcal{D} (\sigma_{{-E}}, \sigma_{{G-}})$. Moreover, these five density-matrix elements form a closed subsystem of differential equations:
\begin{align}
\begin{cases}
\frac{d \rho'_{11}}{dt} =  \gamma_{-} \left(- \rho'_{11} + \rho'_{33}\right) \\
\frac{d \rho'_{22}}{dt} = \gamma_{+} \left(- \rho'_{22} + \rho'_{33}\right) \\
\frac{d \rho'_{33}}{dt} = - \left(\gamma_{+} + \gamma_{-}\right) \rho'_{33}  \\
\frac{d \rho'_{12}}{dt} = \frac{1}{2}\left(- \gamma_{+} - \gamma_{-}\right) \rho'_{12} \\ 
\frac{d \rho'_{21}}{dt} = \frac{1}{2} \left(- \gamma_{+} - \gamma_{-}\right) \rho'_{21}
\end{cases}.
\label{eqn:system_differential_equations_rho}
 \end{align}
Therefore, the off-diagonal dissipators do not affect $\left< \sigma_0^{\dagger} \sigma_0 \right>$ or $\left< \sigma_1^{\dagger} \sigma_1 \right>$ here through their coupling to other density-matrix elements. This finding shows that to accurately describe the population dynamics of the two-emitter chain it suffices to model the four diagonal dissipators  $\mathcal{D} (\sigma_{{G+}})$, $\mathcal{D} (\sigma_{{+E}})$, $\mathcal{D} (\sigma_{{G-}})$, $\mathcal{D} (\sigma_{{-E}})$ in our circuits. Furthermore, including the unitary evolution term $-i[H,\rho]$ in the master equation \autoref{app:dissipative_ME_all_eight_dissipators} also does not couple neither of the five matrix elements of interest, namely $\rho'_{11}$,  $\rho'_{12}$, $\rho'_{21}$, $\rho'_{22}$ and  $\rho'_{33}$, with the rest and therefore does not affect the reasoning above. In fact, if we consider the complete master equation
\begin{align}
\frac{d \rho}{dt} &= -i[H,\rho] + \gamma_{+} \left( \mathcal{D} (\sigma_{{G+}}) + \mathcal{D} (\sigma_{{+E}}) \right) \left[ \rho \right] + 
\gamma_{+} \left(  \mathcal{D} (\sigma_{{G+}}, \sigma_{{+E}})  +  \mathcal{D} (\sigma_{{+E}}, \sigma_{{G+}}) \right) \left[ \rho \right] + \nonumber \\
&+ \gamma_{-} \left( \mathcal{D} (\sigma_{{G-}}) + \mathcal{D} (\sigma_{{-E}}) \right) \left[ \rho \right] -
\gamma_{-} \left( \mathcal{D} (\sigma_{{G-}}, \sigma_{{-E}}) + \mathcal{D} (\sigma_{{-E}}, \sigma_{{G-}}) \right) \left[ \rho \right],
\label{app:full_ME_all_eight_dissipators}
\end{align}
and compute the reduced density operator matrix elements,
\begin{align}
\frac{d}{dt} \begin{pmatrix}\rho'_{00} & \rho'_{01} & \rho'_{02} & \rho'_{03}\\\rho'_{10} & \rho'_{11} & \rho'_{12} & \rho'_{13}\\\rho'_{20} & \rho'_{21} & \rho'_{22} & \rho'_{23}\\\rho'_{30} & \rho'_{31} & \rho'_{32} & \rho'_{33}\end{pmatrix} = \nonumber  \\ =
\resizebox{0.96\textwidth}{!}{$
\begin{pmatrix}\gamma_{+} \rho'_{22} + \gamma_{-} \rho'_{11} & \left[ i\left(\tilde{\omega} - \tilde{g}_{0,1} \right) - \frac{\gamma_{-}}{2} \right] \rho'_{01} - \gamma_{-} \rho'_{13} & \left[ i\left(\tilde{\omega} + \tilde{g}_{0,1} \right) - \frac{\gamma_{+}}{2} \right] \rho'_{02} + \gamma_{+} \rho'_{23} & \left( 2i \tilde{\omega} - \frac{\gamma_{+} + \gamma_{-}}{2} \right)\rho'_{03} \\
\left[ -i\left(\tilde{\omega} - \tilde{g}_{0,1} \right) -\frac{\gamma_{-}}{2} \right] \rho'_{10} - \gamma_{-} \rho'_{31}  & \gamma_{-} \left(\rho'_{33} - \rho'_{11}\right) & \left(2i\tilde{g}_{0,1} - \frac{\gamma_{+} + \gamma_{-}}{2} \right) \rho'_{12} & \left[ i(\tilde{\omega} + \tilde{g}_{0,1})- \frac{ \gamma_{+} }{2} - \gamma_{-}\right] \rho'_{13}\\
\left[ -i\left(\tilde{\omega} + \tilde{g}_{0,1} \right) - \frac{\gamma_{+}}{2} \right] \rho'_{20} + \gamma_{+} \rho'_{32}  & \left(-2i\tilde{g}_{0,1} - \frac{\gamma_{+} + \gamma_{-}}{2} \right) \rho'_{21} & \gamma_{+} \left(\rho'_{33} - \rho'_{22} \right) & \left[ i(\tilde{\omega} - \tilde{g}_{0,1}) - \frac{ \gamma_{-} }{2} - \gamma_{+}\right] \rho'_{23} \\ 
\left( -2i \tilde{\omega} - \frac{\gamma_{+} + \gamma_{-}}{2} \right)\rho'_{30} & \left[ -i(\tilde{\omega} + \tilde{g}_{0,1})- \frac{ \gamma_{+} }{2} - \gamma_{-}\right] \rho'_{31}& \left[-i \left(\tilde{\omega} - \tilde{g}_{0,1} \right) - \gamma_{+} - \frac{\gamma_{-} }{2} \right] \rho'_{32} &  - \left(\gamma_{+} + \gamma_{-}\right) \rho'_{33}\end{pmatrix} $},
\label{app:complete_ME}
\end{align}
we observe the same decoupling of the matrix elements  $\rho'_{11}$,  $\rho'_{12}$, $\rho'_{21}$, $\rho'_{22}$ and  $\rho'_{33}$ from the rest.

\begin{figure*}[tbp]
    \centering
    \def\svgwidth{\textwidth}
    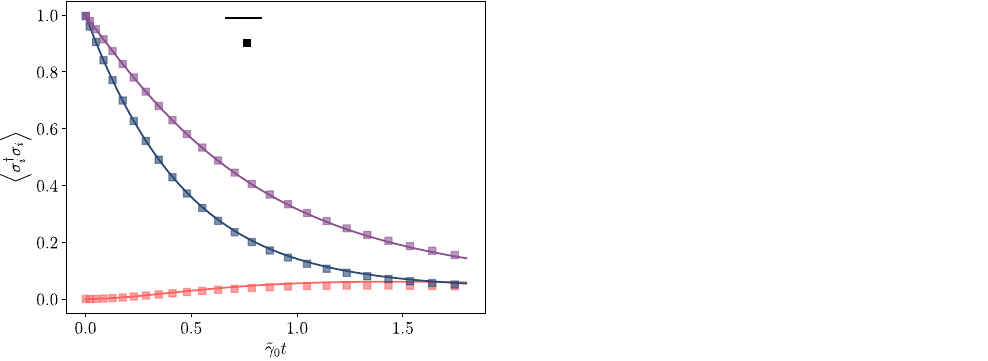
    \caption{\footnotesize Analysis of the validity of omitting the off-diagonal dissipators for $n = 3$ emitters using classical computational methods. (a) Time evolution of the excited-state population of the three emitters, labeled $q_0$ (red), $q_1$ (blue) and $q_2$ (purple), under the Markovian equation in \autoref{eqn:gksl_equation} with $\tilde{H} = 0$ and $n=3$ (solid lines) and under the (purely dissipative) master equation including only diagonal dissipators (squares). (b) Time evolution of the absolute error $\left| \rho^{(diag)}_{ii} -  \rho^{(full)}_{ii} \right|$, $i = 1, 2, ..., 7$, where $\rho^{(full)}$ is the density operator evolved under  \autoref{eqn:gksl_equation} with $\tilde{H} = 0$ and $n=3$ and $\rho^{(diag)}$ is the density operator evolved under the exclusive action of the diagonal dissipators. The absolute error associated to each density matrix is represented by a different colored circle, as indicated in the legend. The initial state of the three-emitter system is $\ket{\psi(0)}_S =  \ket{011}_S$. All emitters and couplings are assumed identical and the effective parameters used are $\tilde{\omega} = 1.2045$eV, $\tilde{g}_{i,i+1} = 4.5$ meV, $\tilde{\gamma} = 9$meV, $\tilde{\gamma}_{i,i+1} = 9$meV, $\forall i = 0, 1, 2$.}
    \label{fig:validity_omission_off_diagonals}
\end{figure*}

The finding that the off-diagonal dissipators can be neglected to study population dynamics is no longer exact for $n>2$ emitters. However, we discuss next that this effect can be expected to be small. We consider specifically the case of $n = 3$ and compute the time evolution of the excited state population of each emitter, $\left< \sigma_i^{\dag} \sigma_i \right>$ (i = 0,1,2), in terms of the elements of the reduced density matrix of the three emitters $\rho_{ij}$ ($i,j = 0, 1, ..., 7$) as
\begin{align}
\frac{d}{dt} \left< \sigma_0^{\dag} \sigma_0 \right> \equiv \frac{d}{dt} \mathrm{Tr} \left(  \sigma_0^{\dag} \sigma_0 \rho \right) =
\frac{d}{dt} \left( \rho_{44}+ \rho_{55} + \rho_{66} + \rho_{77} \right),
\label{eqn:exp_value_sigma0_n_3}
\end{align} 
\begin{align}
\frac{d}{dt} \left< \sigma_1^{\dag} \sigma_1 \right> \equiv \frac{d}{dt} \mathrm{Tr} \left(  \sigma_1^{\dag} \sigma_1 \rho \right) = 
\frac{d}{dt} \left( \rho_{22}+ \rho_{33} + \rho_{66} + \rho_{77} \right),
\label{eqn:exp_value_sigma1_n_3}
\end{align} 
\begin{align}
\frac{d}{dt} \left< \sigma_2^{\dag} \sigma_2 \right> \equiv \frac{d}{dt} \mathrm{Tr} \left(  \sigma_2^{\dag} \sigma_2 \rho \right) = 
\frac{d}{dt} \left( \rho_{11}+ \rho_{33} + \rho_{55} + \rho_{77} \right).
\label{eqn:exp_value_sigma2_n_3}
\end{align} 
We show in \autoref{fig:validity_omission_off_diagonals}a the time evolution of the excited-state populations of every emitter, when considering only the purely-diagonal terns (squares) and the (exclusively-dissipative) master equation in \autoref{eqn:gksl_equation} with $\tilde{H}=0$ and $n = 3$ (solid lines), for the initial state $\ket{\psi(0)}_S =  \ket{011}_S$. Overall, there is a good agreement between the two approaches, specially for qubits $q_1$ (blue) and $q_2$ (purple).

We plot in \autoref{fig:validity_omission_off_diagonals}b the time evolution of the error $\left| \rho_{ii}^{(diag)} -  \rho_{ii}^{(full)} \right|$ between the density matrix elements obtained using the diagonal dissipators approach ($\rho^{(diag)}$) and the density matrix elements obtained using the (exclusively-dissipative) master equation in \autoref{eqn:gksl_equation} with $\tilde{H}=0$ and $n = 3$ ($\rho^{(full)}$). For the initial state considered, elements $\rho_{33}$, $\rho_{66}$, $\rho_{77}$ almost perfectly agree in both approaches, with errors remaining always under $10^{-5}$, while elements $\rho_{11}$, $\rho_{22}$ and $\rho_{44}$ present errors of the order of $10^{-2}$. Taking this small error into consideration, we conclude that neglecting the off-diagonal dissipators serves as a good approach for approximately capturing population dynamics. More importantly, this approximation enables a direct connection to amplitude damping channels and facilitates the construction of the quantum circuits, as discussed in \autoref{two_qubit_disspators}.

\subsection{Justification of the quantum circuit for cross-dissipation}
\label{app:justification_circuit}
In this subsection we derive the core quantum circuit (shown in \autoref{fig:two_emitters_model}c) that models the four diagonal dissipators $\mathcal{D} (\sigma_{{G+}})$, $\mathcal{D} (\sigma_{{+E}})$, $\mathcal{D} (\sigma_{{G-}})$, $\mathcal{D} (\sigma_{{-E}})$, following a similar procedure as for the single-qubit dissipator case, discussed in \autoref{sec:level2a}. 

First of all, we solve the system of linear differential equations of the density matrix elements in \autoref{eqn:system_differential_equations_rho}, corresponding to the purely-dissipative dynamics. The analytical solution of these equations is given by
\begin{align}
\begin{cases}
\rho'_{11} (t) =  \rho'_{11}(0) e^{-\gamma_{-}t} + \rho'_{33} (0) \frac{\gamma_{-}}{\gamma_{+}} e^{-\gamma_{-}t}  \left(  1 -  e^{-\gamma_{+} t} \right), \\
\rho'_{22} (t) =  \rho'_{22}(0) e^{-\gamma_{+}t} + \rho'_{33} (0) \frac{\gamma_{+}}{\gamma_{-}} e^{-\gamma_{+}t} \left( 1  -  e^{-\gamma_{-} t} \right), \\
\rho'_{33} (t) =  \rho'_{33} (0)  e^{-(\gamma_{-} +\gamma_{+} ) t},  \\
\rho'_{12} (t) =  \rho'_{12} (0)  e^{-\frac{(\gamma_{-} +\gamma_{+} )}{2} t}, \\ 
\rho'_{21} (t) =  \rho'_{21} (0)  e^{-\frac{(\gamma_{-} +\gamma_{+} )}{2} t}.
\end{cases}
\label{eqn:analytical_solution_system_differential_equations_rho}
 \end{align}

We show next that this time evolution can be approximated at first order in $t$ with the circuit in \autoref{fig:two_emitters_model}c. First, we verify that this circuit accomplishes the same transformations as the quantum channel $\varphi$ defined through the set of transformation equations in \autoref{eqn:conjoined_channel_transformation_equations}, repeated here for convenience

\begin{align}
\left| G \right>_S \otimes \left|00 \right>_A &\to \left| G \right>_S \otimes \left|00 \right>_A, \nonumber \\
\left| \Lambda_- \right>_S \otimes \left|00 \right>_A & \to \sqrt{1- p_{-}} \left| \Lambda_- \right>_S \otimes \left|00 \right>_A + \sqrt{p_{-}} \left| G \right>_S \otimes \left|01 \right>_A, \nonumber \\
\left| \Lambda_+ \right>_S \otimes \left|00 \right>_A &\to \sqrt{1- p_{+}} \left| \Lambda_+ \right>_S \otimes \left|00 \right>_A + \sqrt{p_{+}} \left| G \right>_S \otimes \left|10 \right>_A, \nonumber \\
\left| E \right>_S \otimes \left|00 \right>_A &\to \sqrt{1- p_{-}}\sqrt{1- p_{+}} \left| E \right>_S \otimes \left|00 \right>_A +  \sqrt{p_{+}}\sqrt{1- p_{-}} \left| \Lambda_+ \right>_S \otimes \left|01 \right>_A +  \nonumber \\ &+ \sqrt{p_{-}}\sqrt{1- p_{+}} \left| \Lambda_- \right>_S \otimes \left|10 \right>_A + \sqrt{p_{-}}\sqrt{p_{+}} \left| G \right>_S \otimes \left|11 \right>_A ,
\label{eqn:app_conjoined_channel_transformation_equations}
\end{align}
with the identification  $\sin^2 \theta_{\pm} = p_{\pm}$. To show this equivalence, we take as an example the last state $\left| E \right>_S \otimes \left|00 \right>_A$, corresponding to initializing both system qubits that represent the quantum emitters to the excited state and each ancilla qubit, to the zero state. In the circuit in \autoref{fig:two_emitters_model}c, the basis change gate $P$ allows to work in the interaction basis $\left\lbrace \ket{G}, \ket{\Lambda_-}, \ket{\Lambda_+}, \ket{E} \right\rbrace$. Let $U$ be the unitary transformation representing the part of the circuit sandwiched between the $P$ and $P^{\dagger}$ gates. $U$ can be decomposed as a sequence $U = U_5 U_4 U_3 U_2 U_1 U_0$, where $ U_0 = CCRy_{q_0, q_1 \to a_0} (\theta_{+} - \theta_{-})$, $U_1 = CCRy_{q_0, q_1 \to a_1} (\theta_{-} - \theta_{+})$, $U_2 = CRy_{q_0 \to a_0} (\theta_{-})$, $U_3 = CRy_{q_1 \to a_1} (\theta_{+})$, $U_4 =CX_{a_0 \to q_0} $ and $U_5 =CX_{a_1 \to q_2} $. Applying every gate step-by-step yields

\begin{align}
\left| E \right>_S \otimes \left|00 \right>_A \xrightarrow{P_{q_0, q_1}} \ket{11}_S \otimes \ket{00}_A \xrightarrow{U_0}   \ket{11}_S \otimes Ry (\theta_{+} - \theta_{-}) \ket{0}_{a_0} \otimes \ket{0}_{a_1} \nonumber \\
\xrightarrow{U_1} \ket{11}_S \otimes Ry (\theta_{+} - \theta_{-}) \ket{0}_{a_0} \otimes Ry (\theta_{-} - \theta_{+}) \ket{0}_{a_1} \nonumber \\
\xrightarrow{U_2} \ket{11}_S \otimes Ry (\theta_{+}) \ket{0}_{a_0} \otimes Ry (\theta_{-} - \theta_{+}) \ket{0}_{a_1} \xrightarrow{U_3} \ket{11}_S \otimes Ry (\theta_{+}) \ket{0}_{a_0} \otimes Ry (\theta_{-}) \ket{0}_{a_1}  = \nonumber \\
= \ket{11}_S \otimes \left( \cos \theta_{+} \ket{0}_{a_0} + \sin \theta_{+} \ket{1}_{a_0}  \right) \otimes \left( \cos \theta_{-} \ket{0}_{a_1} + \sin \theta_{-} \ket{1}_{a_1}  \right)  =  \nonumber \\
= \ket{11}_S \otimes \left( \cos \theta_{+} \cos \theta_{-} \ket{00}_{A} +\cos \theta_{+}\sin \theta_{-} \ket{01}_{A}  + \sin \theta_{+} \cos \theta_{-} \ket{10}_{A}  +  \sin \theta_{+} \sin \theta_{-} \ket{11}_{A} \right)
\nonumber \\
\xrightarrow{U_4}  \left( \cos \theta_{+} \cos \theta_{-} \ket{11}_S \otimes \ket{00}_{A} +\cos \theta_{+}\sin \theta_{-} \ket{11}_S \otimes \ket{01}_{A}  + \sin \theta_{+} \cos \theta_{-} \ket{01}_S \otimes \ket{10}_{A}  +  \sin \theta_{+} \sin \theta_{-} \ket{01}_S \otimes \ket{11}_{A} \right) \nonumber \\
\xrightarrow{U_5}  \left( \cos \theta_{+} \cos \theta_{-} \ket{11}_S \otimes \ket{00}_{A} +\cos \theta_{+}\sin \theta_{-} \ket{10}_S \otimes \ket{01}_{A}  + \sin \theta_{+} \cos \theta_{-} \ket{01}_S \otimes \ket{10}_{A}  +  \sin \theta_{+} \sin \theta_{-} \ket{00}_S \otimes \ket{11}_{A} \right)
\label{eqn:app_example_state} 
\end{align}

This procedure can be repeated with the rest of input states to reveal that the identification $\sin^2 \theta_{\pm} = p_{\pm}$ recovers the transformation equations in \autoref{eqn:app_conjoined_channel_transformation_equations}. 

Last, we need to show that the transformation given by \autoref{eqn:app_conjoined_channel_transformation_equations} is equivalent to solving the master equation. Based on \autoref{eqn:app_conjoined_channel_transformation_equations}, we define the isometry $W: \mathcal{H}_S \to \mathcal{H}_S \otimes \mathcal{H}_A$ in the interaction basis as
\begin{align}
W \left| G \right>_S &= \left| G \right>_S \otimes \left|00 \right>_A, \nonumber \\
W \left| \Lambda_- \right>_S &= \sqrt{1- p_{-}} \left| \Lambda_- \right>_S \otimes \left|00 \right>_A + \sqrt{p_{-}} \left| G \right>_S \otimes \left|01 \right>_A, \nonumber \\
W \left| \Lambda_+ \right>_S &= \sqrt{1- p_{+}} \left| \Lambda_+ \right>_S \otimes \left|00 \right>_A + \sqrt{p_{+}} \left| G \right>_S \otimes \left|10 \right>_A, \nonumber \\
W \left| E \right>_S &= \sqrt{1- p_{-}}\sqrt{1- p_{+}} \left| E \right>_S \otimes \left|00 \right>_A + \sqrt{p_{+}}\sqrt{1- p_{-}} \left| \Lambda_+ \right>_S \otimes \left|01 \right>_A +  \nonumber \\ &+ \sqrt{p_{-}}\sqrt{1- p_{+}} \left| \Lambda_- \right>_S \otimes \left|10 \right>_A +  \sqrt{p_{-}}\sqrt{p_{+}} \left| G \right>_S \otimes \left|11 \right>_A.
\label{eqn:app_isometric_representation}
\end{align}
From the isometry $W$ we can now compute a set of Kraus operators, $\left\lbrace M_j \right\rbrace_{j=0}^{3}$, representing the quantum channel $\varphi$ in \autoref{eqn:app_conjoined_channel_transformation_equations} using the relation $W = \sum_{j=0}^{3} M_j \otimes \ket{e_j}_{A}$, where $\left\lbrace \ket{e_j} \right\rbrace_{j=0}^{3} = \left\lbrace \ket{00}, \ket{01}, \ket{10}, \ket{11} \right\rbrace$ is the two-qubit computational basis in $\mathcal{H}_A$. With this relation, we obtain $M_j = \left( \mathbf{1}_S \otimes \bra{e_j}_A \right) W$, for $j=0,1,2,3$. In matrix form,

\begin{align}
M_0 &= \begin{pmatrix}1 & 0 & 0 & 0\\0 & \sqrt{1 - p_{-}} & 0 & 0\\0 & 0 & \sqrt{1 - p_{+}} & 0\\0 & 0 & 0 & \sqrt{1 - p_{+}} \sqrt{1 - p_{-}}\end{pmatrix}, \nonumber \\
M_1 &= \begin{pmatrix}0 & \sqrt{p_{-}} & 0 & 0\\0 & 0 & 0 & 0\\0 & 0 & 0 & \sqrt{p_{+}} \sqrt{1 - p_{-}}\\0 & 0 & 0 & 0\end{pmatrix}, \nonumber \\
M_2 &= \begin{pmatrix}0 & 0 & \sqrt{p_{+}} & 0\\0 & 0 & 0 & \sqrt{p_{-}} \sqrt{1 - p_{+}}\\0 & 0 & 0 & 0\\0 & 0 & 0 & 0\end{pmatrix}, \nonumber \\
M_3 &= \begin{pmatrix}0 & 0 & 0 & \sqrt{p_{+}} \sqrt{p_{-}}\\0 & 0 & 0 & 0\\0 & 0 & 0 & 0\\0 & 0 & 0 & 0\end{pmatrix}.
\label{eqn:app_kraus_representation}
\end{align}

Using $\rho' (t) = \varphi[\rho' (0)] = \sum_{i=0}^{3} M_i \rho' (0) M_i^{\dagger}$ (first line in \autoref{eqn:kraus_decomposition}), we obtain the evolution of the density matrix under the channel $\varphi$ defined in \autoref{eqn:app_conjoined_channel_transformation_equations} (and thus the circuit in \autoref{fig:two_emitters_model}c). In particular, since we are interested in population dynamics, we write explicitly the equations of the elements  $\rho'_{11}$,  $\rho'_{12}$, $\rho'_{21}$, $\rho'_{22}$ and  $\rho'_{33}$. From \autoref{eqn:kraus_decomposition}, we obtain

\begin{align}
\begin{cases}
\rho'_{11} (t) =  \left(1 - p_{-}\right) \rho'_{11}{\left(0 \right)} +  p_{-} \left(1 - p_{+}\right) \rho'_{33}{\left(0 \right)} \\
\rho'_{22} (t) =  \left(1 - p_{+}\right) \rho'_{22}{\left(0 \right)} + p_{+} \left(1 - p_{-}\right) \rho'_{33}{\left(0 \right)}  \\
\rho'_{33} (t) = \left(1 - p_{+}\right) \left(1 - p_{-}\right) \rho'_{33}{\left(0 \right)}  \\
\rho'_{12} (t) =  \sqrt{1 - p_{+}} \sqrt{1 - p_{-}} \rho'_{12}{\left(0 \right)} \\ 
\rho'_{21} (t) =  \sqrt{1 - p_{+}} \sqrt{1 - p_{-}} \rho'_{21}{\left(0 \right)}
\end{cases}.
\label{eqn:app_kraus_evolution_rho}
\end{align}
This expression is equivalent to first order in $t$ to \autoref{eqn:analytical_solution_system_differential_equations_rho} with the identification $ p_{\pm} \approx 1 - e^{-\gamma_{\pm} t}$. Indeed, using the Taylor expansion of the exponential and square root functions $e^{-\gamma_{\pm} t} = 1 - \gamma_{\pm} t + \mathcal{O} \left( t^2 \right)$ and $\sqrt{1 - p_{\pm}} = 1 - \frac{\gamma_{\pm}t}{2} + \mathcal{O} \left( t^2 \right)$, \autoref{eqn:app_kraus_evolution_rho} reproduces the first-order expansion of the analytical solution in \autoref{eqn:analytical_solution_system_differential_equations_rho}. This first-order agreement is sufficient in the context of Trotterization, where each circuit realizes the evolution over a sufficiently small time step $t/k$.

\section{Generalization of the core circuit to $n$-qubit chains}
\label{appendix:generalization_to_n_qubits}

As mentioned in \autoref{sec:level2c}, we can use the circuit in \autoref{fig:two_emitters_model}c as the core unit for generalizing our approach to the cases of $n$-qubit chains. We expand this point in this appendix. The $n$-qubit chain would evolve according to the ME in \autoref{eqn:gksl_equation}

\begin{align}
\frac{d\rho}{dt} = -i [\tilde{H}, \rho] +\sum_{i=0}^{n-1} \tilde{\gamma}_i\mathcal{D} (\sigma_i) [\rho] + 
\sum_{i=0}^{n-2} \tilde{\gamma}_{i,i+1}\mathcal{D} (\sigma_i, \sigma_{i+1}) [\rho] + \sum_{i=0}^{n-2} \tilde{\gamma}_{i+1,i}\mathcal{D} (\sigma_{i+1}, \sigma_{i}) [\rho],
\label{eqn:app_gksl_equation}
\end{align}

with the effective Hamiltonian in \autoref{eqn:n_emitters_hamiltonian},

\begin{equation}
\tilde{H} =  \sum_{i=0}^{n-1}  \tilde{\omega}_i \sigma^{\dagger}_i \sigma_i + \sum_{i=0}^{n-2} \tilde{g}_{i, i+1} \left(\sigma^{\dagger}_i\sigma_{i+1} + \sigma_i\sigma_{i+1}^{\dagger} \right)
\label{eqn:app_n_emitters_hamiltonian}
\end{equation}

and the single-body and two-body dissipators, also introduced previously in \autoref{eqn:n_emitters_single_body_dissipators},

\begin{equation}
 \tilde{\gamma}_{i,j} \mathcal{D} (\sigma_i, \sigma_j) [\rho]  = \tilde{\gamma}_{i,j} \left( \sigma_i \rho \sigma^{\dagger}_j - \frac{1}{2} \left\lbrace \sigma^{\dagger}_j \sigma_i, \rho \right\rbrace\right).
\label{eqn:app_n_emitters_single_body_dissipators}
\end{equation}

Again, we assume identical qubits for simplicity to obtain $\tilde{\omega}_i = \tilde{\omega}_j \equiv \tilde{\omega}$, $\tilde{\gamma}_i = \tilde{\gamma}_j \equiv \tilde{\gamma}$ ($i,j = 0,1,...,n-1$) and $\tilde{g}_{i,i+1} = \tilde{g}_{j,j+1} \equiv \tilde{g}$, $\tilde{\gamma}_{i,i+1} = \tilde{\gamma}_{j,j+1} \equiv \tilde{\gamma}_{0,1}$ ($i,j = 0,1,...,n-2$). Both the Hamiltonian and dissipators in \autoref{eqn:app_n_emitters_hamiltonian} and \autoref{eqn:app_n_emitters_single_body_dissipators} decompose into local terms that act non-trivially on at most two qubits (nearest-neighbor interaction). For Trotterization, it is convenient to edge-color the chain, i.e., divide the two-body terms into two groups (even and odd), so that all terms within each group commute:

\begin{equation}
\tilde{H} = \sum_{i=0}^{n-1} \tilde{H}^{(0)}_{i} + \sum_{\substack{j = 0\\ \text{j even}}}^{n-2} \tilde{H}^{(int)}_{j,j+1} + \sum_{\substack{j = 1\\ \text{j odd}}}^{n-2} \tilde{H}^{(int)}_{j,j+1},
\label{eqn:app_n_emitters_hamiltonian_rewritten}
\end{equation}

\begin{align}
\mathcal{D} [\rho] = \tilde{\gamma} \sum_{i=0}^{n-1} \mathcal{D} (\sigma_i) [\rho] + 
\tilde{\gamma}_{0,1} \sum_{\substack{j = 0\\ \text{j even}}}^{n-2}  \left( \mathcal{D} (\sigma_j, \sigma_{j+1}) + \mathcal{D} (\sigma_{j+1}, \sigma_{j})   \right) [\rho] + 
\tilde{\gamma}_{0,1} \sum_{\substack{j = 0\\ \text{j odd}}}^{n-2}  \left( \mathcal{D} (\sigma_j, \sigma_{j+1}) +  \mathcal{D} (\sigma_{j+1}, \sigma_{j})   \right) [\rho],
\label{eqn:app_n_emitters_dissipators_rewritten}
\end{align}
where we have defined the free-body Hamiltonian $\tilde{H}^{(0)}_{i} = \tilde{\omega} \sigma^{\dagger}_i \sigma_i$ and the nearest-neighbor interaction Hamiltonian $\tilde{H}^{(int)}_{j,j+1} = \tilde{g} \left(\sigma^{\dagger}_j \sigma_{j+1} + \sigma_j \sigma_{j+1}^{\dagger} \right)$.

We discuss next the implementation of each part of the dynamics (unitary and dissipative) as quantum circuits separately.

\subsection{\label{unitary_dynamics_appendix} Unitary dynamics}
Each summand in both the free-body and the interaction Hamiltonians (\autoref{eqn:app_n_emitters_hamiltonian_rewritten}) can be rewritten in terms of Pauli-X, Y and Z operators, using $\sigma \equiv \frac{1}{2} (X - i Y)$, which results in

\begin{align}
\tilde{H}^{(0)}_{j} =  \frac{\tilde{\omega}}{2} \left( \mathbf{1}_j - Z_j \right), 
\label{eqn:app_rewrite_single_body_hamiltonians_pauli_basis}
\end{align}
\begin{align}
\tilde{H}^{(int)}_{j,j+1} =  \frac{\tilde{g}}{2} \left( X_j X_{j+1} + Y_j Y_{j+1} \right),
\label{eqn:app_rewrite_two_body_hamiltonians_pauli_basis}
\end{align}
where $X_k$, $Y_k$, $Z_k$ and $\mathbf{1}_k$ denote the Pauli-X, -Y, -Z and identity operators acting on the $k$-th qubit, respectively. The decomposition of the Hamiltonians in the Pauli basis in \autoref{eqn:app_rewrite_single_body_hamiltonians_pauli_basis} and \autoref{eqn:app_rewrite_two_body_hamiltonians_pauli_basis} helps to identify single and two-qubit rotation gates.

We can now obtain an approximation of the unitary evolution operator $U(t) = \exp (-i\tilde{H}t)$ using the Trotter-Suzuki decomposition as

\begin{align}
U(t) \approx \prod_{p = 0}^{k} \prod_{\substack{l = 1\\ \text{l odd}}}^{n-2} \exp \left( {-i\tilde{H}^{(int)}_{l,l+1} \frac{t}{k}} \right) \prod_{\substack{m = 0\\ \text{m even}}}^{n-2} \exp \left( {-i\tilde{H}^{(int)}_{m,m+1} \frac{t}{k}} \right) \prod_{j = 0}^{n-1} \exp \left( {-i \tilde{H}^{(0)}_{j} \frac{t}{k}} \right),
\end{align}
where $k$ is the number of Trotter steps and thus $t/k$ is the duration of each of these steps. Using  the Pauli decomposition of the Hamiltonian we can identify each of these exponentials with quantum gates,

\begin{align}
\exp \left( {-i\tilde{H}^{(int)}_{j,j+1} \frac{t}{k}}\right) = \exp \left( {-i\frac{\tilde{g}t}{2k} \left( X_j X_{j+1} + Y_j Y_{j+1} \right)} \right) \approx 
R_{XX, q_j, q_{j+1}} \left( \alpha \right) R_{YY, q_j, q_{j+1}} \left( \alpha \right),
\end{align}
\begin{align}
\exp \left( {-i\tilde{H}^{(0)}_{j} \frac{t}{k}} \right) = \exp \left( -i\frac{\tilde{\omega}t}{2k} \left( \mathbf{1} - Z_j \right) \right) = R_{Z, q_j} \left( \beta \right),
\end{align}
where we have defined the angles $\alpha = \frac{\tilde{g} t}{k}$, $\beta = \frac{\tilde{\omega}t}{k}$ and used that $R_{Z, q_j} \left(\theta \right) \equiv \exp \left(  -i \frac{\theta}{2} Z_j \right)$, $R_{XX, q_j, q_{j+1}} (\theta) \equiv \exp \left(  -i \frac{\theta}{2} X_j X_{j+1} \right)$ and $R_{YY, q_j, q_{j+1}} (\theta) \equiv \exp \left(  -i \frac{\theta}{2} Y_j Y_{j+1} \right)$. We obtain

\begin{align}
U(t) &\approx \prod_{p = 0}^{k} \prod_{\substack{l = 1\\ \text{l odd}}}^{n-2}  R_{XX, q_l, q_{l+1}} \left( \alpha \right) R_{YY, q_l, q_{l+1}} \left( \alpha \right) 
\prod_{\substack{m = 0\\ \text{m even}}}^{n-2}  R_{XX, q_m, q_{m+1}} \left( \alpha \right) R_{YY, q_m, q_{m+1}} \left( \alpha \right) \prod_{j = 0}^{n-1} R_{Z, q_j} \left( \beta \right) \nonumber \\ 
&\equiv \prod_{l = 0}^{k} \prod_{\substack{l = 1\\ \text{l odd}}}^{n-2} U^{(int)}_{q_l, q_{l+1}} \prod_{\substack{m = 0\\ \text{m even}}}^{n-2}  U^{(int)}_{q_m, q_{m+1}} \prod_{j = 0}^{n-1} U^{(0)}_{q_j}.
\label{eqn:unitary_total}
\end{align}

\begin{figure*}[tbp]
    \centering
    \def\svgwidth{\textwidth}
    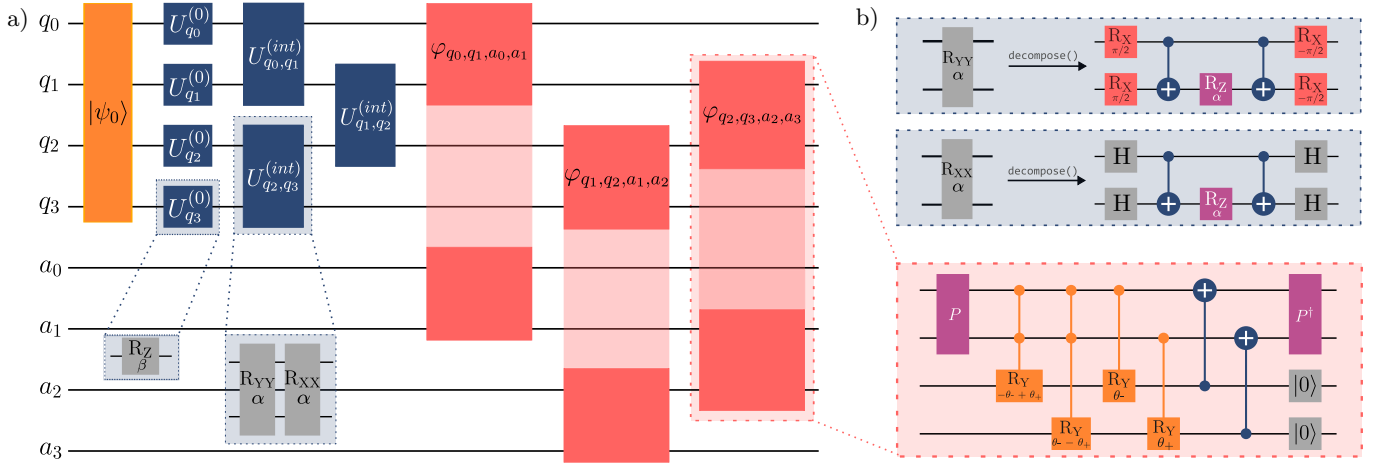
    \caption{\footnotesize Quantum circuit modelling the population dynamics of a four-emitter chain evolving under the master equation in \autoref{eqn:app_gksl_equation}. a) Schematic circuit representation for one Trotter step. The four quantum emitters are represented by the system qubits $q_0$, $q_1$, $q_2$, $q_3$. For every system qubit $q_i$ representing a quantum emitter, an ancillary qubit $a_i$ is used to model the dissipative part of the dynamics. The circuit works as follows: first, we initialize the qubits to the state $\ket{\psi_0} = \ket{q_0 q_1 q_2 q_3} \otimes \ket{a_0 a_1 a_2 a_3} = \ket{\psi} \otimes \ket{0000}$ (orange box labeled $\ket{\psi_0}$), with $\ket{\psi}$ the initial state of the quantum emitters. Then, a layer of $R_{Z, q_j} \left( \beta \right)$ gates ($j = 0,1,2,3$), with $\beta = \tilde{\omega}t/k$ is applied to the system qubits, representing the unitary evolution operator $U^{(0)}_{q_j}$ associated to the free Hamiltonian $\tilde{H}^{(0)}_{j}$ (one-qubit boxes in dark blue, labeled $U^{(0)}_{q_j} $, $j=0,1,2,3$). Afterwards, we apply the interaction layer $\prod_{\substack{m = 0\\ \text{m even}}}^{2}  R_{XX, q_m, q_{m+1}} \left( \alpha \right) R_{YY, q_m, q_{m+1}} \left( \alpha \right)$ associated to the interaction Hamiltonian $\tilde{H}^{(int)}_{j,j+1}$ with even $j$ (two-qubit dark blue boxes labeled $U^{(int)}_{q_0, q_{1}}$ and $U^{(int)}_{q_2, q_{3}}$, acting on qubit pairs $(q_0, q_1)$ and $(q_2, q_3)$, respectively). An equivalent layer is repeated right after for odd $j$ (two-qubit dark blue box labeled  $U^{(int)}_{q_1, q_{2}}$, acting on qubit pair $(q_1, q_2)$). Finally, we apply the dissipative part of the dynamics through the conjoined quantum channel $\varphi$ defined in \autoref{eqn:conjoined_channel_transformation_equations}. Similarly to the unitary evolution approach, we apply this channel first in parallel to the qubit-ancilla pairs $(q_0, q_1,a_0,a_1)$ and $(q_2, q_3, a_2, a_3)$, as represented by the first two red boxes, labeled $\varphi_{q_0,q_1,a_0,a_1}$ and $\varphi_{q_2,q_3,a_2,a_3}$. Last, we apply the quantum channel to the remaining pair $(q_1, q_2,a_1,a_2)$, as represented by the last red box, labeled  $\varphi_{q_1, q_2,a_1,a_2}$. b) Decomposition of the two-qubit interaction gates $R_{YY}  \left( \alpha \right)$, $R_{XX}  \left( \alpha \right)$ into one-qubit rotations and $CX$ gates (top two panels shaded in gray). Both decompositions were obtained using the method \texttt{decompose()} in \texttt{Qiskit}. The bottom panel, shaded in light red, corresponds to the decomposition of the decay channel $\varphi$, given by the transformation equations in \autoref{eqn:conjoined_channel_transformation_equations} and modeled using the original circuit derived in  \autoref{fig:two_emitters_model}c (in the final simulations and quantum experiments, the circuit in \autoref{fig:dynamic_circuits}c is used instead).}
    \label{fig:generalization_schematic}
\end{figure*}

The circuit that implements one Trotter step of the operator in \autoref{eqn:unitary_total}, for four emitters, is shown in \autoref{fig:generalization_schematic}a as the ensemble of dark blue boxes (with the decomposition of the interaction gates $R_{XX, q_j, q_{j+1}} \left( \alpha \right)$ and $R_{YY, q_j, q_{j+1}} \left( \alpha \right)$ into single and two-qubit gates shown by the gray boxes below and at the top of \autoref{fig:generalization_schematic}b).

\subsection{Dissipative dynamics}
Regarding the dissipative parts of the dynamics, \autoref{eqn:app_n_emitters_dissipators_rewritten} can be written in an equivalent form
\begin{align}
\mathcal{D} [\rho] = \sum_{\substack{j = 0\\ \text{j even}}}^{n-2}  \left(  \tilde{\gamma}\mathcal{D} (\sigma_j)  + \tilde{\gamma} \mathcal{D} (\sigma_{j+1}) + \tilde{\gamma}_{j,j+1}\mathcal{D} (\sigma_j, \sigma_{j+1}) \tilde{\gamma}_{j+1,j}\mathcal{D} (\sigma_{j+1}, \sigma_{j})   \right)  [\rho] +  \nonumber  \\
+ \sum_{\substack{j = 0\\ \text{j odd}}}^{n-2}  \left( \tilde{\gamma} \mathcal{D} (\sigma_j) + \tilde{\gamma} \mathcal{D} (\sigma_{j+1}) + \tilde{\gamma}_{j,j+1}\mathcal{D} (\sigma_j, \sigma_{j+1}) +  \tilde{\gamma}_{j+1,j}\mathcal{D} (\sigma_{j+1}, \sigma_{j})   \right)  [\rho].
\label{eqn:app_n_emitters_dissipators_rewritten_pairwise}
\end{align}

For a given $j$, the dissipators $\tilde{\gamma} \mathcal{D}(\sigma_j) [\rho] + \tilde{\gamma} \mathcal{D}(\sigma_{j+1}) [\rho]
  +\allowbreak \tilde{\gamma}_{j,j+1}\mathcal{D}(\sigma_j, \sigma_{j+1}) [\rho] +\allowbreak \tilde{\gamma}_{j+1,j}\mathcal{D}(\sigma_{j+1}, \sigma_j) [\rho]$ in \autoref{eqn:app_n_emitters_dissipators_rewritten_pairwise} are equivalent to \autoref{eqn:ME_two_emitters_app}, and thus, according to the discussion in \autoref{two_qubit_disspators}, can be represented by either the original circuit in \autoref{fig:two_emitters_model}c using $n$ ancilla qubits or the hardware-aware dynamic circuit shown in \autoref{fig:dynamic_circuits}c which involves $\lfloor n/2 \rfloor$ ancilla qubits. Importantly, and regardless of the circuit representation of the channel, within each layer (even or odd $j$), the corresponding channels act on disjoint qubit pairs and therefore can be executed in parallel. This is shown schematically in the red boxes in \autoref{fig:generalization_schematic}a, with the decomposition at the bottom of \autoref{fig:generalization_schematic}b, for the particular choice of the circuit in \autoref{fig:two_emitters_model}c as the representation of the decay channels (we choose this circuit, which is not the one used in the final simulations and quantum experiments, because of its simpler pictorial representation).
  
\begin{figure}[tbp]
    \centering
    \def\svgwidth{0.5\textwidth}
    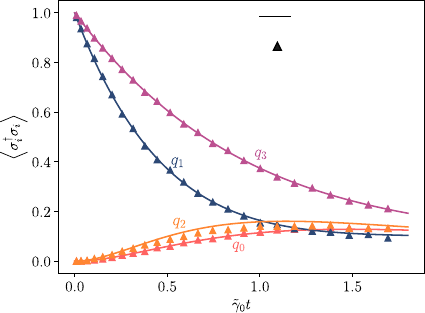
    \caption{\footnotesize Time evolution of the excited-state population of every emitter in a four-emitter chain. The solid lines represent the exact solution of the master equation in \autoref{eqn:gksl_equation}, while the triangles represent the results obtained from executing the circuit in \autoref{fig:generalization_schematic}a in a classical noiseless simulation for $k = 10$ Trotter steps and $N_{shots} = 2\cdot 10^4$. The initial state of the four-emitter system is $\ket{\psi(0)}_S =  \ket{0101}_S$. All emitters are assumed identical and the effective parameters used are $\tilde{\omega} = 1.2045$eV, $\tilde{g}_{i,i+1} = 4.5$ meV, $\tilde{\gamma} = 9$meV, $\tilde{\gamma}_{i,i+1} = 9$meV, $\forall i = 0, 1, 2$.}
    \label{fig:app_validation_circuit}
\end{figure}

The validity of the quantum circuit shown in \autoref{fig:generalization_schematic}a is illustrated in \autoref{fig:app_validation_circuit}. In particular, we show the time evolution of the excited population of every emitter in a four-emitter chain computed in two ways: (i) solving exactly \autoref{eqn:gksl_equation} (solid lines) and (ii) executing a noiseless classical simulation of the circuit in \autoref{fig:two_emitters_model}c for $k = 10$ Trotter steps and $N_{shots} = 2\cdot 10^4$ (triangles). We observe a good agreement between the exact solution and the Trotterized circuit, for a large number of Trotter steps, up to the diagonal approach approximation error discussed in \appref{appendix:omission}.

\section{ \label{app_error_mitigation} Error mitigation techniques and bootstrapping}
Error mitigation techniques are designed to reduce the impact of noise in quantum devices. Since current QPUs are noisy, raw measurement outcomes are often distorted by gate errors, decoherence, and readout errors. Error mitigation methods aim to improve estimates of observables by post-processing the raw measured data, using additional circuits, noise-rescaling procedures, or discarding non-physical outputs. These approaches are essential for extracting reliable results from current quantum experiments, especially when working with deep circuits or complex operations—such as mid-circuit measurements or resets—that can further increase the noise. 

In this section we describe the statistical and error-mitigation procedures used to analyze the data obtained in the quantum experiments. We first introduce the bootstrap resampling method employed to estimate statistical uncertainties from the raw results of the quantum experiments. We then analyze the suitability to our dissipative circuits of two widely used error-mitigation techniques—Zero-Noise Extrapolation (ZNE) and Clifford Data Regression (CDR). We compare different implementations of these techniques, including several Cliffordization rules within CDR, and show that CDR with inverse bias is the most favorable for our system. We implement these procedures from scratch—rather than relying on \texttt{Qiskit}’s built-in functions—to gain full control of each step of the procedure.
 
  \subsection{Bootstrapping}
\label{appendix:bootstrap}
We use bootstrap resampling \cite{Efron1994} to estimate the statistical uncertainty of observables obtained from the raw QPU data. For a given circuit and measurement basis, the QPU returns a finite sample of bitstrings and their respective counts. From this empirical distribution, we generate $B = 10^2$ resampled datasets of size $N_{shots}$ by randomly sampling with replacement from the bitstrings obtained from the original experiment, using the same number of shots as in the original experiment. For each bootstrap sample $b$, we compute the observable of interest, obtaining a set of estimates $\{\langle O \rangle^{(b)}\}_{b=0}^{B-1}$. The mean of this bootstrap distribution is used as the central value (in \autoref{fig:error_mitigation_benchmark} and \autoref{fig:quantum_experiment_results}), while its standard deviation provides an estimate of the uncertainty associated with finite sampling only; it does not account for the uncertainty associated with errors arising from hardware imperfections.

This procedure does not require assuming, for example, a Gaussian or other form for the measured distribution, since the resampling is performed directly from the experimentally observed bitstrings. Throughout the manuscript, shaded uncertainty regions associated with raw results of the quantum experiments are obtained from this bootstrap procedure. As such, they quantify the statistical uncertainty associated with finite sampling of the measured bitstrings, rather than the full effect of hardware noise. When error mitigation is applied, the uncertainty regions are obtained by propagating these bootstrap estimates to obtain the full uncertainty of the mitigated observables (\appref{appendix:constrained_richardson_extrapolation}, \appref{appendix:bised_clifford_data_regression}).

\subsection{Zero-noise extrapolation. Constrained Richardson extrapolation}
\label{appendix:constrained_richardson_extrapolation}
Zero-noise extrapolation (ZNE) is one of the most widely adopted error mitigation strategies for noisy intermediate-scale quantum (NISQ) devices \cite{Giurgica2020}. The main idea behind ZNE is to artificially amplify the noise in a given quantum circuit, measure the corresponding noisy expectation values, and then extrapolate these results back to the zero-noise limit. Noise amplification is typically achieved through gate folding techniques \cite{Temme_2017}, in which selected quantum gates are replaced by equivalent sequences that reproduce the same unitary operation while increasing the effective noise exposure of the circuit. In practice, this can be implemented through full gate folding, where all (unitary) gates $U$ are replaced by the sequence $U U^\dagger U$, or through partial folding strategies that provide finer control over the overall noise amplification. We follow this latter approach, which allows us to explore a variety of noise regimes without excessively increasing circuit depth.

Concretely, in the calculations discussed in \autoref{fig:error_mitigation_benchmark} we follow Ref. \cite{Temme_2017} and implement random partial gate folding of two-qubit layers under four different noise-amplification factors  $\lambda \in \left\lbrace 1.3, 1.6, 2, 2.3 \right\rbrace$. We perform this folding at the transpiled-circuit level, where all the two-qubit gates are translated into single-qubit gates and $CZ$ gates. For a fixed circuit depth, each value of $\lambda$ corresponds to folding a definite number of $CZ$ layers. For example, for the circuit in \autoref{fig:dynamic_circuits}c used for \autoref{fig:error_mitigation_benchmark}b, the values $\lambda \in \left\lbrace 1.3, 1.6, 2, 2.3 \right\rbrace$ correspond to folding 10, 21, 34 and 45 $CZ$ layers, respectively. Since the choice of layers to fold is random, for each $\lambda$ we generate $N_{rand} = 100$ independently-folded circuits and take the average of the resulting expectation values.

Once several noise-amplified expectation values $\{\left<O \right> (\lambda_i) \}$ are obtained, the zero-noise value can be recovered through extrapolation techniques. In this work, we use linear extrapolation, where $\left<O \right>$ is assumed to vary linearly with $\lambda$, as well as second-order polynomial, exponential and Richardson extrapolation \cite{Richardson1911, Temme_2017} (note that we do not plot in \autoref{fig:error_mitigation_benchmark}b the results of all the approaches that we implement). Richardson extrapolation constructs a weighted linear combination of noisy results designed to cancel leading-order noise terms in an expansion in orders of $\lambda$. The first-order Richardson extrapolation corresponds to combining results from multiple noise levels to eliminate the first-order contribution of the noise amplification to the expectation value. 

Concretely, let $\{ \left<O \right> (\lambda_i) \}_{i=0}^{m-1}$ denote the set of $m$ noise-amplified expectation values retrieved. Throughout, we assume that folding produces an effective noise channel whose action on expectation values is analytic in $\lambda$ in a neighborhood of $\lambda=0$. Under this assumption, the observable $ \left<O \right>$ admits an expansion in powers of $\lambda$, 
\begin{equation}
\left<O \right> (\lambda) = \left<O \right>^{*} + c_1 \lambda + c_2\lambda ^2 + \cdots,
\label{app:analytic_expansion}
\end{equation}
where $c_i$ ($i = 0, 1, ...$) are the expansion coefficients and $ \left<O \right>^{*}$ denotes the zero-noise expectation value. First-order Richardson (FOR) extrapolation seeks coefficients $\{ \omega_i \}_{i=0}^{m-1}$ such that the estimator
\begin{equation}
\left<O \right>_{\mathrm{FOR}} = \sum_{i = 0}^{m-1} \omega_i \left<O \right> (\lambda_i)
\label{app:richardson_estimator}
\end{equation}
cancels the contribution from the linear term in the expansion in \autoref{app:analytic_expansion}
\begin{equation}
\sum_{i = 0}^{m-1} \omega_i = 1, \hspace{20mm} \sum_{i = 0}^{m-1} \omega_i  \lambda_i = 0,
\label{app:richardson_constraints}
\end{equation}
so that $\left<O \right>_{\mathrm{FOR}} =  \left<O \right>^{*}  + \mathcal{O} (\lambda^2)$.

More in detail, \autoref{app:richardson_constraints} can be obtained by substituting the expansion in \autoref{app:analytic_expansion} into the different terms in the right-hand side of \autoref{app:richardson_estimator}. \autoref{app:richardson_constraints} results from setting to one the prefactor on the term proportional to $\left<O \right>^{*}$ and cancelling the prefactor of the leading term in $\lambda$. After solving the system of equations for $\omega_i$, we obtain the zero-noise expectation value by applying \autoref{app:richardson_estimator}.

We consider in what follows $\lambda \in \left\lbrace 1.3, 1.6, 2, 2.3 \right\rbrace$. First, we compute $\left<O \right>_{\mathrm{FOR}}$ for each pair of consecutive noise amplification factors $(\lambda_i, \lambda_{i+1})$. The values labeled ZNE(FOR) in \autoref{fig:error_mitigation_benchmark}b (yellow triangles) have been obtained using the first-order Richardson extrapolation approach and averaging the values obtained for $\left<O \right>_{\mathrm{FOR}}$ over all consecutive pairs.

On the other hand, if we use all four noise-amplified expectation values simultaneously in \autoref{app:richardson_estimator}, the system of equations \autoref{app:richardson_constraints} remains undetermined, meaning that the first-order Richardson constraints no longer determine the coefficients uniquely. This freedom can be exploited to impose additional constraints on the coefficient vector. In this context, we reformulate the extrapolation method as a constrained optimization problem \cite{Boyd2004, Matousek2007}, choosing the coefficients  $\{ \omega_i \}_{i=0}^{m-1}$ by minimizing the $l1$-norm of the coefficient vector $\vec{\omega} = (\omega_0, \omega_1, ..., \omega_{m-1})$ subject to the first-order Richardson constraints,
\begin{equation}
\min_{\omega_i} ||\vec{\omega}||_1 \hspace{5mm} \text{subject to} \hspace{5mm} \sum_{i = 0}^{m-1} \omega_i = 1, \hspace{3mm} \sum_{i = 0}^{m-1} \omega_i  \lambda_i = 0,
\end{equation}
which penalizes large extrapolation coefficients and thereby helps reduce the amplification of the noise (a substantial disadvantage of ZNE). From now on, we call this method constrained first-order Richardson extrapolation with $l1$-norm minimization. The same formulation can be applied to higher-order Richardson extrapolations provided that the extrapolation order is smaller than the number of noise-amplified results. The values labeled ZNE(CFOR-$l1$) in \autoref{fig:error_mitigation_benchmark}b (yellow squares) have been obtained using this first-order Richardson extrapolation with $l1$-norm minimization approach.

The statistical deviation of the noise-amplified expectation values $\{ \left<O \right> (\lambda_i) \}_{i=0}^{m-1}$ is estimated by bootstrap resampling, as discussed in \appref{appendix:bootstrap}. These bootstrap deviations are then propagated through the extrapolation procedure to obtain the statistical uncertainty of the zero-noise expectation value. For linear extrapolation, the statistical deviation is extracted from the covariance matrix of the fitted parameters. For Richardson extrapolation, the zero-noise estimate is a linear combination of noise-amplified expectation values (\autoref{app:richardson_estimator}), and its statistical uncertainty is obtained by standard error propagation for a linear estimator,
\begin{equation}
\sigma^2_{\mathrm{FOR}} = \sum_{i = 0}^{m-1} \omega_i^2 \sigma_{i, \mathrm{raw}}^2,
\end{equation}
where $\sigma_{i, \mathrm{raw}}$ is the bootstrap deviation of the raw data associated with $\left<O \right> (\lambda_i)$. Therefore, the extrapolation error regions reported in \autoref{fig:error_mitigation_benchmark}b should be interpreted as propagated statistical uncertainties.

%

As discussed in \autoref{sec:level4a}, independently of the extrapolation method used, ZNE does not introduce any substantial improvement of the quantum experiment results in intermediate-size chains, as shown in \autoref{fig:error_mitigation_benchmark}b for 15 total qubits. For this reason, we do not use any ZNE protocol for the large-scale simulations shown in \autoref{sec:level4b}. 



\subsection{Clifford data regression with biased Cliffordization}
\label{appendix:bised_clifford_data_regression}
Clifford Data Regression (CDR) is an error mitigation technique that leverages the efficient classical simulability of Clifford circuits to obtain information on how to correct the effect of noise on the expectation values of more general, non-Clifford circuits. The central idea is to construct a regression model that maps noisy quantum expectation values, obtained from quantum experiments for a set of related Clifford circuits, to their corresponding noiseless values, computed classically. This mapping is then used to predict the noise-corrected value of the target, non-Clifford circuit. In practice, the method involves generating a collection of “Cliffordized” versions of the original circuit, which preserve as much structure as possible after replacing non-Clifford gates with Clifford counterparts. These modified circuits can be efficiently simulated on classical hardware —by the Gottesman-Knill theorem \cite{gottesman1998, Aaronson_2004}—, enabling the creation of a paired dataset $\{(\langle O \rangle^{(i)}_{\mathrm{noiseless}}, \langle O \rangle^{(i)}_{\mathrm{noisy}})\}_{i=0}^{N-1}$ used to train a regression model whose coefficients capture the relationship between noisy and ideal expectation values,
\begin{equation}
\langle O \rangle_{\mathrm{noisy}} = a + b \langle O \rangle_{\mathrm{noiseless}},
\end{equation}
where $a$ and $b$ are regression parameters obtained by standard least-squares fitting.

Once the regression parameters are determined, the CDR-corrected expectation value for the original (non-Clifford) circuit, $\langle O \rangle_{\mathrm{CDR}}$, is obtained by evaluating the model at the noisy expectation value of the target circuit  $\langle O \rangle_{\mathrm{raw}}$
\begin{equation}
\langle O \rangle_{\mathrm{CDR}} = \frac{ \langle O \rangle_{\mathrm{raw}} }{b} - a.
\label{app:cdr_correction}
\end{equation}

Importantly, the overall performance of Clifford data regression depends heavily on how faithfully the Cliffordized circuits reproduce the noise structure of the original circuit. If the noise structure is very different, the regression can suffer from systematic bias. To mitigate this effect, we tailor the Cliffordization process to maximize the similarity between the original and Cliffordized circuits, as suggested in Ref. \cite{Czarnik2021}. From now on, we refer to this implementation as biased Clifford data regression. In the following we describe the details of this custom Cliffordization process.

The only non-Clifford gates that appear in our circuits after transpilation are (i) measurements, (ii) resets, (iii) conditional blocks and (iv) arbitrary $R_z (\theta)$ rotations. First, measurements do not need to be converted into Clifford gates since they are efficiently classically simulable by the Gottesman-Knill theorem. Keeping these mid-circuit measurements ensures that the training circuits capture the readout errors more faithfully. A similar reasoning applies to the conditional blocks; even though these operations do not belong to the Clifford group, classical control flow operations do not increase the simulation complexity. As long as we transform the contents of the conditional blocks to Clifford gates, the overall process remains classically efficient. As a result, the only operations that need to be converted into Clifford gates are reset operations and arbitrary $R_z (\theta)$ gates. 

We tried different Cliffordization strategies for the reset gates, such as replacing them by $X$ gates, directly removing them from the circuit, or randomly replacing them with Clifford $R_z (\theta_C)$ rotations, where $\theta_C$ is an angle in the set $\{0, \pi/2, \pi, 3\pi/2\}$ for which the gate is Clifford and that is chosen randomly with equal probability. We found that the best performing of these strategies was the random replacement by $R_z (\theta_C)$ gates. 

Last, regarding the $R_z (\theta)$ gates with arbitrary angle $\theta$, they are also replaced by Clifford $R_z (\theta_C)$ gates, with $\theta_C \in \{0, \pi/2, \pi, 3\pi/2\}$. As an unbiased baseline, we choose this angle $\theta_C$ randomly, with equal probability for all values. We then compare this baseline with biased sampling rules that introduce a probabilistic bias that favors Clifford angles $\theta_C \in \{0, \pi/2, \pi, 3\pi/2\}$ closest to the original non-Clifford angle $\theta(t_i)$ of each parametrized rotation gate $R_z (\theta(t_i))$ at each simulation time $t_i$. To this end, for each Clifford angle $\theta_C$ we define the angular distance $d(\theta(t_i), \theta_C)$ as the distance modulo $2\pi$, and assign probability weights $\omega (\theta_C, \theta(t_i))$ to the Clifford angles according to different distributions.

Specifically, we define the probability of choosing a Clifford angle $\theta_C$ conditioned on the original non-Clifford angle $\theta(t_i)$, $p(\theta_C | \theta (t_i))$, as
\begin{equation}
p(\theta_C | \theta (t_i)) = \frac{\omega (\theta_C, \theta(t_i))}{\sum_{\theta'_C \in \{0, \pi/2, \pi, 3\pi/2\}} \omega (\theta'_C, \theta(t_i))},
\end{equation} 
where the weights $\omega (\theta_C, \theta(t_i))$ are defined differently depending on the chosen probability distribution. We consider three choices of weights:
\begin{itemize}
\item The uniform distribution, $\omega (\theta_C, \theta(t_i)) = 1$, $\forall i$, used here as the unbiased Cliffordization baseline.
\item An inverse-distance distribution, $\omega (\theta_C, \theta(t_i)) = 1/(d(\theta(t_i), \theta_C) + \epsilon)$, with $\epsilon = 10^{-6}$ a small regularization parameter introduced to avoid divergences when $d(\theta(t_i), \theta_C) = 0$.
\item A Gaussian distribution centered at the original angle $\omega (\theta_C, \theta(t_i)) = \exp \left[ -d(\theta(t_i), \theta_C)^2/(2\sigma^2) \right]$, with $\sigma = 1$.
\end{itemize}
We found that the inverse-distance distribution results in the better mitigation of our results in \autoref{fig:error_mitigation_benchmark}b, and thus it is the one we use for the larger systems in \autoref{fig:quantum_experiment_results}.

Finally, the uncertainty of the CDR-corrected expectation value in \autoref{app:cdr_correction} is estimated using quadratic error propagation applied to the regression model and the bootstrapped data. More precisely, the variance $\sigma_{\mathrm{CDR}}^2$ is approximated as 
\begin{align}
\sigma_{\mathrm{CDR}}^2 &\approx \left(  \frac{\partial \langle O \rangle_{\mathrm{CDR}}}{\partial a} \right)^2  (\Delta a)^2 + \left(  \frac{\partial \langle O \rangle_{\mathrm{CDR}}}{\partial b} \right)^2  (\Delta b)^2 + \left(  \frac{\partial \langle O \rangle_{\mathrm{CDR}}}{\partial \langle O \rangle_{\mathrm{raw}}} \right)^2  (\Delta \langle O \rangle_{\mathrm{raw}})^2,
\label{app:cdr_error}
\end{align}
where $\Delta a$ and $\Delta b$ are extracted from the covariance matrix of the linear regression and $\Delta \langle O \rangle_{\mathrm{raw}}$ is the standard deviation of the bootstrapped raw data (see \appref{appendix:bootstrap}). The error estimate in \autoref{app:cdr_error} does not account for systematic model bias arising from deviations from linearity, but can provide a good measure of statistical uncertainty for comparing mitigation strategies.

\section{Numerical analysis of the Trotter error}
\label{trotterization_analysis}
Unitary dynamics, i.e., the time evolution dictated by the terms $-i[\tilde{H}, \rho]$ in the effective master equation (e.g. \autoref{eqn:gksl_equation}) are typically implemented as quantum circuits using the Trotter-Suzuki decomposition. Similar decompositions into product formulas can be applied to open quantum system dynamics governed by a Liouvillian $\mathcal{L}$ \cite{PhysRevLett.107.120501}. Our simulations follow a Trotter-like scheme in the sense that the evolution up to time $t$ is implemented by repeating $k$ circuit layers, each corresponding to a short-time evolution over an interval $t/k$. Thus, although the dissipative contribution is realized through CPTP maps based on ancilla dilations and resets rather than through an explicit product decomposition of the full Liouvillian, the overall protocol still defines a Trotterized approximation to the continuous-time dynamics. Increasing the number of Trotter steps improves the accuracy of this approximation at the cost of greater circuit depth.

In this appendix, we estimate the resulting Trotter error numerically by comparing the output of the Trotterized original circuit in \autoref{fig:two_emitters_model}c with the exact solution of the master equation \autoref{eqn:gksl_equation}, both obtained by (noiseless) classical simulations, for different system sizes. This serves two purposes: first, to verify convergence of the circuit dynamics toward the exact solution as $k$ increases; and second, to establish a practical criterion for choosing $k$ in larger systems, where exact diagonalization is no longer feasible.


\begin{figure*}[tbp]
    \centering
    \def\svgwidth{\textwidth}
    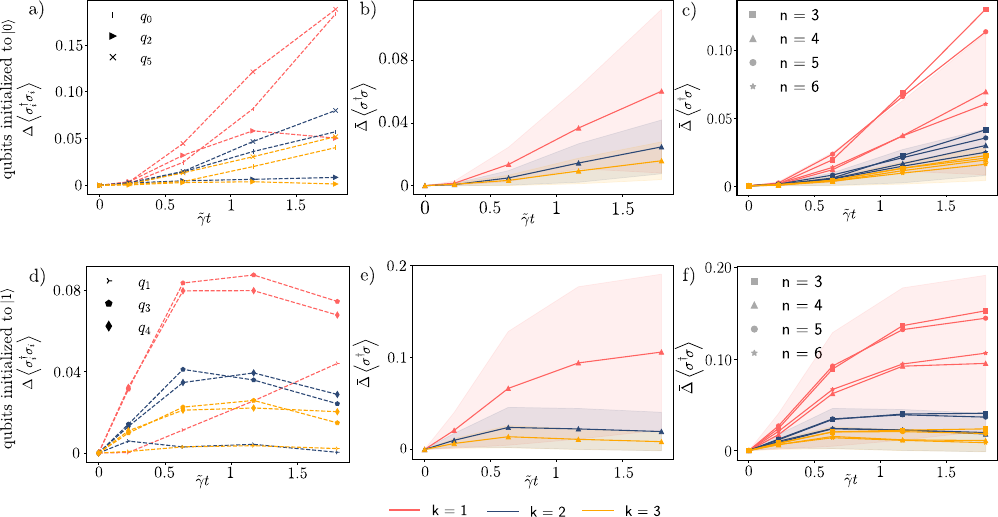
    \caption{\footnotesize Numerical analysis of the absolute Trotter error. This analysis compares the Trotterized circuit in \autoref{fig:two_emitters_model}c with the exact solution of the master equation in \autoref{eqn:gksl_equation}. (a, d) Time evolution of the absolute error of the excited-state population of every emitter in a six-emitter chain, $\Delta \left< \sigma^{\dagger}_i \sigma_i \right> = \left| \left< \sigma^{\dagger}_i \sigma_i \right>_{QC} - \left< \sigma^{\dagger}_i \sigma_i \right>_{exact} \right|$, $i = 0,1,...,5$, where $\left< \sigma^{\dagger}_i \sigma_i \right>_{QC}$ is the expectation value obtained from the classical noiseless simulations of the Trotterized circuit in \autoref{fig:two_emitters_model}c and $\left< \sigma^{\dagger}_i \sigma_i \right>_{exact}$ is the expectation value obtained from an exact classical computation of the master equation in \autoref{eqn:gksl_equation}. The system qubits are initialized to the state $\ket{\psi(0)}_S = \ket{010110}$. The results are plotted for $k = 1$ (red), $k=2$ (dark blue) and $k=3$ (yellow) Trotter steps and a different marker is used for each of the six qubits, as indicated in the legends. Panels (a) and (d) show the error for emitters initialized to the state $\ket{0}$ and $\ket{1}$, respectively. (b, e) Time evolution of the average absolute error $\bar{\Delta} \left< \sigma^{\dagger} \sigma \right>$ of the excited-state population of emitters of a six-emitter chain initialized to (b) the $\ket{0}$ state and (e) the $\ket{1}$ state among $N_{states} = 10$ randomized initial states, for $k = 1$ (red), $k=2$ (dark blue) and $k = 3$ (yellow) Trotter steps.  For each randomization, a random set of qubits, between two and four, are initialized to the $\ket{1}$ state, while leaving the rest at the $\ket{0}$ state. The shaded regions represent the standard deviation of the distribution. (c, f) Time evolution of the scaling of the average absolute Trotter error $\bar{\Delta} \left< \sigma^{\dagger} \sigma \right>$ with the number of emitters for $k = 1$ (red), $ k =2$ (dark blue) and $k = 3$ (yellow) Trotter steps and $N_{states} = 10$. The system sizes considered are $n=3$ (small circles), $n=4$ (dots), $n=5$ (big circles) and $n=6$ (upside down triangles). Panels (c) and (f) correspond to emitters initialized in $\ket{0}$ and $\ket{1}$, respectively. For every panel, all emitters and couplings are considered identical and the effective parameters used are $\tilde{\omega} = 1.2045$eV, $\tilde{g}_{i,i+1} = 4.5$ meV, $\tilde{\gamma} = 9$meV, $\tilde{\gamma}_{i,i+1} = 9$meV, $\forall i = 0, 1, ..., n - 2$.}
     \label{fig:trotter_analysis}
\end{figure*}

\Autoref{fig:trotter_analysis} summarizes this analysis. Panels a and d show the time evolution of the absolute Trotter error comparing the output of the Trotterized original circuit in \autoref{fig:two_emitters_model}c ($\left< \sigma^{\dagger}_i \sigma_i \right>_{QC}$) with the output of the exact classical solution of the master equation in \autoref{eqn:gksl_equation} ($\left< \sigma^{\dagger}_i \sigma_i \right>_{exact}$),
\begin{equation}
\Delta \left< \sigma^{\dagger}_i \sigma_i \right> = \left| \left< \sigma^{\dagger}_i \sigma_i \right>_{QC} - \left< \sigma^{\dagger}_i \sigma_i \right>_{exact} \right|,
\label{app_absolute_error}
\end{equation}
for each emitter in a six-emitter chain initialized to the state $\left| \psi (0) \right> = \left| 010110 \right>$. The dashed curves correspond to $k = 1$ (red curve), $k = 2$ (dark blue curve) and $k = 3$ (yellow curve) Trotter steps. The error decreases with increasing $k$, showing that the Trotterized circuit approaches the exact dynamics. At the same time, \autoref{fig:trotter_analysis}a and \autoref{fig:trotter_analysis}d reveal a clear dependence on the initial qubit state: the maximum error is typically smaller for qubits initialized in $\ket{1}$ than qubits initialized in $\ket{0}$. 

To analyze this dependence more systematically, we repeat the analysis over $N_{states} = 10$ randomly sampled initial states and compute the average error $\bar{\Delta} \left< \sigma^{\dagger} \sigma \right>$ separately for qubits initialized in $\ket{0}$ and in $\ket{1}$. In each realization, between two and four qubits are initialized in $\ket{1}$, while the remaining qubits are prepared in $\ket{0}$. The resulting averages for $k=1$ (red curve), $k=2$ (dark blue curve) and $k=3$ (yellow curve) Trotter steps, shown for qubits initialized in $\ket{0}$ in \autoref{fig:trotter_analysis}b and at $\ket{1}$ in \autoref{fig:trotter_analysis}e, confirm the same trend of decreasing error as $k$ increases, with shaded regions indicating the corresponding standard deviations.

We next examine the dependence of the Trotter error on system size by repeating the previous analysis for emitter chains comprising $n = 3, 4, 5, 6$ emitters. The results, shown in \autoref{fig:trotter_analysis}c for qubits initialized in $\ket{0}$ and in \autoref{fig:trotter_analysis}f for qubits initialized in $\ket{1}$, indicate that over the time range considered the error does not show any clear trend on $n$, and again decreases with increasing number of Trotter steps $k$. These results suggests that the analysis of the Trotter error of these chains can be used in the study of larger ones. 

More precisely, the error scaling for $n = 6$ (the largest chain that we solved by exact diagonalization) in \autoref{fig:trotter_analysis}f always stays under this $\Delta \left< \sigma^{\dagger} \sigma \right> = 0.016$ threshold. This particular value is extracted from the Trotter error corresponding to the maximum number of Trotter steps used in the quantum experiments, which in our case is $k =3$, at the longest simulation time $\tilde{\gamma} t = 1.8$. In order to ensure that the Trotter error always stays under this threshold, while minimizing the circuit depth per time step, we use the results in \autoref{fig:trotter_analysis}c, f to set the number of Trotter steps that we implement for each time instant in the quantum experiments in \autoref{fig:quantum_experiment_results} of the main text. Concretely, we use one Trotter step ($k=1$) for instants $t$ satisfying $\tilde{\gamma}t \leq 0.69$, two Trotter steps ($k=2$) for $0.68 < \tilde{\gamma}t \leq 1.26$ and three Trotter steps ($k=3$) for $1.26 < \tilde{\gamma}t \leq 1.8$.

\section{Matrix product state simulations with Monte Carlo methods}
\label{app:mcwf-tebd}
In this section we describe the classical simulation method used in \autoref{sec:level4} to obtain results, for system sizes beyond exact brute-force simulation, which are compared to the results of the quantum experiments. Instead of evolving the density matrix of the system $\rho$ directly, which requires computational resources that scale exponentially with the number of quantum emitters, we exploit the circuit representation introduced in \autoref{sec:level2}, which implements each step of the evolution by coupling the system qubits to an ancilla register followed by reset operations on the ancilla qubits. The main idea behind our approach is to combine a simulation in terms of matrix product states (MPS) \cite{Schollwoeck2011, Orus2014} with a quantum-trajectory (Monte-Carlo) \cite{Zoller1992} sampling of the reset operations. 

To combine these two methodologies, we consider that, as shown in \autoref{fig:generalization_schematic}a, one step of the time evolution circuit acting on the system register $S$ and the ancilla register $A$ can be separated into:
\begin{enumerate}
\item A unitary circuit sequence acting on $S$ representing coherent evolution.
\item A sequence of controlled gates acting on the system and the ancilla qubits implementing the dissipative channels.
\item Reset operations on the ancilla register $A$, implemented internally in \texttt{Qiskit} using a measurement on the computational basis followed by a conditional $X$ gate.
\end{enumerate}
According to the open quantum system theoretical framework, provided that the ancilla register $A$ is initialized in a fixed state at each step and discarded after interacting with the system register $S$, this sequence of operations implements a completely positive trace-preserving (CPTP) map $\varepsilon$ on the system register alone. Rather than simulating the reduced density matrix dynamics, we perform a stochastic unraveling \cite{Molmer1993} in terms of pure-state trajectories. Each trajectory involves representing the system-ancilla state $\ket{\psi}$ as an MPS. To evolve this MPS in time, we apply the circuit gates sequentially using tensor-network contractions with adequate truncation of the bond dimension within the time-evolving block decimation algorithm (TEBD) \cite{Vidal2004}. Concretely, and inspired by \texttt{Qiskit}'s implementation of reset operations, we model resets as projective measurements in the computational basis, followed by a sampling of the outcomes and a conditional application of an $X$ gate conditioned on the measurement outcome (see below for details). Repeating this procedure $N_{traj}$ times generates a sequence of pure states $\{ \ket{\psi_m} \}_{i=0}^{N_{traj}-1}$. From this ensemble of states, we can recover the expectation value of an observable $O$ by averaging over the $N_{traj}$ trajectories:
\begin{equation}
\left< O \right> \approx \frac{1}{N_{traj}} \sum_{i = 0}^{N_{traj}-1} \bra{\psi_i} O \ket{\psi_i}.
\label{app_expectation_value_MPSMC}
\end{equation}

This simulation strategy is related to recent work that has established a tensor-network trajectory method for open-system dynamics \cite{Sander2025}, as both combine an MPS representation with stochastic unraveling. In our approach, the simulation is closely aligned with the circuit realization used in the quantum experiments, because it is formulated directly at the circuit level: instead of unraveling a Liouvillian through non-Hermitian evolution and quantum jumps, we simulate the explicit ancilla-assisted implementation of the CPTP map and sample the measurement outcomes of the ancilla qubits.

We implement this method in Julia using the \texttt{ITensors} and \texttt{ITensorMPS} packages. We use as reference circuit the circuit in \autoref{fig:two_emitters_model}c and arrange the $2n$ physical sites of the MPS by allocating one ancilla pair after each system qubit pair. Then, for each time instant $t_i$, we initialize the MPS to the initial state at $t_0 = 0$ and build all the necessary gates for the evolution $t_0 \to t_i$ as matrix product operators (MPOs). For each trajectory, we update the MPS by contracting it layer-wise with the corresponding MPOs following the edge-coloring scheme discussed in \appref{appendix:generalization_to_n_qubits} and implement the reset operations on the ancillas using Monte Carlo sampling. More concretely, the steps involved in the simulation of one trajectory are the following:
\begin{enumerate}
\item Apply the single-qubit $R_z (\beta)$ gates on all system qubits.
\item Apply the two-qubit $R_{YY} (\alpha)$ and $R_{XX} (\alpha)$ gates on every ``even'' bond, followed by ``odd'' bonds.
\item Apply the dissipative channel $\varphi$ on ``even'' bonds: two-qubit basis change gates $P$, followed by the double-controlled $R_y$ gate $CCR_y (\theta_{+} - \theta_{-})$, the double-controlled $R_y$ gate $CCR_y (\theta_{-} - \theta_{+})$, a controlled $R_y$ gate $CR_y (\theta_{-})$, a controlled $R_y$ gate $CR_y (\theta_{+})$ and two $CX$ gates, all acting between a qubit pair and its associated ancilla pair.
\item Move the orthogonality center to an ancilla site, i.e., bring the MPS into canonical form with respect to that site, so that the local tensor directly determines the amplitudes needed to compute measurement probabilities.
\item Compute the probabilities of outcomes \(z \in \{0,1\}\) by projecting the ancilla site tensor onto the local basis states \(\ket{0}\) and \(\ket{1}\), and applying the Born rule $p(z)  = \bra{z} \rho_a \ket{z}$.
\item Sample a measurement outcome \(z\) according to these probabilities.
\item Project the ancilla site onto the corresponding post-measurement state using the projector $\ket{z} \bra{z}$.
\item If $z  = 1$, apply an $X$ gate so that the ancilla site ends in $\ket{0}$.
\item Normalize the MPS.
\item Repeat steps 4-9 to the rest of ancilla sites.
\item Repeat steps 3-10 to the ``odd'' bonds.
\item For a number of Trotter steps $k>1$, repeat steps 1-11 $k-1$ times.
\item Compute the expectation value $\left< \sigma_i^{\dagger} \sigma_i \right>$ on all physical sites representing system qubits and store the result.
\end{enumerate}
After simulating all $N_{traj}$ trajectories, we average over all the expectation values obtained using \autoref{app_expectation_value_MPSMC}. Throughout this work we refer to this method as Monte Carlo Time-Evolving Block-Decimation algorithm (MCTEBD).

To validate the results obtained from the quantum experiments in \autoref{fig:quantum_experiment_results} we run the MCTEBD algorithm for a maximum bond dimension $\chi = 50$ and $N_{traj} = 100$. We compare the results for $\max (\chi) = 50$, $\max (\chi) = 100$ and $\max (\chi) = 200$ and find that $\max (\chi) = 50$ is enough to faithfully capture the time dynamics of our systems, which indicates low entanglement growth along the chain. We attribute his limited entanglement to (i) the particular system parameters we use and (ii) the limited Trotter step sizes we consider $k \in \left\lbrace 1, 2, 3 \right\rbrace$.

\section{Device characterization}
\label{app:device_characterization}
\begin{figure*}[tbp]
    \centering
    
    \def\svgwidth{0.9\textwidth}
    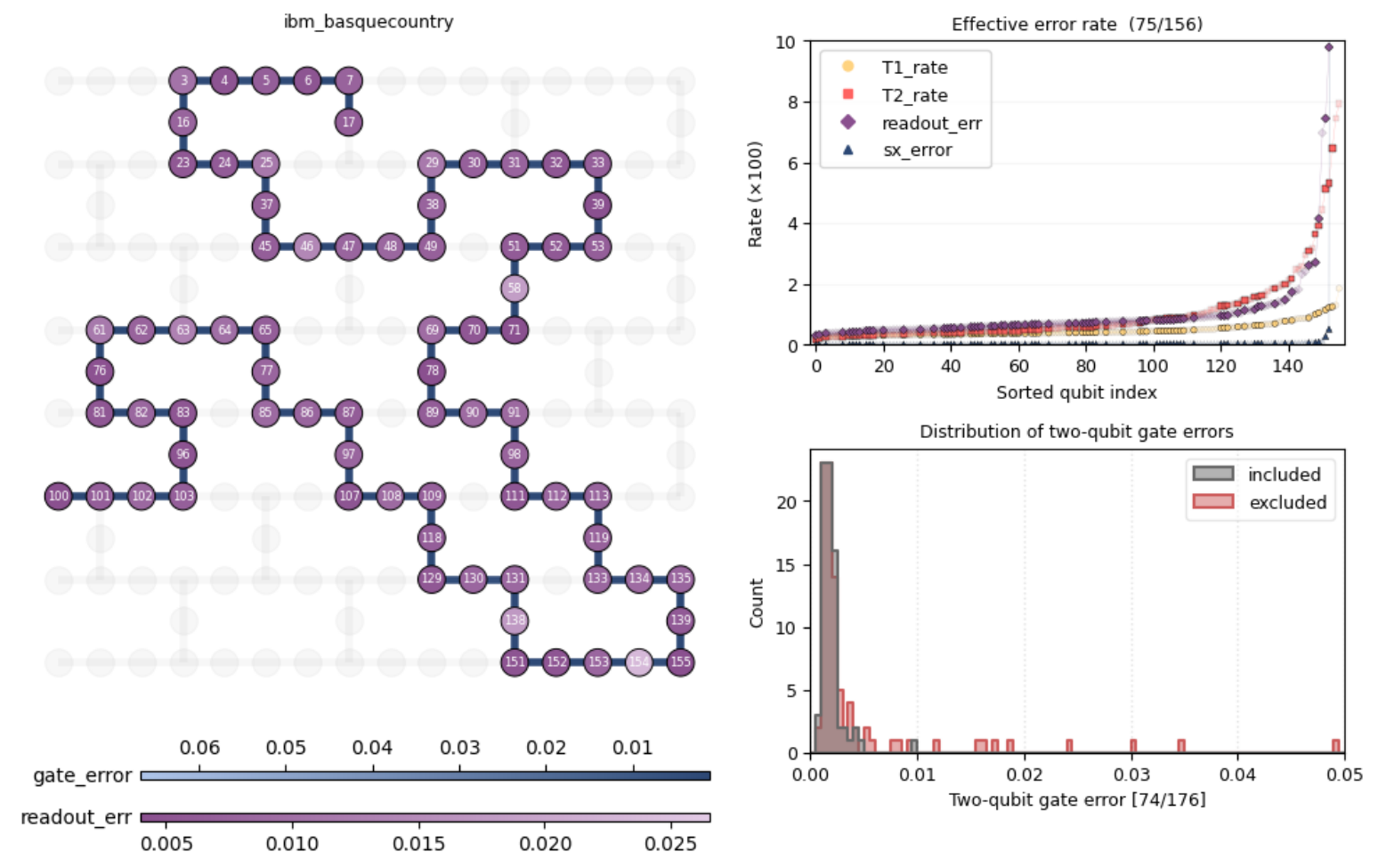

    \vspace{0.5em}

    \def\svgwidth{0.9\textwidth}
    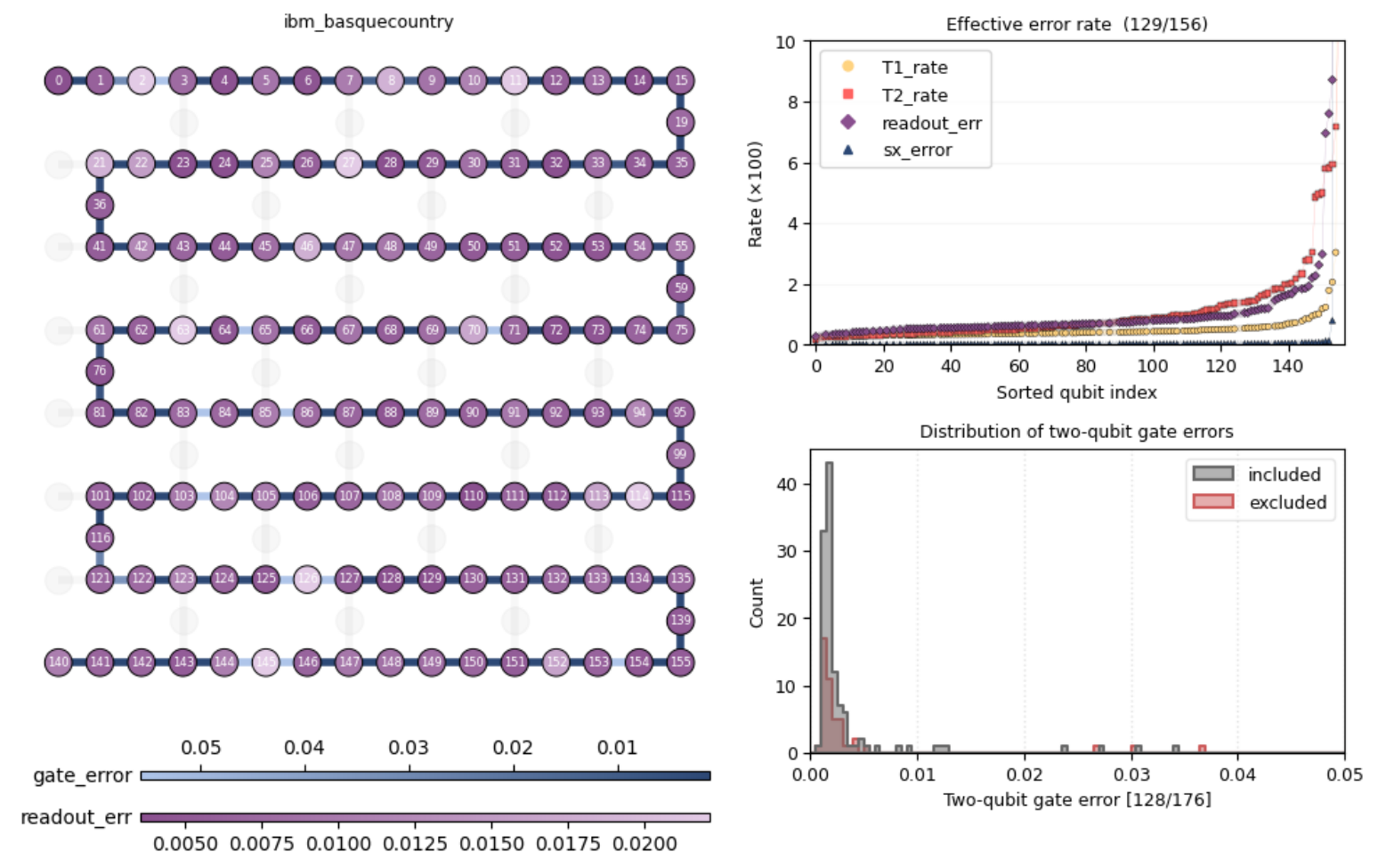

    \caption{\footnotesize Characterization of the IBM Heron r2 quantum processing unit \texttt{ibm\_basquecountry} at the time of execution. (a,d) Readout error rates and two‑qubit (2Q) gate error rates for the device. Two‑qubit gate errors are the primary criterion for the layout selection protocol used for the 75-qubit experiment, as they typically dominate the overall circuit error. (b,e) Per‑qubit coherence ($T_1/T_2$‑derived rates), readout errors, and single‑qubit (1Q) gate error rates, sorted in ascending order. This ordered representation allows a direct visual assessment of how many qubits exceed a given error threshold for each metric, facilitating comparison across error sources. This comparison highlights how the layout selection preferentially retains lower‑error two‑qubit interactions. (c,f) Histogram of two‑qubit gate error rates, separating gates that are included in the selected layout from those that are excluded.}
    \label{fig:device_characterization}
\end{figure*}

All hardware experiments were performed on the IBM Heron r2 quantum processing unit, \texttt{ibm\_basquecountry}, a 156‑qubit superconducting transmon device arranged in a heavy‑hex lattice. The benchmarking data reported here corresponds to the two largest hardware executions reported in this work, which underpin the results shown in \autoref{fig:quantum_experiment_results} of the main text. For the experiment involving 75 total qubits (\autoref{fig:quantum_experiment_results}a,b,c) qubit layouts were selected in real time immediately prior to execution. In particular, we benchmarked candidate one‑dimensional chains by directly estimating the effective Error Per Layer Gate (EPLG) using layer fidelity experiments \cite{mckay2023} and chose the chain minimizing this metric. This dynamic layout selection strategy ensured that, for each run, the chosen subgraph reflected the best available qubits at execution time in terms of readout fidelity and two‑qubit gate performance. The result of the calibration is shown in panels a, b and c of \autoref{fig:device_characterization}. At the time of execution, the processor operated at a CLOPS of 320k. The median single‑shot readout error across the device was $0.67$\%, while the median single‑qubit coherence times were $T_2 = 155.8 \mu$s and $T_1 = 254.9 \mu$s. The median error rate of the calibrated single‑qubit $\sqrt{X}$ ($SX$) gate was $2.22\cdot 10^{-4}$, and the median two‑qubit controlled‑$Z$ ($CZ$) gate error rate was $0.175$\%.

The second execution corresponds to \autoref{fig:quantum_experiment_results}d of the main text, where a total of 129 qubits has been used. For this quantum experiment we use the longest possible chain that we identified as compatible with our circuit. The corresponding device state is reported in panels d, e and f of \autoref{fig:device_characterization}. During this run, the processor also operated at $320$k CLOPS. The median readout error was $0.71$\%, and the median coherence times were $T_2 = 172.4 \mu$s and $T_1 = 251.9 \mu$s. The median $SX$ gate error rate was $2.41\cdot 10^{-4}$, while the median CZ gate error rate was $0.188$\%. Overall, these metrics indicate a highly comparable noise profile between the two executions, with only modest variations consistent with routine day‑to‑day device recalibrations.

\clearpage
\twocolumngrid

\bibliographystyle{apsrev4-2}
\bibliography{bibs.bib}

\end{document}

%% file: two_column_fig_0.pdf_tex
\begingroup%
  \makeatletter%
  \providecommand\color[2][]{%
    \errmessage{(Inkscape) Color is used for the text in Inkscape, but the package 'color.sty' is not loaded}%
    \renewcommand\color[2][]{}%
  }%
  \providecommand\transparent[1]{%
    \errmessage{(Inkscape) Transparency is used (non-zero) for the text in Inkscape, but the package 'transparent.sty' is not loaded}%
    \renewcommand\transparent[1]{}%
  }%
  \providecommand\rotatebox[2]{#2}%
  \newcommand*\fsize{\dimexpr\f@size pt\relax}%
  \newcommand*\lineheight[1]{\fontsize{\fsize}{#1\fsize}\selectfont}%
  \ifx\svgwidth\undefined%
    \setlength{\unitlength}{465.60936635bp}%
    \ifx\svgscale\undefined%
      \relax%
    \else%
      \setlength{\unitlength}{\unitlength * \real{\svgscale}}%
    \fi%
  \else%
    \setlength{\unitlength}{\svgwidth}%
  \fi%
  \global\let\svgwidth\undefined%
  \global\let\svgscale\undefined%
  \makeatother%
  \begin{picture}(1,0.65768014)%
    \lineheight{1}%
    \setlength\tabcolsep{0pt}%
    \put(0.00892036,0.28283614){\color[rgb]{0,0,0}\makebox(0,0)[lt]{\lineheight{1.25}\smash{\begin{tabular}[t]{l}c)\end{tabular}}}}%
    \put(0.37680078,0.28283614){\color[rgb]{0,0,0}\makebox(0,0)[lt]{\lineheight{1.25}\smash{\begin{tabular}[t]{l}d)\end{tabular}}}}%
    \put(0.69246258,0.28283613){\color[rgb]{0,0,0}\makebox(0,0)[lt]{\lineheight{1.25}\smash{\begin{tabular}[t]{l}e)\end{tabular}}}}%
    \put(0,0){\includegraphics[width=\unitlength,page=1]{two_column_fig_0.pdf}}%
    \put(0.15624935,0.40033725){\color[rgb]{0,0,0}\makebox(0,0)[lt]{\lineheight{1.25}\smash{\begin{tabular}[t]{l}Bath (B)\end{tabular}}}}%
    \put(0.11333092,0.62484358){\color[rgb]{0,0,0}\makebox(0,0)[lt]{\lineheight{1.25}\smash{\begin{tabular}[t]{l}System of interest (S)\end{tabular}}}}%
    \put(0.15087454,0.28338167){\color[rgb]{0,0,0}\makebox(0,0)[lt]{\lineheight{1.25}\smash{\begin{tabular}[t]{l}Single-qubit decay\end{tabular}}}}%
    \put(0.42604712,0.28346757){\color[rgb]{0,0,0}\makebox(0,0)[lt]{\lineheight{1.25}\smash{\begin{tabular}[t]{l}Bath-enhanced qubit decay\end{tabular}}}}%
    \put(0.75430197,0.28337307){\color[rgb]{0,0,0}\makebox(0,0)[lt]{\lineheight{1.25}\smash{\begin{tabular}[t]{l}Collective qubit decay\end{tabular}}}}%
    \put(0.7007004,0.30922641){\color[rgb]{0,0,0}\makebox(0,0)[lt]{\lineheight{1.25}\smash{\begin{tabular}[t]{l}layers\end{tabular}}}}%
    \put(0.4132355,0.42173448){\color[rgb]{0,0,0}\makebox(0,0)[lt]{\lineheight{1.25}\smash{\begin{tabular}[t]{l}ancillas\end{tabular}}}}%
    \put(0.00475771,0.64273705){\color[rgb]{0,0,0}\makebox(0,0)[lt]{\lineheight{1.25}\smash{\begin{tabular}[t]{l}a)\end{tabular}}}}%
    \put(0,0){\includegraphics[width=\unitlength,page=2]{two_column_fig_0.pdf}}%
    \put(0.37815374,0.64261016){\color[rgb]{0,0,0}\makebox(0,0)[lt]{\lineheight{1.25}\smash{\begin{tabular}[t]{l}b)\end{tabular}}}}%
    \put(0,0){\includegraphics[width=\unitlength,page=3]{two_column_fig_0.pdf}}%
  \end{picture}%
\endgroup%

%% file: one_column_fig_1.pdf_tex
\begingroup%
  \makeatletter%
  \providecommand\color[2][]{%
    \errmessage{(Inkscape) Color is used for the text in Inkscape, but the package 'color.sty' is not loaded}%
    \renewcommand\color[2][]{}%
  }%
  \providecommand\transparent[1]{%
    \errmessage{(Inkscape) Transparency is used (non-zero) for the text in Inkscape, but the package 'transparent.sty' is not loaded}%
    \renewcommand\transparent[1]{}%
  }%
  \providecommand\rotatebox[2]{#2}%
  \newcommand*\fsize{\dimexpr\f@size pt\relax}%
  \newcommand*\lineheight[1]{\fontsize{\fsize}{#1\fsize}\selectfont}%
  \ifx\svgwidth\undefined%
    \setlength{\unitlength}{206.4752774bp}%
    \ifx\svgscale\undefined%
      \relax%
    \else%
      \setlength{\unitlength}{\unitlength * \real{\svgscale}}%
    \fi%
  \else%
    \setlength{\unitlength}{\svgwidth}%
  \fi%
  \global\let\svgwidth\undefined%
  \global\let\svgscale\undefined%
  \makeatother%
  \begin{picture}(1,1.18659104)%
    \lineheight{1}%
    \setlength\tabcolsep{0pt}%
    \put(0,0){\includegraphics[width=\unitlength,page=1]{one_column_fig_1.pdf}}%
    \put(0.17191,0.06648439){\color[rgb]{0,0,0}\makebox(0,0)[lt]{\lineheight{1.25}\smash{\begin{tabular}[t]{l}0\end{tabular}}}}%
    \put(0.39927466,0.06620067){\color[rgb]{0,0,0}\makebox(0,0)[lt]{\lineheight{1.25}\smash{\begin{tabular}[t]{l}2\end{tabular}}}}%
    \put(0.624041,0.06524461){\color[rgb]{0,0,0}\makebox(0,0)[lt]{\lineheight{1.25}\smash{\begin{tabular}[t]{l}4\end{tabular}}}}%
    \put(0.84945766,0.06576599){\color[rgb]{0,0,0}\makebox(0,0)[lt]{\lineheight{1.25}\smash{\begin{tabular}[t]{l}6\end{tabular}}}}%
    \put(0,0){\includegraphics[width=\unitlength,page=2]{one_column_fig_1.pdf}}%
    \put(0.68304999,0.53710969){\color[rgb]{0.10196078,0.10196078,0.10196078}\makebox(0,0)[lt]{\lineheight{1.25}\smash{\begin{tabular}[t]{l}Quantum circuit\end{tabular}}}}%
    \put(0.68340806,0.58379256){\color[rgb]{0.10196078,0.10196078,0.10196078}\makebox(0,0)[lt]{\lineheight{1.25}\smash{\begin{tabular}[t]{l}Kraus\end{tabular}}}}%
    \put(0.11251349,0.70857253){\color[rgb]{0,0,0}\makebox(0,0)[lt]{\lineheight{1.25}\smash{\begin{tabular}[t]{l}1\end{tabular}}}}%
    \put(0.08199828,0.59244025){\color[rgb]{0,0,0}\makebox(0,0)[lt]{\lineheight{1.25}\smash{\begin{tabular}[t]{l}0.8\end{tabular}}}}%
    \put(0.08155761,0.47463667){\color[rgb]{0,0,0}\makebox(0,0)[lt]{\lineheight{1.25}\smash{\begin{tabular}[t]{l}0.6\end{tabular}}}}%
    \put(0.08097652,0.35824561){\color[rgb]{0,0,0}\makebox(0,0)[lt]{\lineheight{1.25}\smash{\begin{tabular}[t]{l}0.4\end{tabular}}}}%
    \put(0.08280445,0.23953859){\color[rgb]{0,0,0}\makebox(0,0)[lt]{\lineheight{1.25}\smash{\begin{tabular}[t]{l}0.2\end{tabular}}}}%
    \put(0.11150124,0.12506258){\color[rgb]{0,0,0}\makebox(0,0)[lt]{\lineheight{1.25}\smash{\begin{tabular}[t]{l}0\end{tabular}}}}%
    \put(0.68152532,0.63962257){\color[rgb]{0.10196078,0.10196078,0.10196078}\makebox(0,0)[lt]{\lineheight{1.25}\smash{\begin{tabular}[t]{l}Master equation\end{tabular}}}}%
    \put(0,0){\includegraphics[width=\unitlength,page=3]{one_column_fig_1.pdf}}%
    \put(-0.00139453,1.13957277){\color[rgb]{0,0,0}\makebox(0,0)[lt]{\lineheight{1.25}\smash{\begin{tabular}[t]{l}a)\end{tabular}}}}%
    \put(0.00402607,0.7206658){\color[rgb]{0,0,0}\makebox(0,0)[lt]{\lineheight{1.25}\smash{\begin{tabular}[t]{l}c)\end{tabular}}}}%
    \put(0.49394162,1.13746947){\color[rgb]{0,0,0}\makebox(0,0)[lt]{\lineheight{1.25}\smash{\begin{tabular}[t]{l}b)\end{tabular}}}}%
    \put(0,0){\includegraphics[width=\unitlength,page=4]{one_column_fig_1.pdf}}%
  \end{picture}%
\endgroup%

%% file: one_column_fig_2.pdf_tex
\begingroup%
  \makeatletter%
  \providecommand\color[2][]{%
    \errmessage{(Inkscape) Color is used for the text in Inkscape, but the package 'color.sty' is not loaded}%
    \renewcommand\color[2][]{}%
  }%
  \providecommand\transparent[1]{%
    \errmessage{(Inkscape) Transparency is used (non-zero) for the text in Inkscape, but the package 'transparent.sty' is not loaded}%
    \renewcommand\transparent[1]{}%
  }%
  \providecommand\rotatebox[2]{#2}%
  \newcommand*\fsize{\dimexpr\f@size pt\relax}%
  \newcommand*\lineheight[1]{\fontsize{\fsize}{#1\fsize}\selectfont}%
  \ifx\svgwidth\undefined%
    \setlength{\unitlength}{201.91413459bp}%
    \ifx\svgscale\undefined%
      \relax%
    \else%
      \setlength{\unitlength}{\unitlength * \real{\svgscale}}%
    \fi%
  \else%
    \setlength{\unitlength}{\svgwidth}%
  \fi%
  \global\let\svgwidth\undefined%
  \global\let\svgscale\undefined%
  \makeatother%
  \begin{picture}(1,1.1703484)%
    \lineheight{1}%
    \setlength\tabcolsep{0pt}%
    \put(0,0){\includegraphics[width=\unitlength,page=1]{one_column_fig_2.pdf}}%
    \put(-0.00130292,1.13691835){\color[rgb]{0,0,0}\makebox(0,0)[lt]{\lineheight{1.25}\smash{\begin{tabular}[t]{l}a)\end{tabular}}}}%
    \put(-0.00124808,0.80829752){\color[rgb]{0,0,0}\makebox(0,0)[lt]{\lineheight{1.25}\smash{\begin{tabular}[t]{l}b)\end{tabular}}}}%
    \put(0.00017024,0.4071243){\color[rgb]{0,0,0}\makebox(0,0)[lt]{\lineheight{1.25}\smash{\begin{tabular}[t]{l}c)\end{tabular}}}}%
    \put(0,0){\includegraphics[width=\unitlength,page=2]{one_column_fig_2.pdf}}%
  \end{picture}%
\endgroup%

%% file: two_column_fig_3.pdf_tex
\begingroup%
  \makeatletter%
  \providecommand\color[2][]{%
    \errmessage{(Inkscape) Color is used for the text in Inkscape, but the package 'color.sty' is not loaded}%
    \renewcommand\color[2][]{}%
  }%
  \providecommand\transparent[1]{%
    \errmessage{(Inkscape) Transparency is used (non-zero) for the text in Inkscape, but the package 'transparent.sty' is not loaded}%
    \renewcommand\transparent[1]{}%
  }%
  \providecommand\rotatebox[2]{#2}%
  \newcommand*\fsize{\dimexpr\f@size pt\relax}%
  \newcommand*\lineheight[1]{\fontsize{\fsize}{#1\fsize}\selectfont}%
  \ifx\svgwidth\undefined%
    \setlength{\unitlength}{463.37395135bp}%
    \ifx\svgscale\undefined%
      \relax%
    \else%
      \setlength{\unitlength}{\unitlength * \real{\svgscale}}%
    \fi%
  \else%
    \setlength{\unitlength}{\svgwidth}%
  \fi%
  \global\let\svgwidth\undefined%
  \global\let\svgscale\undefined%
  \makeatother%
  \begin{picture}(1,0.5655285)%
    \lineheight{1}%
    \setlength\tabcolsep{0pt}%
    \put(0,0){\includegraphics[width=\unitlength,page=1]{two_column_fig_3.pdf}}%
    \put(0.0353257,0.37836107){\color[rgb]{0,0,0}\makebox(0,0)[lt]{\lineheight{1.25}\smash{\begin{tabular}[t]{l}50\end{tabular}}}}%
    \put(0.03574262,0.33022109){\color[rgb]{0,0,0}\makebox(0,0)[lt]{\lineheight{1.25}\smash{\begin{tabular}[t]{l}25\end{tabular}}}}%
    \put(0.06841411,0.29336049){\color[rgb]{0,0,0}\makebox(0,0)[lt]{\lineheight{1.25}\smash{\begin{tabular}[t]{l}2\end{tabular}}}}%
    \put(0.15444221,0.29356253){\color[rgb]{0,0,0}\makebox(0,0)[lt]{\lineheight{1.25}\smash{\begin{tabular}[t]{l}4\end{tabular}}}}%
    \put(0.23941383,0.29306687){\color[rgb]{0,0,0}\makebox(0,0)[lt]{\lineheight{1.25}\smash{\begin{tabular}[t]{l}6\end{tabular}}}}%
    \put(0.32557636,0.29338189){\color[rgb]{0,0,0}\makebox(0,0)[lt]{\lineheight{1.25}\smash{\begin{tabular}[t]{l}8\end{tabular}}}}%
    \put(0.0251972,0.52576129){\color[rgb]{0,0,0}\makebox(0,0)[lt]{\lineheight{1.25}\smash{\begin{tabular}[t]{l}125\end{tabular}}}}%
    \put(0.02540275,0.47627602){\color[rgb]{0,0,0}\makebox(0,0)[lt]{\lineheight{1.25}\smash{\begin{tabular}[t]{l}100\end{tabular}}}}%
    \put(0.0343666,0.4273899){\color[rgb]{0,0,0}\makebox(0,0)[lt]{\lineheight{1.25}\smash{\begin{tabular}[t]{l}75\end{tabular}}}}%
    \put(0.01767421,0.36644727){\color[rgb]{0.10196078,0.10196078,0.10196078}\rotatebox{90}{\makebox(0,0)[lt]{\lineheight{1.25}\smash{\begin{tabular}[t]{l}two-qubit depth\end{tabular}}}}}%
    \put(0.11125458,0.53224464){\color[rgb]{0.10196078,0.10196078,0.10196078}\makebox(0,0)[lt]{\lineheight{1.25}\smash{\begin{tabular}[t]{l}ISA circuit\end{tabular}}}}%
    \put(0.23024652,0.53736865){\color[rgb]{1,0.52156863,0.19215686}\makebox(0,0)[lt]{\lineheight{1.25}\smash{\begin{tabular}[t]{l}Original (Fig. 2c)\end{tabular}}}}%
    \put(0.23761481,0.41753455){\color[rgb]{0.7372549,0.31372549,0.56470588}\makebox(0,0)[lt]{\lineheight{1.25}\smash{\begin{tabular}[t]{l}Dynamic (Fig. 3b)\end{tabular}}}}%
    \put(0.18364234,0.36772995){\color[rgb]{0.17254902,0.28235294,0.45882353}\makebox(0,0)[lt]{\lineheight{1.25}\smash{\begin{tabular}[t]{l}Hardware-aware dynamic\end{tabular}}}}%
    \put(0.12329198,0.27091661){\color[rgb]{0.10196078,0.10196078,0.10196078}\makebox(0,0)[lt]{\lineheight{1.25}\smash{\begin{tabular}[t]{l}number of emitters,\end{tabular}}}}%
    \put(0,0){\includegraphics[width=\unitlength,page=2]{two_column_fig_3.pdf}}%
    \put(0.11138201,0.51045544){\color[rgb]{0.10196078,0.10196078,0.10196078}\makebox(0,0)[lt]{\lineheight{1.25}\smash{\begin{tabular}[t]{l}virtual circuit\end{tabular}}}}%
    \put(0,0){\includegraphics[width=\unitlength,page=3]{two_column_fig_3.pdf}}%
    \put(0.31320831,0.33782056){\color[rgb]{0.17254902,0.28235294,0.45882353}\makebox(0,0)[lt]{\lineheight{1.29999995}\smash{\begin{tabular}[t]{l}(Fig. 3c)\end{tabular}}}}%
    \put(0.00659941,0.54358359){\color[rgb]{0,0,0}\makebox(0,0)[lt]{\lineheight{1.25}\smash{\begin{tabular}[t]{l}a)\end{tabular}}}}%
    \put(0,0){\includegraphics[width=\unitlength,page=4]{two_column_fig_3.pdf}}%
    \put(0.40113823,0.54161545){\color[rgb]{0,0,0}\makebox(0,0)[lt]{\lineheight{1.25}\smash{\begin{tabular}[t]{l}b)\end{tabular}}}}%
    \put(0.00851286,0.14580073){\color[rgb]{0,0,0}\makebox(0,0)[lt]{\lineheight{1.25}\smash{\begin{tabular}[t]{l}c)\end{tabular}}}}%
    \put(0.5942163,0.52491498){\color[rgb]{0,0,0}\makebox(0,0)[lt]{\lineheight{1.25}\smash{\begin{tabular}[t]{l}If   \end{tabular}}}}%
    \put(0,0){\includegraphics[width=\unitlength,page=5]{two_column_fig_3.pdf}}%
    \put(0.70513342,0.5243615){\color[rgb]{0,0,0}\makebox(0,0)[lt]{\lineheight{1.25}\smash{\begin{tabular}[t]{l}Else\end{tabular}}}}%
    \put(0,0){\includegraphics[width=\unitlength,page=6]{two_column_fig_3.pdf}}%
    \put(0.44682653,0.23312058){\color[rgb]{0,0,0}\makebox(0,0)[lt]{\lineheight{1.25}\smash{\begin{tabular}[t]{l}H\end{tabular}}}}%
    \put(0.57837053,0.23258329){\color[rgb]{0,0,0}\makebox(0,0)[lt]{\lineheight{1.25}\smash{\begin{tabular}[t]{l}T\end{tabular}}}}%
    \put(0.64406701,0.20050106){\color[rgb]{0,0,0}\makebox(0,0)[lt]{\lineheight{1.25}\smash{\begin{tabular}[t]{l}T\end{tabular}}}}%
    \put(0.84162926,0.26578781){\color[rgb]{0,0,0}\makebox(0,0)[lt]{\lineheight{1.25}\smash{\begin{tabular}[t]{l}T\end{tabular}}}}%
    \put(0.7102656,0.23239347){\color[rgb]{0,0,0}\makebox(0,0)[lt]{\lineheight{1.25}\smash{\begin{tabular}[t]{l}T\end{tabular}}}}%
    \put(0.74249873,0.23286654){\color[rgb]{0,0,0}\makebox(0,0)[lt]{\lineheight{1.25}\smash{\begin{tabular}[t]{l}H\end{tabular}}}}%
    \put(0,0){\includegraphics[width=\unitlength,page=7]{two_column_fig_3.pdf}}%
    \put(0.38327742,0.15729135){\color[rgb]{0,0,0}\makebox(0,0)[lt]{\lineheight{1.25}\smash{\begin{tabular}[t]{l}Else\end{tabular}}}}%
    \put(0.65254851,0.11520871){\color[rgb]{0,0,0}\makebox(0,0)[lt]{\lineheight{1.25}\smash{\begin{tabular}[t]{l}Else\end{tabular}}}}%
    \put(0,0){\includegraphics[width=\unitlength,page=8]{two_column_fig_3.pdf}}%
    \put(0.31196337,0.15687798){\color[rgb]{0,0,0}\makebox(0,0)[lt]{\lineheight{1.25}\smash{\begin{tabular}[t]{l}If   \end{tabular}}}}%
    \put(0,0){\includegraphics[width=\unitlength,page=9]{two_column_fig_3.pdf}}%
    \put(0.58112822,0.11457745){\color[rgb]{0,0,0}\makebox(0,0)[lt]{\lineheight{1.25}\smash{\begin{tabular}[t]{l}If   \end{tabular}}}}%
    \put(0,0){\includegraphics[width=\unitlength,page=10]{two_column_fig_3.pdf}}%
  \end{picture}%
\endgroup%

%% file: one_column_fig_4.pdf_tex
\begingroup%
  \makeatletter%
  \providecommand\color[2][]{%
    \errmessage{(Inkscape) Color is used for the text in Inkscape, but the package 'color.sty' is not loaded}%
    \renewcommand\color[2][]{}%
  }%
  \providecommand\transparent[1]{%
    \errmessage{(Inkscape) Transparency is used (non-zero) for the text in Inkscape, but the package 'transparent.sty' is not loaded}%
    \renewcommand\transparent[1]{}%
  }%
  \providecommand\rotatebox[2]{#2}%
  \newcommand*\fsize{\dimexpr\f@size pt\relax}%
  \newcommand*\lineheight[1]{\fontsize{\fsize}{#1\fsize}\selectfont}%
  \ifx\svgwidth\undefined%
    \setlength{\unitlength}{205.84884043bp}%
    \ifx\svgscale\undefined%
      \relax%
    \else%
      \setlength{\unitlength}{\unitlength * \real{\svgscale}}%
    \fi%
  \else%
    \setlength{\unitlength}{\svgwidth}%
  \fi%
  \global\let\svgwidth\undefined%
  \global\let\svgscale\undefined%
  \makeatother%
  \begin{picture}(1,1.16952306)%
    \lineheight{1}%
    \setlength\tabcolsep{0pt}%
    \put(0,0){\includegraphics[width=\unitlength,page=1]{one_column_fig_4.pdf}}%
    \put(0.11417235,1.11800135){\color[rgb]{0.10196078,0.10196078,0.10196078}\makebox(0,0)[lt]{\lineheight{1.25}\smash{\begin{tabular}[t]{l}Raw results\end{tabular}}}}%
    \put(0.10723832,0.92444447){\color[rgb]{0.10196078,0.10196078,0.10196078}\makebox(0,0)[lt]{\lineheight{1.25}\smash{\begin{tabular}[t]{l}Strategy (iv)\end{tabular}}}}%
    \put(0.11662059,1.07316219){\color[rgb]{0.10196078,0.10196078,0.10196078}\makebox(0,0)[lt]{\lineheight{1.25}\smash{\begin{tabular}[t]{l}Strategy (i)\end{tabular}}}}%
    \put(0.11056476,1.02450042){\color[rgb]{0.10196078,0.10196078,0.10196078}\makebox(0,0)[lt]{\lineheight{1.25}\smash{\begin{tabular}[t]{l}Strategy (ii)\end{tabular}}}}%
    \put(0.10612762,0.97617251){\color[rgb]{0.10196078,0.10196078,0.10196078}\makebox(0,0)[lt]{\lineheight{1.25}\smash{\begin{tabular}[t]{l}Strategy (iii)\end{tabular}}}}%
    \put(0.54073902,0.82954969){\color[rgb]{0.10196078,0.10196078,0.10196078}\makebox(0,0)[lt]{\lineheight{1.25}\smash{\begin{tabular}[t]{l}qubits\end{tabular}}}}%
    \put(0,0){\includegraphics[width=\unitlength,page=2]{one_column_fig_4.pdf}}%
    \put(0.7515475,0.87226957){\color[rgb]{0,0,0}\makebox(0,0)[lt]{\lineheight{1.25}\smash{\begin{tabular}[t]{l}8\end{tabular}}}}%
    \put(0.86839268,1.10702265){\color[rgb]{0,0,0}\makebox(0,0)[lt]{\lineheight{1.25}\smash{\begin{tabular}[t]{l}0.4\end{tabular}}}}%
    \put(0.86912751,1.05723372){\color[rgb]{0,0,0}\makebox(0,0)[lt]{\lineheight{1.25}\smash{\begin{tabular}[t]{l}0.3\end{tabular}}}}%
    \put(0.86970732,1.00742366){\color[rgb]{0,0,0}\makebox(0,0)[lt]{\lineheight{1.25}\smash{\begin{tabular}[t]{l}0.2\end{tabular}}}}%
    \put(0.86970732,0.95807122){\color[rgb]{0,0,0}\makebox(0,0)[lt]{\lineheight{1.25}\smash{\begin{tabular}[t]{l}0.1\end{tabular}}}}%
    \put(0.79759189,0.87366822){\color[rgb]{0,0,0}\makebox(0,0)[lt]{\lineheight{1.25}\smash{\begin{tabular}[t]{l}9\end{tabular}}}}%
    \put(0.70437746,0.87275657){\color[rgb]{0,0,0}\makebox(0,0)[lt]{\lineheight{1.25}\smash{\begin{tabular}[t]{l}7\end{tabular}}}}%
    \put(0.65836127,0.87231936){\color[rgb]{0,0,0}\makebox(0,0)[lt]{\lineheight{1.25}\smash{\begin{tabular}[t]{l}6\end{tabular}}}}%
    \put(0.61240647,0.87322582){\color[rgb]{0,0,0}\makebox(0,0)[lt]{\lineheight{1.25}\smash{\begin{tabular}[t]{l}5\end{tabular}}}}%
    \put(0.56431446,0.87414659){\color[rgb]{0,0,0}\makebox(0,0)[lt]{\lineheight{1.25}\smash{\begin{tabular}[t]{l}4\end{tabular}}}}%
    \put(0.52020454,0.87508509){\color[rgb]{0,0,0}\makebox(0,0)[lt]{\lineheight{1.25}\smash{\begin{tabular}[t]{l}3\end{tabular}}}}%
    \put(0.47486699,0.87452301){\color[rgb]{0,0,0}\makebox(0,0)[lt]{\lineheight{1.25}\smash{\begin{tabular}[t]{l}2\end{tabular}}}}%
    \put(0.42729187,0.87367734){\color[rgb]{0,0,0}\makebox(0,0)[lt]{\lineheight{1.25}\smash{\begin{tabular}[t]{l}1\end{tabular}}}}%
    \put(0.38177408,0.87461584){\color[rgb]{0,0,0}\makebox(0,0)[lt]{\lineheight{1.25}\smash{\begin{tabular}[t]{l}0\end{tabular}}}}%
    \put(0,0){\includegraphics[width=\unitlength,page=3]{one_column_fig_4.pdf}}%
    \put(0.18825984,0.14714262){\color[rgb]{0,0,0}\makebox(0,0)[lt]{\lineheight{1.25}\smash{\begin{tabular}[t]{l}0\end{tabular}}}}%
    \put(0.38001039,0.14807861){\color[rgb]{0,0,0}\makebox(0,0)[lt]{\lineheight{1.25}\smash{\begin{tabular}[t]{l}0.5\end{tabular}}}}%
    \put(0.60096609,0.14642598){\color[rgb]{0,0,0}\makebox(0,0)[lt]{\lineheight{1.25}\smash{\begin{tabular}[t]{l}1\end{tabular}}}}%
    \put(0.79099817,0.1479288){\color[rgb]{0,0,0}\makebox(0,0)[lt]{\lineheight{1.25}\smash{\begin{tabular}[t]{l}1.5\end{tabular}}}}%
    \put(0.09431293,0.21436821){\color[rgb]{0,0,0}\makebox(0,0)[lt]{\lineheight{1.25}\smash{\begin{tabular}[t]{l}1.5\end{tabular}}}}%
    \put(0.0938261,0.39072454){\color[rgb]{0,0,0}\makebox(0,0)[lt]{\lineheight{1.25}\smash{\begin{tabular}[t]{l}2.5\end{tabular}}}}%
    \put(0.09336915,0.56143432){\color[rgb]{0,0,0}\makebox(0,0)[lt]{\lineheight{1.25}\smash{\begin{tabular}[t]{l}3.5\end{tabular}}}}%
    \put(0.09536653,0.7369901){\color[rgb]{0,0,0}\makebox(0,0)[lt]{\lineheight{1.25}\smash{\begin{tabular}[t]{l}4.5\end{tabular}}}}%
    \put(0.04232322,0.07104847){\color[rgb]{0,0,0}\makebox(0,0)[lt]{\lineheight{1.25}\smash{\begin{tabular}[t]{l}ZNE (linear)\end{tabular}}}}%
    \put(0.04194246,0.01186881){\color[rgb]{0,0,0}\makebox(0,0)[lt]{\lineheight{1.25}\smash{\begin{tabular}[t]{l}ZNE (FOR)\end{tabular}}}}%
    \put(0.33969265,0.07104847){\color[rgb]{0,0,0}\makebox(0,0)[lt]{\lineheight{1.25}\smash{\begin{tabular}[t]{l}ZNE (CFOR-l1)\end{tabular}}}}%
    \put(0.34102831,0.01093028){\color[rgb]{0,0,0}\makebox(0,0)[lt]{\lineheight{1.25}\smash{\begin{tabular}[t]{l}CDR (inverse)\end{tabular}}}}%
    \put(0.68404908,0.01186881){\color[rgb]{0,0,0}\makebox(0,0)[lt]{\lineheight{1.25}\smash{\begin{tabular}[t]{l}CDR (no bias)\end{tabular}}}}%
    \put(0.68538113,0.07104847){\color[rgb]{0,0,0}\makebox(0,0)[lt]{\lineheight{1.25}\smash{\begin{tabular}[t]{l}CDR (Gaussian)\end{tabular}}}}%
    \put(0.24825267,0.71575737){\color[rgb]{0,0,0}\makebox(0,0)[lt]{\lineheight{1.25}\smash{\begin{tabular}[t]{l}Noiseless\end{tabular}}}}%
    \put(0,0){\includegraphics[width=\unitlength,page=4]{one_column_fig_4.pdf}}%
    \put(0.40001968,0.53318382){\color[rgb]{0,0,0}\makebox(0,0)[lt]{\lineheight{1.25}\smash{\begin{tabular}[t]{l}Raw\end{tabular}}}}%
    \put(0,0){\includegraphics[width=\unitlength,page=5]{one_column_fig_4.pdf}}%
    \put(0.49974664,0.74144409){\color[rgb]{0,0,0}\makebox(0,0)[lt]{\lineheight{1.25}\smash{\begin{tabular}[t]{l}0.15\end{tabular}}}}%
    \put(0.49960918,0.64212934){\color[rgb]{0,0,0}\makebox(0,0)[lt]{\lineheight{1.25}\smash{\begin{tabular}[t]{l}0.05\end{tabular}}}}%
    \put(0.4884609,0.53991515){\color[rgb]{0,0,0}\makebox(0,0)[lt]{\lineheight{1.25}\smash{\begin{tabular}[t]{l}-0.05\end{tabular}}}}%
    \put(0.5826509,0.46749489){\color[rgb]{0,0,0}\makebox(0,0)[lt]{\lineheight{1.25}\smash{\begin{tabular}[t]{l}0\end{tabular}}}}%
    \put(0.67702504,0.46641889){\color[rgb]{0,0,0}\makebox(0,0)[lt]{\lineheight{1.25}\smash{\begin{tabular}[t]{l}0.5\end{tabular}}}}%
    \put(0.78811876,0.46690158){\color[rgb]{0,0,0}\makebox(0,0)[lt]{\lineheight{1.25}\smash{\begin{tabular}[t]{l}1\end{tabular}}}}%
    \put(0.88083811,0.46641889){\color[rgb]{0,0,0}\makebox(0,0)[lt]{\lineheight{1.25}\smash{\begin{tabular}[t]{l}1.5\end{tabular}}}}%
    \put(0,0){\includegraphics[width=\unitlength,page=6]{one_column_fig_4.pdf}}%
    \put(0.001092,1.13673201){\color[rgb]{0,0,0}\makebox(0,0)[lt]{\lineheight{1.25}\smash{\begin{tabular}[t]{l}a)\end{tabular}}}}%
    \put(-0.0012242,0.77206168){\color[rgb]{0,0,0}\makebox(0,0)[lt]{\lineheight{1.25}\smash{\begin{tabular}[t]{l}b)\end{tabular}}}}%
  \end{picture}%
\endgroup%

%% file: one_line_merge_fig_5_6.pdf_tex
\begingroup%
  \makeatletter%
  \providecommand\color[2][]{%
    \errmessage{(Inkscape) Color is used for the text in Inkscape, but the package 'color.sty' is not loaded}%
    \renewcommand\color[2][]{}%
  }%
  \providecommand\transparent[1]{%
    \errmessage{(Inkscape) Transparency is used (non-zero) for the text in Inkscape, but the package 'transparent.sty' is not loaded}%
    \renewcommand\transparent[1]{}%
  }%
  \providecommand\rotatebox[2]{#2}%
  \newcommand*\fsize{\dimexpr\f@size pt\relax}%
  \newcommand*\lineheight[1]{\fontsize{\fsize}{#1\fsize}\selectfont}%
  \ifx\svgwidth\undefined%
    \setlength{\unitlength}{476.48920495bp}%
    \ifx\svgscale\undefined%
      \relax%
    \else%
      \setlength{\unitlength}{\unitlength * \real{\svgscale}}%
    \fi%
  \else%
    \setlength{\unitlength}{\svgwidth}%
  \fi%
  \global\let\svgwidth\undefined%
  \global\let\svgscale\undefined%
  \makeatother%
  \begin{picture}(1,0.29315934)%
    \lineheight{1}%
    \setlength\tabcolsep{0pt}%
    \put(0.4006461,0.27899593){\color[rgb]{0,0,0}\makebox(0,0)[lt]{\lineheight{1.25}\smash{\begin{tabular}[t]{l}b)\end{tabular}}}}%
    \put(-0.00055876,0.27874202){\color[rgb]{0,0,0}\makebox(0,0)[lt]{\lineheight{1.25}\smash{\begin{tabular}[t]{l}a)\end{tabular}}}}%
    \put(0.40066217,0.14202913){\color[rgb]{0,0,0}\makebox(0,0)[lt]{\lineheight{1.25}\smash{\begin{tabular}[t]{l}c)\end{tabular}}}}%
    \put(0,0){\includegraphics[width=\unitlength,page=1]{one_line_merge_fig_5_6.pdf}}%
    \put(0.46690469,0.01864113){\color[rgb]{0,0,0}\makebox(0,0)[lt]{\lineheight{1.25}\smash{\begin{tabular}[t]{l}0\end{tabular}}}}%
    \put(0.45185334,0.03724647){\color[rgb]{0,0,0}\makebox(0,0)[lt]{\lineheight{1.25}\smash{\begin{tabular}[t]{l}0\end{tabular}}}}%
    \put(0.52754592,0.01773923){\color[rgb]{0,0,0}\makebox(0,0)[lt]{\lineheight{1.25}\smash{\begin{tabular}[t]{l}1\end{tabular}}}}%
    \put(0.5877086,0.0179177){\color[rgb]{0,0,0}\makebox(0,0)[lt]{\lineheight{1.25}\smash{\begin{tabular}[t]{l}2\end{tabular}}}}%
    \put(0,0){\includegraphics[width=\unitlength,page=2]{one_line_merge_fig_5_6.pdf}}%
    \put(0.68996715,0.07833675){\color[rgb]{0,0,0}\makebox(0,0)[lt]{\lineheight{1.25}\smash{\begin{tabular}[t]{l}MC-TEBD\end{tabular}}}}%
    \put(0.68996337,0.05171392){\color[rgb]{0,0,0}\makebox(0,0)[lt]{\lineheight{1.25}\smash{\begin{tabular}[t]{l}CDR (inverse bias)\end{tabular}}}}%
    \put(0,0){\includegraphics[width=\unitlength,page=3]{one_line_merge_fig_5_6.pdf}}%
    \put(0.66449225,0.01871621){\color[rgb]{0,0,0}\makebox(0,0)[lt]{\lineheight{1.25}\smash{\begin{tabular}[t]{l}0\end{tabular}}}}%
    \put(0.74423192,0.0184312){\color[rgb]{0,0,0}\makebox(0,0)[lt]{\lineheight{1.25}\smash{\begin{tabular}[t]{l}0.5\end{tabular}}}}%
    \put(0.83938539,0.01704384){\color[rgb]{0,0,0}\makebox(0,0)[lt]{\lineheight{1.25}\smash{\begin{tabular}[t]{l}1\end{tabular}}}}%
    \put(0.91976203,0.01833603){\color[rgb]{0,0,0}\makebox(0,0)[lt]{\lineheight{1.25}\smash{\begin{tabular}[t]{l}1.5\end{tabular}}}}%
    \put(0,0){\includegraphics[width=\unitlength,page=4]{one_line_merge_fig_5_6.pdf}}%
    \put(0.68952993,0.10250196){\color[rgb]{0,0,0}\makebox(0,0)[lt]{\lineheight{1.25}\smash{\begin{tabular}[t]{l}Raw\end{tabular}}}}%
    \put(0.60156251,0.2776706){\color[rgb]{0,0,0}\makebox(0,0)[lt]{\lineheight{1.25}\smash{\begin{tabular}[t]{l}d)\end{tabular}}}}%
    \put(0,0){\includegraphics[width=\unitlength,page=5]{one_line_merge_fig_5_6.pdf}}%
    \put(0.09434567,0.08148121){\color[rgb]{0,0,0}\makebox(0,0)[lt]{\lineheight{1.25}\smash{\begin{tabular}[t]{l}MC-TEBD\end{tabular}}}}%
    \put(0.09434189,0.05485838){\color[rgb]{0,0,0}\makebox(0,0)[lt]{\lineheight{1.25}\smash{\begin{tabular}[t]{l}CDR (inverse bias)\end{tabular}}}}%
    \put(0,0){\includegraphics[width=\unitlength,page=6]{one_line_merge_fig_5_6.pdf}}%
    \put(0.0643998,0.02088816){\color[rgb]{0,0,0}\makebox(0,0)[lt]{\lineheight{1.25}\smash{\begin{tabular}[t]{l}0\end{tabular}}}}%
    \put(0.14454047,0.02060164){\color[rgb]{0,0,0}\makebox(0,0)[lt]{\lineheight{1.25}\smash{\begin{tabular}[t]{l}0.5\end{tabular}}}}%
    \put(0.23845035,0.01956553){\color[rgb]{0,0,0}\makebox(0,0)[lt]{\lineheight{1.25}\smash{\begin{tabular}[t]{l}1\end{tabular}}}}%
    \put(0.31975962,0.01992964){\color[rgb]{0,0,0}\makebox(0,0)[lt]{\lineheight{1.25}\smash{\begin{tabular}[t]{l}1.5\end{tabular}}}}%
    \put(0,0){\includegraphics[width=\unitlength,page=7]{one_line_merge_fig_5_6.pdf}}%
  \end{picture}%
\endgroup%

%% file: two_column_fig_omission.pdf_tex
\begingroup%
  \makeatletter%
  \providecommand\color[2][]{%
    \errmessage{(Inkscape) Color is used for the text in Inkscape, but the package 'color.sty' is not loaded}%
    \renewcommand\color[2][]{}%
  }%
  \providecommand\transparent[1]{%
    \errmessage{(Inkscape) Transparency is used (non-zero) for the text in Inkscape, but the package 'transparent.sty' is not loaded}%
    \renewcommand\transparent[1]{}%
  }%
  \providecommand\rotatebox[2]{#2}%
  \newcommand*\fsize{\dimexpr\f@size pt\relax}%
  \newcommand*\lineheight[1]{\fontsize{\fsize}{#1\fsize}\selectfont}%
  \ifx\svgwidth\undefined%
    \setlength{\unitlength}{473.07738777bp}%
    \ifx\svgscale\undefined%
      \relax%
    \else%
      \setlength{\unitlength}{\unitlength * \real{\svgscale}}%
    \fi%
  \else%
    \setlength{\unitlength}{\svgwidth}%
  \fi%
  \global\let\svgwidth\undefined%
  \global\let\svgscale\undefined%
  \makeatother%
  \begin{picture}(1,0.36492194)%
    \lineheight{1}%
    \setlength\tabcolsep{0pt}%
    \put(0.00268633,0.34921203){\color[rgb]{0,0,0}\makebox(0,0)[lt]{\lineheight{1.25}\smash{\begin{tabular}[t]{l}a)\end{tabular}}}}%
    \put(0.4994639,0.34990123){\color[rgb]{0,0,0}\makebox(0,0)[lt]{\lineheight{1.25}\smash{\begin{tabular}[t]{l}b)\end{tabular}}}}%
    \put(0,0){\includegraphics[width=\unitlength,page=1]{two_column_fig_omission.pdf}}%
    \put(0.29489459,0.34217983){\color[rgb]{0,0,0}\makebox(0,0)[lt]{\lineheight{1.29999995}\smash{\begin{tabular}[t]{l}Including all dissipators\end{tabular}}}}%
    \put(0.28322682,0.31746137){\color[rgb]{0,0,0}\makebox(0,0)[lt]{\lineheight{1.29999995}\smash{\begin{tabular}[t]{l}Only diagonal dissipators\end{tabular}}}}%
    \put(0,0){\includegraphics[width=\unitlength,page=2]{two_column_fig_omission.pdf}}%
  \end{picture}%
\endgroup%

%% file: two_column_fig_7.pdf_tex
\begingroup%
  \makeatletter%
  \providecommand\color[2][]{%
    \errmessage{(Inkscape) Color is used for the text in Inkscape, but the package 'color.sty' is not loaded}%
    \renewcommand\color[2][]{}%
  }%
  \providecommand\transparent[1]{%
    \errmessage{(Inkscape) Transparency is used (non-zero) for the text in Inkscape, but the package 'transparent.sty' is not loaded}%
    \renewcommand\transparent[1]{}%
  }%
  \providecommand\rotatebox[2]{#2}%
  \newcommand*\fsize{\dimexpr\f@size pt\relax}%
  \newcommand*\lineheight[1]{\fontsize{\fsize}{#1\fsize}\selectfont}%
  \ifx\svgwidth\undefined%
    \setlength{\unitlength}{469.17006125bp}%
    \ifx\svgscale\undefined%
      \relax%
    \else%
      \setlength{\unitlength}{\unitlength * \real{\svgscale}}%
    \fi%
  \else%
    \setlength{\unitlength}{\svgwidth}%
  \fi%
  \global\let\svgwidth\undefined%
  \global\let\svgscale\undefined%
  \makeatother%
  \begin{picture}(1,0.36867626)%
    \lineheight{1}%
    \setlength\tabcolsep{0pt}%
    \put(0,0){\includegraphics[width=\unitlength,page=1]{two_column_fig_7.pdf}}%
    \put(-0.00061385,0.35086229){\color[rgb]{0,0,0}\makebox(0,0)[lt]{\lineheight{1.25}\smash{\begin{tabular}[t]{l}a)\end{tabular}}}}%
    \put(0,0){\includegraphics[width=\unitlength,page=2]{two_column_fig_7.pdf}}%
    \put(0.62581915,0.35166252){\color[rgb]{0,0,0}\makebox(0,0)[lt]{\lineheight{1.25}\smash{\begin{tabular}[t]{l}b)\end{tabular}}}}%
    \put(0,0){\includegraphics[width=\unitlength,page=3]{two_column_fig_7.pdf}}%
    \put(0.95611558,0.24931377){\color[rgb]{0,0,0}\makebox(0,0)[lt]{\lineheight{1.29999995}\smash{\begin{tabular}[t]{l}H\end{tabular}}}}%
    \put(0.95611558,0.21451602){\color[rgb]{0,0,0}\makebox(0,0)[lt]{\lineheight{1.29999995}\smash{\begin{tabular}[t]{l}H\end{tabular}}}}%
    \put(0.8155229,0.24931377){\color[rgb]{0,0,0}\makebox(0,0)[lt]{\lineheight{1.29999995}\smash{\begin{tabular}[t]{l}H\end{tabular}}}}%
    \put(0.8155229,0.21451602){\color[rgb]{0,0,0}\makebox(0,0)[lt]{\lineheight{1.29999995}\smash{\begin{tabular}[t]{l}H\end{tabular}}}}%
    \put(0,0){\includegraphics[width=\unitlength,page=4]{two_column_fig_7.pdf}}%
  \end{picture}%
\endgroup%

%% file: one_column_validation.pdf_tex
\begingroup%
  \makeatletter%
  \providecommand\color[2][]{%
    \errmessage{(Inkscape) Color is used for the text in Inkscape, but the package 'color.sty' is not loaded}%
    \renewcommand\color[2][]{}%
  }%
  \providecommand\transparent[1]{%
    \errmessage{(Inkscape) Transparency is used (non-zero) for the text in Inkscape, but the package 'transparent.sty' is not loaded}%
    \renewcommand\transparent[1]{}%
  }%
  \providecommand\rotatebox[2]{#2}%
  \newcommand*\fsize{\dimexpr\f@size pt\relax}%
  \newcommand*\lineheight[1]{\fontsize{\fsize}{#1\fsize}\selectfont}%
  \ifx\svgwidth\undefined%
    \setlength{\unitlength}{203.74002772bp}%
    \ifx\svgscale\undefined%
      \relax%
    \else%
      \setlength{\unitlength}{\unitlength * \real{\svgscale}}%
    \fi%
  \else%
    \setlength{\unitlength}{\svgwidth}%
  \fi%
  \global\let\svgwidth\undefined%
  \global\let\svgscale\undefined%
  \makeatother%
  \begin{picture}(1,0.73499771)%
    \lineheight{1}%
    \setlength\tabcolsep{0pt}%
    \put(0.72034327,0.68104856){\color[rgb]{0.10196078,0.10196078,0.10196078}\makebox(0,0)[lt]{\lineheight{1.25}\smash{\begin{tabular}[t]{l}Exact solution\end{tabular}}}}%
    \put(0.72548747,0.61381687){\color[rgb]{0.10196078,0.10196078,0.10196078}\makebox(0,0)[lt]{\lineheight{1.25}\smash{\begin{tabular}[t]{l}Quantum circuit\end{tabular}}}}%
    \put(0,0){\includegraphics[width=\unitlength,page=1]{one_column_validation.pdf}}%
  \end{picture}%
\endgroup%

%% file: two_column_fig_8.pdf_tex
\begingroup%
  \makeatletter%
  \providecommand\color[2][]{%
    \errmessage{(Inkscape) Color is used for the text in Inkscape, but the package 'color.sty' is not loaded}%
    \renewcommand\color[2][]{}%
  }%
  \providecommand\transparent[1]{%
    \errmessage{(Inkscape) Transparency is used (non-zero) for the text in Inkscape, but the package 'transparent.sty' is not loaded}%
    \renewcommand\transparent[1]{}%
  }%
  \providecommand\rotatebox[2]{#2}%
  \newcommand*\fsize{\dimexpr\f@size pt\relax}%
  \newcommand*\lineheight[1]{\fontsize{\fsize}{#1\fsize}\selectfont}%
  \ifx\svgwidth\undefined%
    \setlength{\unitlength}{478.47219392bp}%
    \ifx\svgscale\undefined%
      \relax%
    \else%
      \setlength{\unitlength}{\unitlength * \real{\svgscale}}%
    \fi%
  \else%
    \setlength{\unitlength}{\svgwidth}%
  \fi%
  \global\let\svgwidth\undefined%
  \global\let\svgscale\undefined%
  \makeatother%
  \begin{picture}(1,0.51737133)%
    \lineheight{1}%
    \setlength\tabcolsep{0pt}%
    \put(0,0){\includegraphics[width=\unitlength,page=1]{two_column_fig_8.pdf}}%
  \end{picture}%
\endgroup%

%% file: 75_qubits.pdf_tex
\begingroup%
  \makeatletter%
  \providecommand\color[2][]{%
    \errmessage{(Inkscape) Color is used for the text in Inkscape, but the package 'color.sty' is not loaded}%
    \renewcommand\color[2][]{}%
  }%
  \providecommand\transparent[1]{%
    \errmessage{(Inkscape) Transparency is used (non-zero) for the text in Inkscape, but the package 'transparent.sty' is not loaded}%
    \renewcommand\transparent[1]{}%
  }%
  \providecommand\rotatebox[2]{#2}%
  \newcommand*\fsize{\dimexpr\f@size pt\relax}%
  \newcommand*\lineheight[1]{\fontsize{\fsize}{#1\fsize}\selectfont}%
  \ifx\svgwidth\undefined%
    \setlength{\unitlength}{742.5bp}%
    \ifx\svgscale\undefined%
      \relax%
    \else%
      \setlength{\unitlength}{\unitlength * \real{\svgscale}}%
    \fi%
  \else%
    \setlength{\unitlength}{\svgwidth}%
  \fi%
  \global\let\svgwidth\undefined%
  \global\let\svgscale\undefined%
  \makeatother%
  \begin{picture}(1,0.62929293)%
    \lineheight{1}%
    \setlength\tabcolsep{0pt}%
    \put(0,0){\includegraphics[width=\unitlength,page=1]{75_qubits.pdf}}%
    \put(0.01399692,0.60707379){\color[rgb]{0,0,0}\makebox(0,0)[lt]{\lineheight{1.29999995}\smash{\begin{tabular}[t]{l}a)\end{tabular}}}}%
    \put(0.5200633,0.60665646){\color[rgb]{0,0,0}\makebox(0,0)[lt]{\lineheight{1.29999995}\smash{\begin{tabular}[t]{l}b)\end{tabular}}}}%
    \put(0.5259027,0.31174486){\color[rgb]{0,0,0}\makebox(0,0)[lt]{\lineheight{1.29999995}\smash{\begin{tabular}[t]{l}c)\end{tabular}}}}%
  \end{picture}%
\endgroup%

%% file: 129_qubits.pdf_tex
\begingroup%
  \makeatletter%
  \providecommand\color[2][]{%
    \errmessage{(Inkscape) Color is used for the text in Inkscape, but the package 'color.sty' is not loaded}%
    \renewcommand\color[2][]{}%
  }%
  \providecommand\transparent[1]{%
    \errmessage{(Inkscape) Transparency is used (non-zero) for the text in Inkscape, but the package 'transparent.sty' is not loaded}%
    \renewcommand\transparent[1]{}%
  }%
  \providecommand\rotatebox[2]{#2}%
  \newcommand*\fsize{\dimexpr\f@size pt\relax}%
  \newcommand*\lineheight[1]{\fontsize{\fsize}{#1\fsize}\selectfont}%
  \ifx\svgwidth\undefined%
    \setlength{\unitlength}{742.5bp}%
    \ifx\svgscale\undefined%
      \relax%
    \else%
      \setlength{\unitlength}{\unitlength * \real{\svgscale}}%
    \fi%
  \else%
    \setlength{\unitlength}{\svgwidth}%
  \fi%
  \global\let\svgwidth\undefined%
  \global\let\svgscale\undefined%
  \makeatother%
  \begin{picture}(1,0.62929293)%
    \lineheight{1}%
    \setlength\tabcolsep{0pt}%
    \put(0,0){\includegraphics[width=\unitlength,page=1]{129_qubits.pdf}}%
    \put(0.01978404,0.59974994){\color[rgb]{0,0,0}\makebox(0,0)[lt]{\lineheight{1.29999995}\smash{\begin{tabular}[t]{l}d)\end{tabular}}}}%
    \put(0.52584463,0.59368933){\color[rgb]{0,0,0}\makebox(0,0)[lt]{\lineheight{1.29999995}\smash{\begin{tabular}[t]{l}e)\end{tabular}}}}%
    \put(0.52584463,0.31288125){\color[rgb]{0,0,0}\makebox(0,0)[lt]{\lineheight{1.29999995}\smash{\begin{tabular}[t]{l}f)\end{tabular}}}}%
  \end{picture}%
\endgroup%

%% file: bibs.bib
@article{Feynman1982,
  author  = {Feynman, R. P.},
  title   = {Simulating physics with computers},
  journal = {International Journal of Theoretical Physics},
  year    = {1982},
  volume  = {21},
  number  = {6–7},
  pages   = {467--488},
  doi     = {10.1007/BF02650179},
}

@article{Lloyd1996,
author = {Seth Lloyd },
title = {Universal Quantum Simulators},
journal = {Science},
volume = {273},
number = {5278},
pages = {1073-1078},
year = {1996},
doi = {10.1126/science.273.5278.1073},
URL = {https://www.science.org/doi/abs/10.1126/science.273.5278.1073},
eprint = {https://www.science.org/doi/pdf/10.1126/science.273.5278.1073}}

@book{scully1997quantum,
  author    = {Marlan O. Scully and M. Suhail Zubairy},
  title     = {Quantum Optics},
  year      = {1997},
  publisher = {Cambridge University Press},
  address   = {Cambridge},
  isbn      = {9780521435956},
  doi       = {10.1017/CBO9780511813993}
}

@book{weiss2012quantum,
  title={Quantum Dissipative Systems},
  author={Weiss, U.},
  isbn={9789814374910},
  lccn={2012418311},
  series={G - Reference,Information and Interdisciplinary Subjects Series},
  url={https://books.google.es/books?id=qgfuFZxvGKQC},
  year={2012},
  publisher={World Scientific}
}

@article{Akihito2009,
author = {Akihito Ishizaki  and Graham R. Fleming },
title = {Theoretical examination of quantum coherence in a photosynthetic system at physiological temperature},
journal = {Proceedings of the National Academy of Sciences},
volume = {106},
number = {41},
pages = {17255-17260},
year = {2009},
doi = {10.1073/pnas.0908989106},
URL = {https://www.pnas.org/doi/abs/10.1073/pnas.0908989106},
eprint = {https://www.pnas.org/doi/pdf/10.1073/pnas.0908989106}}

@article{Dicke1954,
  title = {Coherence in Spontaneous Radiation Processes},
  author = {Dicke, R. H.},
  journal = {Phys. Rev.},
  volume = {93},
  issue = {1},
  pages = {99--110},
  numpages = {0},
  year = {1954},
  month = {Jan},
  publisher = {American Physical Society},
  doi = {10.1103/PhysRev.93.99},
  url = {https://link.aps.org/doi/10.1103/PhysRev.93.99}
}

@article{Lehmberg1970,
  title = {Radiation from an $N$-Atom System. I. General Formalism},
  author = {Lehmberg, R. H.},
  journal = {Phys. Rev. A},
  volume = {2},
  issue = {3},
  pages = {883--888},
  numpages = {0},
  year = {1970},
  month = {Sep},
  publisher = {American Physical Society},
  doi = {10.1103/PhysRevA.2.883},
  url = {https://link.aps.org/doi/10.1103/PhysRevA.2.883}
}

@article{Gross1982,
title = {Superradiance: An essay on the theory of collective spontaneous emission},
journal = {Physics Reports},
volume = {93},
number = {5},
pages = {301-396},
year = {1982},
issn = {0370-1573},
doi = {https://doi.org/10.1016/0370-1573(82)90102-8},
url = {https://www.sciencedirect.com/science/article/pii/0370157382901028},
author = {M. Gross and S. Haroche}
}

@article{Shore1993,
  title={The jaynes-cummings model},
  author={Shore, Bruce W and Knight, Peter L},
  journal={Journal of Modern Optics},
  volume={40},
  number={7},
  pages={1195--1238},
  year={1993},
  publisher={Taylor \& Francis}
}

@article{Purcell1946,
  title = {Spontaneous Emission Probabilities at Radio Frequencies},
  author = {Purcell, E. M.},
  journal = {Physical Review},
  volume = {69},
  pages = {681},
  year = {1946},
  doi = {10.1103/PhysRev.69.681}
}

@InProceedings{Zurek2001,
author="Paz, J. P.
and Zurek, W. H.",
editor="Kaiser, R.
and Westbrook, C.
and David, F.",
title="Environment-Induced Decoherence and the Transition from Quantum to Classical",
booktitle="Coherent atomic matter waves",
year="2001",
publisher="Springer Berlin Heidelberg",
address="Berlin, Heidelberg",
pages="533--614",
isbn="978-3-540-45338-3"
}

@article{Rivas2014,
doi = {10.1088/0034-4885/77/9/094001},
url = {https://doi.org/10.1088/0034-4885/77/9/094001},
year = {2014},
month = {aug},
publisher = {IOP Publishing},
volume = {77},
number = {9},
pages = {094001},
author = {Rivas, {\'A}ngel and Huelga, Susana F and Plenio, Martin B},
title = {Quantum non-Markovianity: characterization, quantification and detection},
journal = {Reports on Progress in Physics}
}

@article{trotter1959,
  title={On the product of semi-groups of operators},
  author={Trotter, Hale F.},
  journal={Proceedings of the American Mathematical Society},
  volume={10},
  number={4},
  pages={545--551},
  year={1959}
}

@article{suzuki1976,
  title={General theory of higher-order decomposition of exponential operators and symplectic integrators},
  author={Suzuki, Masuo},
  journal={Communications in Mathematical Physics},
  volume={51},
  pages={183--190},
  year={1976}
}

@article{kandala2017,
  title={Hardware-efficient variational quantum eigensolver for small molecules and quantum magnets},
  author={Kandala, Abhinav and Mezzacapo, Antonio and Temme, Kristan and Takita, Maika and Chow, Jerry M. and Gambetta, Jay M.},
  journal={Nature},
  volume={549},
  pages={242--246},
  year={2017}
}

@article{arute2020,
  title={Quantum supremacy using a programmable superconducting processor},
  author={Arute, Frank and Arya, Kunal and Babbush, Ryan and others},
  journal={Nature},
  volume={574},
  pages={505--510},
  year={2019}
}

@misc{Hartnett2026,
      title={Fast, accurate, high-resolution simulation of large-scale Fermi-Hubbard models on a digital quantum processor}, 
      author={Gavin S. Hartnett and Khadijeh Sona Najafi and Aleksei Khindanov and Haoran Liao and Michael Schutzman and Michael R. Hush and Michael J. Biercuk and Yuval Baum},
      year={2026},
      eprint={2605.04025},
      archivePrefix={arXiv},
      primaryClass={quant-ph},
      url={https://arxiv.org/abs/2605.04025}, 
}

@misc{Alam2025,
      title={Fermionic dynamics on a trapped-ion quantum computer beyond exact classical simulation}, 
      author={Faisal Alam and Jan Lukas Bosse and Ieva Čepaitė and Adrian Chapman and Laura Clinton and Marcos Crichigno and Elizabeth Crosson and Toby Cubitt and Charles Derby and Oliver Dowinton and Norhan Eassa and Paul K. Faehrmann and Steve Flammia and Brian Flynn and Filippo Maria Gambetta and Raúl García-Patrón and Max Hunter-Gordon and Glenn Jones and Abhishek Khedkar and Joel Klassen and Michael Kreshchuk and Edward Harry McMullan and Lana Mineh and Ashley Montanaro and Caterina Mora and John J. L. Morton and Alberto Nocera and Dhrumil Patel and Pete Rolph and Raul A. Santos and James R. Seddon and Evan Sheridan and Wilfrid Somogyi and Marika Svensson and Niam Vaishnav and Sabrina Yue Wang and Gethin Wright and Eli Chertkov and Henrik Dreyer and Michael Foss-Feig},
      year={2025},
      eprint={2510.26300},
      archivePrefix={arXiv},
      primaryClass={quant-ph},
      url={https://arxiv.org/abs/2510.26300}, 
}

@misc{Alam2025_2,
      title={Programmable digital quantum simulation of 2D Fermi-Hubbard dynamics using 72 superconducting qubits}, 
      author={Faisal Alam and Jan Lukas Bosse and Ieva Čepaitė and Adrian Chapman and Laura Clinton and Marcos Crichigno and Elizabeth Crosson and Toby Cubitt and Charles Derby and Oliver Dowinton and Paul K. Faehrmann and Steve Flammia and Brian Flynn and Filippo Maria Gambetta and Raúl García-Patrón and Max Hunter-Gordon and Glenn Jones and Abhishek Khedkar and Joel Klassen and Michael Kreshchuk and Edward Harry McMullan and Lana Mineh and Ashley Montanaro and Caterina Mora and John J. L. Morton and Dhrumil Patel and Pete Rolph and Raul A. Santos and James R. Seddon and Evan Sheridan and Wilfrid Somogyi and Marika Svensson and Niam Vaishnav and Sabrina Yue Wang and Gethin Wright},
      year={2025},
      eprint={2510.26845},
      archivePrefix={arXiv},
      primaryClass={quant-ph},
      url={https://arxiv.org/abs/2510.26845}, 
}

@article{barends2015,
  title   = {Digital quantum simulation of fermionic models with a superconducting circuit},
  author  = {Barends, R. and Lamata, L. and Kelly, J. and Garc{\'i}a-{\'A}lvarez, L. and Fowler, A. G. and Megrant, A. and Jeffrey, E. and White, T. C. and Sank, D. and Mutus, J. Y. and Campbell, B. and Chen, Yu and Chen, Z. and Chiaro, B. and Dunsworth, A. and Hoi, I.-C. and Neill, C. and O'Malley, P. J. J. and Quintana, C. and Roushan, P. and Vainsencher, A. and Wenner, J. and Solano, E. and Martinis, J. M.},
  journal = {Nature Communications},
  volume  = {6},
  pages   = {7654},
  year    = {2015},
  doi     = {10.1038/ncomms8654}
}

@article{martinez2016,
  title   = {Real-time dynamics of lattice gauge theories with a few-qubit quantum computer},
  author  = {Martinez, Esteban A. and Muschik, Christine A. and Schindler, Philipp and Nigg, Daniel and Erhard, Alexander and Heyl, Markus and Hauke, Philipp and Dalmonte, Marcello and Monz, Thomas and Zoller, Peter and Blatt, Rainer},
  journal = {Nature},
  volume  = {534},
  pages   = {516--519},
  year    = {2016},
  doi     = {10.1038/nature18318}
}

@misc{Cobos2025,
      title={Real-Time Dynamics in a (2+1)-D Gauge Theory: The Stringy Nature on a Superconducting Quantum Simulator}, 
      author={Jesús Cobos and Joana Fraxanet and César Benito and Francesco di Marcantonio and Pedro Rivero and Kornél Kapás and Miklós Antal Werner and Örs Legeza and Alejandro Bermudez and Enrique Rico},
      year={2025},
      eprint={2507.08088},
      archivePrefix={arXiv},
      primaryClass={quant-ph},
      url={https://arxiv.org/abs/2507.08088}, 
}

@article{Farrell2024,
  title = {Quantum simulations of hadron dynamics in the Schwinger model using 112 qubits},
  author = {Farrell, Roland C. and Illa, Marc and Ciavarella, Anthony N. and Savage, Martin J.},
  journal = {Phys. Rev. D},
  volume = {109},
  issue = {11},
  pages = {114510},
  numpages = {52},
  year = {2024},
  month = {Jun},
  publisher = {American Physical Society},
  doi = {10.1103/PhysRevD.109.114510},
  url = {https://link.aps.org/doi/10.1103/PhysRevD.109.114510}
}

@misc{Xu2026,
      title={Observation of glueball excitations and string breaking in a $2+1$D $\mathbb{Z}_2$ lattice gauge theory on a trapped-ion quantum computer}, 
      author={Kaidi Xu and Umberto Borla and Kevin Hemery and Rohan Joshi and Henrik Dreyer and Enrico Rinaldi and Jad C. Halimeh},
      year={2026},
      eprint={2604.07435},
      archivePrefix={arXiv},
      primaryClass={hep-lat},
      url={https://arxiv.org/abs/2604.07435}, 
}

@article{Gonzalez2025,
  title     = {Observation of string breaking on a (2 + 1)D Rydberg quantum simulator},
  author    = {Gonz{\'a}lez-Cuadra, Daniel and Hamdan, Majd and Zache, Torsten V. and Braverman, Boris and Kornja{\v{c}}a, Milan and Lukin, Alexander and Cant{\'u}, Sergio H. and Liu, Fangli and Wang, Sheng-Tao and Keesling, Alexander and Lukin, Mikhail D. and Zoller, Peter and Bylinskii, Alexei},
  journal   = {Nature},
  volume    = {642},
  pages     = {321--326},
  year      = {2025},
  doi       = {10.1038/s41586-025-09051-6},
  url       = {https://doi.org/10.1038/s41586-025-09051-6}
}

@article{Cochran2025,
  title     = {Visualizing dynamics of charges and strings in (2 + 1)D lattice gauge theories},
  author    = {Cochran, T. A. and Jobst, B. and Rosenberg, E. and others},
  journal   = {Nature},
  volume    = {642},
  pages     = {315--320},
  year      = {2025},
  doi       = {10.1038/s41586-025-08999-9},
  url       = {https://doi.org/10.1038/s41586-025-08999-9}
}

@article{Switzer2026,
  author  = {Switzer, Eric D. and Robertson, Niall F. and Keenan, Nathan and Rodr{\'{\i}}guez-Alcaraz, {\'A}ngel and D'Urbano, Andrea and Pokharel, Bibek and Rahman, Talat S. and Shtanko, Oles and Zhuk, Sergiy and Lorente, Nicol{\'a}s},
  title   = {Realization of two-dimensional discrete time crystals with anisotropic Heisenberg coupling},
  journal = {Nature Communications},
  year    = {2026},
  volume  = {17},
  pages   = {605},
  doi     = {10.1038/s41467-025-67787-1},
}

@article{kim2023,
  title     = {Evidence for the utility of quantum computing before fault tolerance},
  author    = {Kim, Youngseok and Eddins, Andrew and Anand, Sajant and Wei, Ken Xuan and van den Berg, Ewout and Rosenblatt, Sami and Nayfeh, Hasan and Wu, Yantao and Zaletel, Michael and Temme, Kristan and Kandala, Abhinav},
  journal   = {Nature},
  volume    = {618},
  pages     = {500--505},
  year      = {2023},
  doi       = {10.1038/s41586-023-06096-3},
  url       = {https://doi.org/10.1038/s41586-023-06096-3}
}

@article{Shtanko2025,
  title     = {Uncovering local integrability in quantum many-body dynamics},
  author    = {Shtanko, Oles and Wang, Derek S. and Zhang, Haimeng and Harle, Nikhil and Seif, Alireza and Movassagh, Ramis and Minev, Zlatko},
  journal   = {Nature Communications},
  volume    = {16},
  number    = {1},
  pages     = {2552},
  year      = {2025},
  doi       = {10.1038/s41467-025-57623-x},
  url       = {https://doi.org/10.1038/s41467-025-57623-x}
}

@article{Fischer2026,
  title     = {Dynamical simulations of many-body quantum chaos on a quantum computer},
  author    = {Fischer, Laurin E. and Leahy, Matea and Eddins, Andrew and Keenan, Nathan and Ferracin, Davide and Rossi, Matteo A. C. and Kim, Youngseok and He, Andre and Pietracaprina, Francesca and Sokolov, Boris and Dooley, Shane and Zimbor{\'a}s, Zolt{\'a}n and Tacchino, Francesco and Maniscalco, Sabrina and Goold, John and Garc{\'i}a-P{\'e}rez, Guillermo and Tavernelli, Ivano and Kandala, Abhinav and Filippov, Sergey N.},
  journal   = {Nature Physics},
  volume    = {22},
  pages     = {302--307},
  year      = {2026},
  doi       = {10.1038/s41567-025-03144-9},
  url       = {https://doi.org/10.1038/s41567-025-03144-9}
}

@article{lindblad1976,
  title={On the generators of quantum dynamical semigroups},
  author={Lindblad, G{\"o}ran},
  journal={Communications in Mathematical Physics},
  volume={48},
  pages={119--130},
  year={1976}
}

@article{gorini1976,
  title={Completely positive dynamical semigroups of $N$-level systems},
  author={Gorini, Vittorio and Kossakowski, Andrzej and Sudarshan, E. C. G.},
  journal={Journal of Mathematical Physics},
  volume={17},
  pages={821--825},
  year={1976}
}

@article{daley2014,
  title={Quantum trajectories and open many-body quantum systems},
  author={Daley, Andrew J.},
  journal={Advances in Physics},
  volume={63},
  pages={77--149},
  year={2014}
}

@article{childs2019,
  title={Toward the first quantum simulation with quantum speedup},
  author={Childs, Andrew M. and Su, Yuan and Tran, Minh C. and Wiebe, Nathan and Zhu, Shuchen},
  journal={Proceedings of the National Academy of Sciences},
  volume={115},
  pages={9456--9461},
  year={2018}
}

@article{yuan2019,
  title={Theory of variational quantum simulation},
  author={Yuan, Xin and Endo, Suguru and Zhao, Qisheng and Li, Ying and Benjamin, Simon C.},
  journal={Quantum},
  volume={3},
  pages={191},
  year={2019}
}

@article{PRXQuantum.5.030339,
  title = {Efficient Long-Range Entanglement Using Dynamic Circuits},
  author = {B\"aumer, Elisa and Tripathi, Vinay and Wang, Derek S. and Rall, Patrick and Chen, Edward H. and Majumder, Swarnadeep and Seif, Alireza and Minev, Zlatko K.},
  journal = {PRX Quantum},
  volume = {5},
  issue = {3},
  pages = {030339},
  numpages = {20},
  year = {2024},
  month = {Aug},
  publisher = {American Physical Society},
  doi = {10.1103/PRXQuantum.5.030339},
  url = {https://link.aps.org/doi/10.1103/PRXQuantum.5.030339}
}

@article{Buhrman_2024,
   title={State preparation by shallow circuits using feed forward},
   volume={8},
   ISSN={2521-327X},
   url={http://dx.doi.org/10.22331/q-2024-12-09-1552},
   DOI={10.22331/q-2024-12-09-1552},
   journal={Quantum},
   publisher={Verein zur Forderung des Open Access Publizierens in den Quantenwissenschaften},
   author={Buhrman, Harry and Folkertsma, Marten and Loff, Bruno and Neumann, Niels M. P.},
   year={2024},
   month=dec, pages={1552} }

@article{Cao2025,
  title={Measurement-driven quantum advantages in shallow circuits},
  author={Chenfeng Cao and Jens Eisert},
  journal={Physical Review Letters},
  year={2025},
  url={https://api.semanticscholar.org/CorpusID:278394412}
}

@article{Temme_2017,
   title={Error Mitigation for Short-Depth Quantum Circuits},
   volume={119},
   ISSN={1079-7114},
   url={http://dx.doi.org/10.1103/PhysRevLett.119.180509},
   DOI={10.1103/physrevlett.119.180509},
   number={18},
   journal={Physical Review Letters},
   publisher={American Physical Society (APS)},
   author={Temme, Kristan and Bravyi, Sergey and Gambetta, Jay M.},
   year={2017},
   month=nov }

@article{Richardson1911,
  author  = {Richardson, Lewis Fry},
  title   = {The Approximate Arithmetical Solution by Finite Differences of Physical Problems Involving Differential Equations},
  journal = {Philosophical Transactions of the Royal Society A},
  volume  = {210},
  pages   = {307--357},
  year    = {1911},
  doi     = {10.1098/rsta.1911.0009}
}

@book{Boyd2004,
  author    = {Boyd, Stephen and Vandenberghe, Lieven},
  title     = {Convex Optimization},
  publisher = {Cambridge University Press},
  year      = {2004}
}

@article{Czarnik2021,
  doi = {10.22331/q-2021-11-26-592},
  url = {https://doi.org/10.22331/q-2021-11-26-592},
  title = {Error mitigation with {C}lifford quantum-circuit data},
  author = {Czarnik, Piotr and Arrasmith, Andrew and Coles, Patrick J. and Cincio, Lukasz},
  journal = {{Quantum}},
  issn = {2521-327X},
  publisher = {{Verein zur F{\"{o}}rderung des Open Access Publizierens in den Quantenwissenschaften}},
  volume = {5},
  pages = {592},
  month = nov,
  year = {2021}
}

@article{Lowe2021,
  title = {Unified approach to data-driven quantum error mitigation},
  author = {Lowe, Angus and Gordon, Max Hunter and Czarnik, Piotr and Arrasmith, Andrew and Coles, Patrick J. and Cincio, Lukasz},
  journal = {Phys. Rev. Res.},
  volume = {3},
  issue = {3},
  pages = {033098},
  numpages = {12},
  year = {2021},
  month = {Jul},
  publisher = {American Physical Society},
  doi = {10.1103/PhysRevResearch.3.033098},
  url = {https://link.aps.org/doi/10.1103/PhysRevResearch.3.033098}
}

@article{Strikis2021,
  title = {Learning-Based Quantum Error Mitigation},
  author = {Strikis, Armands and Qin, Dayue and Chen, Yanzhu and Benjamin, Simon C. and Li, Ying},
  journal = {PRX Quantum},
  volume = {2},
  issue = {4},
  pages = {040330},
  numpages = {30},
  year = {2021},
  month = {Nov},
  publisher = {American Physical Society},
  doi = {10.1103/PRXQuantum.2.040330},
  url = {https://link.aps.org/doi/10.1103/PRXQuantum.2.040330}
}

@article{Weaving2025,
  doi = {10.22331/q-2025-05-05-1732},
  url = {https://doi.org/10.22331/q-2025-05-05-1732},
  title = {Accurately {S}imulating the {T}ime {E}volution of an {I}sing {M}odel with {E}cho {V}erified {C}lifford {D}ata {R}egression on a {S}uperconducting {Q}uantum {C}omputer},
  author = {Weaving, Tim and Ralli, Alexis and Love, Peter J. and Succi, Sauro and Coveney, Peter V.},
  journal = {{Quantum}},
  issn = {2521-327X},
  publisher = {{Verein zur F{\"{o}}rderung des Open Access Publizierens in den Quantenwissenschaften}},
  volume = {9},
  pages = {1732},
  month = may,
  year = {2025}
}

@article{Giurgica2020,
  title={Digital zero noise extrapolation for quantum error mitigation},
  author={Tudor Giurgică-Tiron and Yousef Hindy and Ryan Larose and Andrea Mari and William J. Zeng},
  journal={2020 IEEE International Conference on Quantum Computing and Engineering (QCE)},
  year={2020},
  pages={306-316},
  url={https://api.semanticscholar.org/CorpusID:218862807}
}

@article{Viola_1998,
   title={Dynamical suppression of decoherence in two-state quantum systems},
   volume={58},
   ISSN={1094-1622},
   url={http://dx.doi.org/10.1103/PhysRevA.58.2733},
   DOI={10.1103/physreva.58.2733},
   number={4},
   journal={Physical Review A},
   publisher={American Physical Society (APS)},
   author={Viola, Lorenza and Lloyd, Seth},
   year={1998},
   month=oct, pages={2733–2744} }

@book{breuer2002,
  title={The Theory of Open Quantum Systems},
  author={Breuer, Heinz-Peter and Petruccione, Francesco},
  publisher={Oxford University Press},
  year={2002}
}

@book{Nielsen_Chuang_2010, 
place={Cambridge}, 
title={Quantum Computation and Quantum Information: 10th Anniversary Edition}, 
publisher={Cambridge University Press}, 
author={Nielsen, Michael A. and Chuang, Isaac L.}, 
year={2010}}

@book{Kraus1983,
  editor    = {Kraus, Karl and Böhm, A. and Dollard, J. D. and Wootters, W. H.},
  title     = {States, Effects, and Operations: Fundamental Notions of Quantum Theory},
  series    = {Lecture Notes in Physics},
  volume    = {190},
  publisher = {Springer Berlin Heidelberg},
  year      = {1983},
  doi       = {10.1007/3-540-12732-1},
  isbn      = {978-3-540-12732-1},
}

@article{ElGordo2024,
  title = {Tailoring the statistics of light emitted from two interacting quantum emitters},
  author = {Juan-Delgado, Adri\'an and Esteban, Ruben and Nodar, \'Alvaro and Trebbia, Jean-Baptiste and Lounis, Brahim and Aizpurua, Javier},
  journal = {Phys. Rev. Res.},
  volume = {6},
  issue = {2},
  pages = {023207},
  numpages = {26},
  year = {2024},
  month = {May},
  publisher = {American Physical Society},
  doi = {10.1103/PhysRevResearch.6.023207},
  url = {https://link.aps.org/doi/10.1103/PhysRevResearch.6.023207}
}

@article{ElGordo2025,
  title = {Addressing the Correlation of Stokes-Shifted Photons Emitted from Two Quantum Emitters},
  author = {Juan-Delgado, Adri\'an and Trebbia, Jean-Baptiste and Esteban, Ruben and Deplano, Quentin and Tamarat, Philippe and Avriller, R\'emi and Lounis, Brahim and Aizpurua, Javier},
  journal = {Phys. Rev. Lett.},
  volume = {135},
  issue = {16},
  pages = {163602},
  numpages = {7},
  year = {2025},
  month = {Oct},
  publisher = {American Physical Society},
  doi = {10.1103/1z52-p73t},
  url = {https://link.aps.org/doi/10.1103/1z52-p73t}
}

@book{Wilde_2017, place={Cambridge}, edition={2}, title={Quantum Information Theory}, publisher={Cambridge University Press}, author={Wilde, Mark M.}, year={2017}}

@inproceedings{Stinespring1955,
  title={Positive functions on *-algebras},
  author={William F. Stinespring},
  year={1955},
  url={https://api.semanticscholar.org/CorpusID:119642419}
}

@article{Choi1975,
title = {Completely positive linear maps on complex matrices},
journal = {Linear Algebra and its Applications},
volume = {10},
number = {3},
pages = {285-290},
year = {1975},
issn = {0024-3795},
doi = {https://doi.org/10.1016/0024-3795(75)90075-0},
url = {https://www.sciencedirect.com/science/article/pii/0024379575900750},
author = {Man-Duen Choi}
}

@article{DelRe2020,
  author  = {Del Re, Lorenzo and Rost, Brian and Kemper, A. F. and Freericks, J. K.},
  title   = {Driven-dissipative quantum mechanics on a lattice: Simulating a fermionic reservoir on a quantum computer},
  journal = {Physical Review B},
  year    = {2020},
  volume  = {102},
  pages   = {125112},
  doi     = {10.1103/PhysRevB.102.125112},
}

@article{Cattaneo2023,
  title = {Quantum Simulation of Dissipative Collective Effects on Noisy Quantum Computers},
  author = {Cattaneo, Marco and Rossi, Matteo A.C. and Garc\'{\i}a-P\'erez, Guillermo and Zambrini, Roberta and Maniscalco, Sabrina},
  journal = {PRX Quantum},
  volume = {4},
  issue = {1},
  pages = {010324},
  numpages = {31},
  year = {2023},
  month = {Mar},
  publisher = {American Physical Society},
  doi = {10.1103/PRXQuantum.4.010324},
  url = {https://link.aps.org/doi/10.1103/PRXQuantum.4.010324}
}

@article{Vidal2004,
  title = {Efficient Simulation of One-Dimensional Quantum Many-Body Systems},
  author = {Vidal, G.},
  journal = {Physical Review Letters},
  volume = {93},
  pages = {040502},
  year = {2004},
  doi = {10.1103/PhysRevLett.93.040502}
}

@article{Riste2013,
  author       = {Rist\`e, D. and Dukalski, M. and Watson, C. A. and de Lange, G. and Tiggelman, M. J. and Blanter, Y. M. and Lehnert, K. W. and Schouten, R. N. and DiCarlo, L.},
  title        = {Deterministic entanglement of superconducting qubits by parity measurement and feedback},
  journal      = {Nature},
  volume       = {502},
  pages        = {350--354},
  year         = {2013},
  doi          = {10.1038/nature12513}
}

@misc{fossfeig2023,
      title={Experimental demonstration of the advantage of adaptive quantum circuits}, 
      author={Michael Foss-Feig and Arkin Tikku and Tsung-Cheng Lu and Karl Mayer and Mohsin Iqbal and Thomas M. Gatterman and Justin A. Gerber and Kevin Gilmore and Dan Gresh and Aaron Hankin and Nathan Hewitt and Chandler V. Horst and Mitchell Matheny and Tanner Mengle and Brian Neyenhuis and Henrik Dreyer and David Hayes and Timothy H. Hsieh and Isaac H. Kim},
      year={2023},
      eprint={2302.03029},
      archivePrefix={arXiv},
      primaryClass={quant-ph},
      url={https://arxiv.org/abs/2302.03029}, 
}

@article{Javadi-Abhari2024,
  author       = {Ali Javadi-Abhari and Matthew Treinish and Kevin Krsulich and Christopher J. Wood and Jake Lishman and Julien Gacon and Simon Martiel and Paul D. Nation and Lev S. Bishop and Andrew W. Cross and Blake R. Johnson and Jay M. Gambetta},
  title        = {Quantum computing with Qiskit},
  journal      = {arXiv:2405.08810 [quant-ph]},
  year         = {2024},
  doi          = {10.48550/arXiv.2405.08810},
  url          = {https://arxiv.org/abs/2405.08810}
}

@book{Efron1994,
  author    = {Efron, Bradley and Tibshirani, Robert J.},
  title     = {An Introduction to the Bootstrap},
  edition   = {1},
  publisher = {Chapman and Hall/CRC},
  address   = {New York},
  year      = {1994},
  doi       = {10.1201/9780429246593}
}

@article{Schollwoeck2011,
   title={The density-matrix renormalization group in the age of matrix product states},
   volume={326},
   ISSN={0003-4916},
   url={http://dx.doi.org/10.1016/j.aop.2010.09.012},
   DOI={10.1016/j.aop.2010.09.012},
   number={1},
   journal={Annals of Physics},
   publisher={Elsevier BV},
   author={Schollwöck, Ulrich},
   year={2011},
   month=jan, pages={96–192} }

@article{Molmer1992,
  title = {Wave-function approach to dissipative processes in quantum optics},
  author = {Dalibard, Jean and Castin, Yvan and M\o{}lmer, Klaus},
  journal = {Phys. Rev. Lett.},
  volume = {68},
  issue = {5},
  pages = {580--583},
  numpages = {0},
  year = {1992},
  month = {Feb},
  publisher = {American Physical Society},
  doi = {10.1103/PhysRevLett.68.580},
  url = {https://link.aps.org/doi/10.1103/PhysRevLett.68.580}
}

@book{Matousek2007,
  author    = {Ji{\v{r}}{\'\i} Matou{\v{s}}ek and Bernd G{\"a}rtner},
  title     = {Understanding and Using Linear Programming},
  series    = {Universitext},
  publisher = {Springer Berlin Heidelberg},
  address   = {Berlin, Heidelberg},
  year      = {2007},
  isbn      = {978-3-540-30697-9},
  doi       = {10.1007/978-3-540-30717-4}
}

@article{Molmer1993,
author = {Klaus M{\o}lmer and Yvan Castin and Jean Dalibard},
journal = {J. Opt. Soc. Am. B},
number = {3},
pages = {524--538},
publisher = {Optica Publishing Group},
title = {Monte Carlo wave-function method in quantum optics},
volume = {10},
month = {Mar},
year = {1993},
url = {https://opg.optica.org/josab/abstract.cfm?URI=josab-10-3-524},
doi = {10.1364/JOSAB.10.000524},
}

@article{Zoller1992,
  title = {Monte Carlo simulation of the atomic master equation for spontaneous emission},
  author = {Dum, R. and Zoller, P. and Ritsch, H.},
  journal = {Phys. Rev. A},
  volume = {45},
  issue = {7},
  pages = {4879--4887},
  numpages = {0},
  year = {1992},
  month = {Apr},
  publisher = {American Physical Society},
  doi = {10.1103/PhysRevA.45.4879},
  url = {https://link.aps.org/doi/10.1103/PhysRevA.45.4879}
}

@article{Orus2014,
title = {A practical introduction to tensor networks: Matrix product states and projected entangled pair states},
journal = {Annals of Physics},
volume = {349},
pages = {117-158},
year = {2014},
issn = {0003-4916},
doi = {https://doi.org/10.1016/j.aop.2014.06.013},
url = {https://www.sciencedirect.com/science/article/pii/S0003491614001596},
author = {Román Orús}
}

@misc{gottesman1998,
      title={The Heisenberg Representation of Quantum Computers}, 
      author={Daniel Gottesman},
      year={1998},
      eprint={quant-ph/9807006},
      archivePrefix={arXiv},
      primaryClass={quant-ph},
      url={https://arxiv.org/abs/quant-ph/9807006}, 
}

@article{Aaronson_2004,
   title={Improved simulation of stabilizer circuits},
   volume={70},
   ISSN={1094-1622},
   url={http://dx.doi.org/10.1103/PhysRevA.70.052328},
   DOI={10.1103/physreva.70.052328},
   number={5},
   journal={Physical Review A},
   publisher={American Physical Society (APS)},
   author={Aaronson, Scott and Gottesman, Daniel},
   year={2004},
   month=nov }

@article{PhysRevLett.107.120501,
  title = {Dissipative Quantum Church-Turing Theorem},
  author = {Kliesch, M. and Barthel, T. and Gogolin, C. and Kastoryano, M. and Eisert, J.},
  journal = {Phys. Rev. Lett.},
  volume = {107},
  issue = {12},
  pages = {120501},
  numpages = {5},
  year = {2011},
  month = {Sep},
  publisher = {American Physical Society},
  doi = {10.1103/PhysRevLett.107.120501},
  url = {https://link.aps.org/doi/10.1103/PhysRevLett.107.120501}
}

@article{Griffiths1996,
  title = {Semiclassical Fourier Transform for Quantum Computation},
  author = {Griffiths, Robert B. and Niu, Chi-Sheng},
  journal = {Phys. Rev. Lett.},
  volume = {76},
  issue = {17},
  pages = {3228--3231},
  numpages = {0},
  year = {1996},
  month = {Apr},
  publisher = {American Physical Society},
  doi = {10.1103/PhysRevLett.76.3228},
  url = {https://link.aps.org/doi/10.1103/PhysRevLett.76.3228}
}

@article{Baumer2024,
  title = {Quantum Fourier Transform Using Dynamic Circuits},
  author = {B\"aumer, Elisa and Tripathi, Vinay and Seif, Alireza and Lidar, Daniel and Wang, Derek S.},
  journal = {Phys. Rev. Lett.},
  volume = {133},
  issue = {15},
  pages = {150602},
  numpages = {7},
  year = {2024},
  month = {Oct},
  publisher = {American Physical Society},
  doi = {10.1103/PhysRevLett.133.150602},
  url = {https://link.aps.org/doi/10.1103/PhysRevLett.133.150602}
}

@article{Corcoles2021,
  title = {Exploiting Dynamic Quantum Circuits in a Quantum Algorithm with Superconducting Qubits},
  author = {C\'orcoles, A. D. and Takita, Maika and Inoue, Ken and Lekuch, Scott and Minev, Zlatko K. and Chow, Jerry M. and Gambetta, Jay M.},
  journal = {Phys. Rev. Lett.},
  volume = {127},
  issue = {10},
  pages = {100501},
  numpages = {6},
  year = {2021},
  month = {Aug},
  publisher = {American Physical Society},
  doi = {10.1103/PhysRevLett.127.100501},
  url = {https://link.aps.org/doi/10.1103/PhysRevLett.127.100501}
}

@article{Sander2025,
  author    = {Sander, Aaron and Fr{\"o}hlich, Maximilian and Eigel, Martin and Eisert, Jens and Gel{\ss}, Patrick and Hinterm{\"u}ller, Michael and Milbradt, Richard M. and Wille, Robert and Mendl, Christian B.},
  title     = {Large-scale stochastic simulation of open quantum systems},
  journal   = {Nature Communications},
  volume    = {16},
  pages     = {11074},
  year      = {2025},
  doi       = {10.1038/s41467-025-66846-x},
  url       = {https://doi.org/10.1038/s41467-025-66846-x}
}

@misc{mckay2023,
      title={Benchmarking Quantum Processor Performance at Scale}, 
      author={David C. McKay and Ian Hincks and Emily J. Pritchett and Malcolm Carroll and Luke C. G. Govia and Seth T. Merkel},
      year={2023},
      eprint={2311.05933},
      archivePrefix={arXiv},
      primaryClass={quant-ph},
      url={https://arxiv.org/abs/2311.05933}, 
}
